\documentclass{mn_plus_bib}
\usepackage{epsfig}

\def\etal{et al.\ }

\def\flh{5 \log_{10} h}

\newcommand\E[1]{~10^{#1}}

\renewcommand\[{\begin{equation}}
\renewcommand\]{\end{equation}}

\def\nbody{{\protect\it N}-body}

\def\muo{\relax \ifmmode{\mu_0}\else{$\mu_0$}\fi}
\def\mue{\relax \ifmmode{\mu_{\mathrm{e}}}\else{$\mu_{\mathrm{e}}$}\fi}
\def\re{\relax \ifmmode{r_{\mathrm{e}}}\else{$r_{\mathrm{e}}$}\fi}

\def\<{\langle}
\def\>{\rangle}
\def\lsim{\mathrel{\hbox{\rlap{\hbox{\lower4pt\hbox{$\sim$}}}\hbox{$<$}}}}
\def\gsim{\mathrel{\hbox{\rlap{\hbox{\lower4pt\hbox{$\sim$}}}\hbox{$>$}}}}
\def\mod#1{\vert #1 \vert}
\def\avg#1{\langle #1 \rangle}

\def\dex{{~\mathrm{dex}}}

\def\hmpc{h^{-1}\,{\mathrm{Mpc}}}

\def\mpc{{\mathrm{Mpc}}}
\def\gyr{{\mathrm{Gyr}}}
\def\kms{{\mathrm{km}\,\mathrm{s}^{-1}}}

\def\erg{{\mathrm{erg}}}

\def\IRAS{{\it IRAS}}

\def\Omegab{\Omega_B}
\def\Ome0{\Omega_0}
\def\Lam0{\Lambda_0}

\def\lbox{L_{\mathrm{BOX}}}

\def\eg{eg.\ }
\def\ie{ie.\ }

\def\gyr{{\mathrm{Gyr}}}
\def\h1{{\mathrm{H}{\scriptstyle I}}}

\def\kpc{\mathrm{kpc}}
\def\mic{\,\mu {\mathrm{m}}}

\def\mag{{\mathrm{~mag}}}

\def\AA{\hbox{\accent'27A}} 
\def\bband{$B$-band}

\def\kband{$K$-band}
\def\iband{$I$-band}

\def\Mhot{M_{\mathrm{hot}}}
\def\Mcold{M_{\mathrm{cold}}}
\def\rd{r_{\mathrm{D}}}
\def\rb{r_{\mathrm{B}}}
\def\rs{r_{\mathrm{S}}}
\def\tdyn{t_{\mathrm{dyn}}}
\def\LCDM{$\Lambda$CDM}
\def\Msun{M_{\odot}}
\def\Lsun{L_{\odot}}

\def\vc{V_{\mathrm{c}}}
\def\mgal{m_{\mathrm{gal}}}
\def\mhalo{m_{\mathrm{halo}}}

\def\vg{v_{\mathrm{gal}}}

\def\mhot{\Mhot}

\def\vesc{v_{\mathrm{esc}}}

\def\zsun{Z_{\odot}}

\def\tl{\tau_{\lambda}}
\def\al{a_{\lambda}}
\def\ol{\omega_{\lambda}}
\def\ei{\,\mathrm{E}_1}

\def\fbulge{f_{\mathrm{bulge}}}

\def\infinity{\infty}
\def\burstradfac{\kappa}

\def\galics{\textsc{galics}}
\def\errorbar{error bar}
\def\cD{cD}

\seceqnum      

\begin{document}

\title[\galics~\rm{I}]{GALICS I: A hybrid \nbody/semi-analytic model of hierarchical galaxy formation}
\author[Hatton et al.]{Steve Hatton$^{1}$, Julien E. G. Devriendt$^{1,2}$\thanks{devriend@iap.fr}, St\'{e}phane Ninin$^{1}$, Fran\c{c}ois R. Bouchet$^{1}$, \newauthor{Bruno Guiderdoni}$^{1}$, \& Didier Vibert$^{1}$\\
$^1$Institut d'Astrophysique de Paris, 98bis Boulevard Arago, 
75014 Paris, France.\\
$^2$Oxford University, Astrophysics, Keble Road, Oxford OX1 3RH, United Kingdom.
}

\maketitle

\begin{abstract}

This is the first paper of a series that describes the methods and 
basic results 
of the \galics\ model (for {\it Galaxies In Cosmological Simulations}).  
\galics\ is a hybrid model for hierarchical galaxy formation studies, 
combining the outputs of large cosmological \nbody\ simulations 
with simple, semi-analytic recipes to describe the fate of the baryons 
within dark matter halos.  
The simulations produce a detailed merging tree for the dark matter 
halos including complete knowledge of the statistical properties 
arising from the gravitational forces.   
We intend to predict the overall statistical properties of galaxies, with 
special emphasis on the panchromatic spectral energy distribution emitted by 
galaxies in the UV/optical and IR/submm wavelength ranges. 

In this paper, we outline the physically motivated assumptions and key free 
parameters that go into the model, comparing and contrasting with other 
parallel efforts.  We specifically illustrate the success of the model in 
comparison to several datasets, showing how it is able to predict the 
galaxy disc sizes, colours, luminosity functions from the ultraviolet to far 
infrared, the Tully--Fisher and Faber--Jackson relations,  
and the fundamental plane in the local universe.  We also identify certain 
areas where the model fails, or where the assumptions needed 
to succeed are at odds with observations, and pay special attention to understanding 
the effects of the finite resolution of the simulations on the predictions made.  
Other papers in this series will take advantage of different data sets available in the 
literature to extend the study of the limitations and predictive power 
of \galics, with particular emphasis put on high-redshift galaxies.

\end{abstract}

\begin{keywords}
galaxies: evolution  -- 
galaxies: formation 

\end{keywords}

\section{Introduction}
\label{sec:intro}

In the last five years, the discovery of the Cosmic Infrared Background 
\cite{Puget96,Guiderdoni97,Fixsen98,Hauser98} and the faint galaxy 
counts with ISO at $15\mic$ \cite{Elbaz_etal99} and $175\mic$ 
\cite{Puget99}, SCUBA at $850\mic$ \cite{Smail97} and MAMBO at 
$1.3$ mm \cite{Carilli00} have shown that about two-thirds of the 
luminosity budget of galaxies is emitted by dust in the IR/submm 
range.  Whilst the nature of the source that heats up the dust is 
still uncertain, it now seems increasingly plausible that the 
contribution of AGNs to heating is not dominant, and that most of 
the energy is powered by starbursts due to gas inflows triggered 
by close encounters and merging.  The IR/submm wavelength range is 
actually tracking the star formation rate history of the Universe 
more accurately than the UV range, with strong sensitivity to the 
merging phenomenon that is the signpost of hierarchical galaxy 
formation.  
The luminous and ultra-luminous infrared galaxies that contribute 
to the CIB are thought to be the progenitors of the bulges and 
elliptical galaxies in the local Universe.  
The goal of \galics\ is to get a consistent panchromatic 
description of this process and of the luminosity budget of galaxies 
as it appears, for instance, through the faint galaxy counts at 
various wavelengths in the optical, IR and submm.

Various pieces of work have converged to build up a consistent 
description of galaxy formation within the paradigm of hierarchical 
clustering.  Initial density perturbations are gravitationally 
amplified and collapse to form almost relaxed, virialized structures 
called dark matter halos. In all variants of the Cold Dark Matter 
scenario \cite{Peebles82,Blumenthal84}, smaller halos form first, and bigger 
halos form continuously from the collapse of smaller halos.  Gas 
radiates and cools down in the potential wells of the halos 
\cite{WR78}.  Halos have little angular momentum,
and dissipative collapse stops when the cold gas settles in 
rotationally-supported discs \cite{FE80,DSS97,MMW98}.  Star formation at 
galaxy scales can be reasonably described with Schmidt laws 
or more sophisticated recipes \cite{Kennicutt89}.  Models of 
spectrophotometric evolution of the stellar populations produce 
luminosities, spectra, and colours from the star formation rate 
histories and initial mass function \cite{Bruzual81,GRV87,BC93}, and 
are now extended to the modelling of dust thermal emission 
(\citeNP{Mazzei92,Silva98,DGS99}, hereafter DGS).   When they die, stars eject gas, 
heavy elements and energy into their environment.  The energy 
feedback heats up the remaining gas and can produce galactic winds in the 
shallower potential wells that deplete the galaxies and quench subsequent
star formation \cite{DS86}.  Finally, spheroids form from major 
mergers \cite{TT72,Kent81}.  If the spheroid can still be a centre for 
gas cooling, a new disc forms around this bulge.  As a result, 
the morphological and spectral types of galaxies are not fixed once for 
all, but rather evolve as star formation, gas accretion and merging occur.

These ingredients can be put together with some success within a fully 
semi-analytic model (hereafter SAM) that starts from the power 
spectrum of linear fluctuations, and follows the various processes 
right up to spectral energy distributions of stellar populations.  
For instance, under the assumption that each newly-collapsed peak 
produces a halo where a new galaxy forms, therefore neglecting the 
classical `cloud-in-cloud' problem, it is possible to reproduce at 
least qualitatively the main statistical properties of galaxies 
\cite{LS91,LCRVS93,GHBM98,DG00}.  The Extended Press--Schechter 
prescription (EPS, \citeNP{PS74,BKCE91,Bower91,LC93}) is a more 
efficient tool to describe gravitational collapse and estimate 
the merging history trees of dark matter halos.  The fate of the stars, 
gas and heavy elements can also be followed within the hierarchy of merging 
halos with an implementation of these ingredients in the EPS formalism 
\cite{WF91}.  However it is only through Monte--Carlo realizations of 
halo merging history trees \cite{KW93,LC94,SK99} that galaxy merging 
can be followed and hierarchical galaxy formation can be addressed.  
Implementing a simple recipe for dynamical friction of satellite galaxies 
in the potential wells of halos enabled \citeN{KWG93} and \citeN{KGW94} to 
follow the galaxy merging history.  Though the `block model' for hierarchical 
structure formation has been used for some time \cite{Cole94},  most studies 
now involve this type of random realization based on the EPS for the merging history
(\citeN{SP99}, hereafter SP99; \citeN{Cole00}, hereafter CLBF, and papers of these series).  In addition to 
the cosmological parameters ($H_0$, $\Omega_0$, $\Omega_\Lambda$, $\Omegab$, the 
shape of the power spectrum and its normalization $\sigma_8$), the 
semi-analytic method introduces a limited set of free parameters because 
some of the processes have to be addressed phenomenologically: in the 
most general case they can be reduced to a star formation efficiency, a stellar 
feedback efficiency for the ISM/IGM, and a parameter that 
describes our ignorance on the complicated merging processes.  
These parameters are determined by requiring the results to fit certain 
datasets, commonly including the \kband\ luminosity density in the 
local Universe, the number of dwarf galaxies (which are the most 
sensitive objects to stellar feedback) and the number of elliptical 
galaxies (which only form from major mergers in the simplest hierarchical 
scenario).  Once the free parameters are fixed, many predictions can be 
produced and compared to data.  

A large number of papers have been devoted to various aspects of galaxy 
formation and evolution, ranging from the Butcher--Oemler effect 
\cite{Kauffmann95}, the formation of discs and bulges \cite{Kauffmann96_bulge}, 
damped Lyman-$\alpha$ systems \cite{Kauffmann96_dla}, Lyman-break 
galaxies \cite{Baugh_etal98,SPF01}, and the parallel evolutions of quasars and galaxies 
\cite{KH00}.  

However, this approach still suffers from a number of shortcomings.  
First, even if the EPS agrees with \nbody\ simulations 
\cite{EFWD88,LC93,KW93,LC94,SLKD00}, it is clearly a limiting simplification 
of the complex dynamical processes that actually occur.  
The non-linear dynamics is computed with the top-hat model that assumes 
sphericity and homogeneity, and it overestimates the number of halos 
on galactic and group scales \cite{Gross_etal98}.  
The only pieces of spatial and dynamical information that are stored in 
the halo merging history trees are the virial radii and circular velocities.  
There is no information on the spatial distribution and peculiar velocities 
of halos.  As a consequence, the outputs of SAM cannot be used for synthesizing 
realistic mock catalogues and images that take into account spatial correlations, 
whereas there is an increasing need for these catalogues and images to analyse 
current and forthcoming observations and to test data processing techniques.

It is tantalizing to bypass some of these limits by using merging history 
trees produced from large cosmological \nbody\ simulations. The basic idea
is to get a description of the dark matter halo merging history trees,  
which can be computed accurately from simulations, 
and to keep the usual semi-analytic approach to model the more 
uncertain physics of baryons.  This model rests on the assumption that baryons do not alter significantly 
the dynamics of the dark matter, except on the smallest scales.  
Hence the name `hybrid model'.  The result is a more realistic merging 
history for halos, which necessarily reflects on galaxy formation and 
evolution.  The drawback is a loss of flexibility since the values of the 
cosmological parameters $H_0$, $\Omega_0$, and $\Omega_\Lambda$, as well as the 
choice of the power spectrum $P(k)$ and normalization $\sigma_8$, are 
built in the \nbody\ simulation.  However, the value $\Omega_B$, the 
physics of baryons, and the associated free parameters can be changed {\it
ad libitum}, very much as in classical semi-analytic models.  

The first attempts at a hybrid model were proposed by \citeNP{WDEF87} and 
\citeNP{Roukema97}.  
Less than ten snapshots of \nbody\ simulations were considered at 
that time, and this crude time resolution had a heavy impact on the results.  
Moreover, it is also obvious that any numerical simulation has a 
finite mass resolution, and that the unknown fate of baryons in systems below this 
threshold is going to propagate over the threshold in the picture of 
hierarchical clustering.  A partial solution of this problem is to use 
the spatial information of a numerical simulation, but to build halo 
merging history trees from Monte--Carlo realizations of the EPS 
\cite{KNS97,Benson99,Governato_etal01}.  Unfortunately, it becomes 
impossible to follow the evolution of galaxies backwards in the structures, 
and the merging trees are not consistent with the merging histories of the 
individual halos in the simulation.  
Only fully hybrid models keep this record and have fully-consistent 
merging history trees (\citeNP{Munich1}, hereafter KCDW, and following papers).
Moreover, only most recent hybrid models, taking advantage of very high resolution \nbody\ simulations 
to dynamically follow substructure within dark matter halos, are capable of 
accurately tracking properties of individual galaxies within clusters \cite{Springeletal01}. 

Here we propose the \galics\ model of hierarchical galaxy formation 
which is intended to provide a fully panchromatic description of the galaxy 
merging history, similar in spirit to that implemented by \citeNP{Granato_etal00}
in a pure semi-analytic model.  
For that purpose, we will follow chemical and 
spectrophotometric evolution in a consistent way, estimate dust extinction 
and radiation transfer, and compute spectral energy distributions of dust 
thermal emission, following the lines of the \textsc{stardust} model 
(DGS).  

Our main goals are to:
\begin{itemize}
\item present an original hybrid model that is entirely independent 
of previous attempts, and study its successes and failures compared to other models.  
\item present an overall view of galaxy evolution by producing a host of predictions 
from a `standard' reference model 
to compare with observed galaxy properties locally and at high redshift.  
\item produce a panchromatic picture, which is closely linked to the hierarchical 
formation of structure, as galaxy mergers are bright in the infrared.  
\item implement the effects of observational selection criteria that follow as closely  
as possible the actual observational processes.  
\item study the effect of mass and time resolution constraints on these predictions. 
\end{itemize}

This paper (the first in a series) proposes an overall presentation of 
the model and the basic predictions for those statistical properties of 
galaxies in the local Universe that can be more easily compared with other 
works.  In section~\ref{sec:halos} we describe the procedure that has 
been developed to find the halos and build the halo merging history trees 
from the series of output snapshots.  Section~\ref{sec:barycool} presents 
the `recipes' for cooling of hot gas in the halos.  In 
section~\ref{sec:galaxies}, we introduce a simple, standard implementation 
of the dissipative physics of baryons, and the construction of the galaxy 
merging history trees.  Section~\ref{sec:merge} describes the modelling 
of merging events, and how these events are assumed to drive galaxy 
morphologies.  In section~\ref{sec:spectra} we present our methods for 
determining galaxy luminosities, which are largely based on the 
\textsc{stardust} model of DGS.  Section~\ref{sec:freepara} briefly 
summarizes the few free parameters that go into the models.  In 
section~\ref{sec:results}, we show results for the $z=0$ properties of 
galaxies in our reference model, including sizes, colours, luminosity functions, 
and structural relations (Tully--Fisher, Faber--Jackson, Fundamental plane).  
In section~\ref{sec:resolu} we 
investigate in detail the effect that the finite spatial, mass and time 
resolution of our simulations has on the predictions of galaxy properties.  
Section~\ref{sec:discuss} presents a discussion of these first results.
Appendix~\ref{sec:appA} describes the parallelized treecode that we use for large 
cosmological \nbody\ simulations.   
 
Four other papers will complete this description of the model. In paper II 
(Devriendt \etal in preparation), we will explore the sensitivity of our 
reference model to changes in some of the modelling assumptions, and 
study the influence of variations in the choice 
of cosmology, recipes for the baryonic processes, and astrophysical free 
parameters.  This paper will also discuss the evolution of galaxy properties.  
In paper III (Blaizot \etal in preparation) we will focus more specifically
on predicted properties for Lyman-break galaxies at redshift 3.
In paper IV (Devriendt \etal in preparation), we will give 
faint galaxy counts, source counts, and angular correlation functions 
in the UV, optical, IR and submm ranges.  Finally, in paper 
V (Blaizot \etal in preparation), we will focus on the redshift distributions 
of different classes of galaxy, the redshift evolution of the spatial correlation 
function, $\xi(r)$, and statistical bias, $b$, and the influence of 
environment on galaxy properties.  Forthcoming papers of the series will address 
several other issues with an improved modelling of the baryonic processes.

\section{Dark Matter Halos}
\label{sec:halos}
We have used for the simulation a parallel tree-code written and 
optimized for the Cray T3E.  This code, based on hierarchical methods, 
is described in detail in \citeN{Ninin99}.  
We assume that, initially, the baryon 
density field traces that of the dark matter, and thus that galaxy 
formation occurs at local maxima of the underlying density distribution.  
For a given cosmology it can be shown that there is a certain turn-around 
density contrast, above which matter has collapsed into gravitationally 
bound systems which have separated from the expansion of the Universe.  
The precise density contrast at which to define a `virialized' halo 
is a matter of some debate \cite{White00}, and in general depends on cosmology.  
A physically sensible definition is that within the virial radius of 
an object the dark matter is virialized, and external to it material 
is infalling.  This is found to occur at around 200 times the critical 
density, regardless of cosmology.  It is this latter definition we will 
use in this work.  
We refer to regions attaining this overdensity as dark matter halos, 
and it is assumed that all galaxy formation processes take place within 
these halos, and furthermore that there is no communication between these 
halos.  

\subsection{Simulation details}

We outline the details of the treecode approach in appendix~\ref{sec:appA}.  
We have initially simulated a spatially flat, low-density model (\LCDM) with 
a cosmological constant, in a cube of comoving size $\lbox = 100\hmpc$.  The 
amplitude of the power spectrum was set by demanding an approximate agreement 
with the present day abundance of rich clusters \cite{ECF96}.  Initial 
conditions were obtained with the \textsc{grafic} code \cite{Bertschinger95}.
A logarithmic spacing in the expansion parameter, $a$, was used for the 
output file times: we finally had around 100 output files.  

The simulation contains $N_\mathrm{part} = 256^3$ particles.  We ran additional 
simulations at lower resolution ($N_\mathrm{part} = 128^3,64^3$), which we will 
use to test for convergence in the properties of our dark matter halos and 
galaxies (see section~\ref{sec:resolu}).  The full details of our standard 
simulation are summarised in table~\ref{tab:sim_params}.

\begin{table}
\begin{center}
\begin{tabular}{| @{\hspace{0.5cm}} l |c   @{\hspace{0.5cm}}|}
   \hline
    Box size $L$ [Mpc]                        & 150.0           \\
    Particle mass[$10^{11}\Msun$]             & 0.08272         \\
    Omega matter $\Omega_{0}$                 & 0.333           \\
    Omega lambda $\Omega_{\Lambda}$           & 0.667           \\
    Omega curvature $\Omega_\mathrm{c}$       & 0.0             \\
    Hubble parameter $h$                      & 0.667           \\
    Variance $\sigma_{8}$                     & 0.88            \\
    $\Gamma = \Omega_{0} h$                   & 0.22            \\        
    Initial redshift                          & 35.59           \\
    Number of timesteps                       & 19269           \\
    Number of outputs                         & 100             \\
    Spatial resolution (kpc)                  & 29.29           \\
    Work time ($10^3$ h)                      & 56.             \\
    \hline
\end{tabular}
\end{center}
\caption{Simulation parameters used in our standard (\LCDM, $N_\mathrm{part} = 256^3$)  
simulation.}
\label{tab:sim_params}
\end{table}

\subsection{Detection of the halos in the simulation}

Many algorithms exist for detecting halos of dark matter 
in \nbody\ simulations.  Examples include \textsc{denmax}, the 
spherical overdensity algorithm, and the `hop' algorithm.  
None of them, in spite of their complexity, has proved to be 
clearly superior to the simplest of all, the `friend-of-friend' (\textsc{fof})
algorithm \cite{DEFW}.  This algorithm links two particles in the same group 
if their distance is less than a certain linking length.  
\textsc{fof} is very simple, and has the advantage of depending on 
only one parameter, the linking length.  This is usually expressed 
as a fraction of the mean inter-particle separation.  Groups of particles 
are thus identified which are bound by a density contrast 
\[
\delta_\mathrm{thresh} \approx 3 / (2\pi b^3)
\]
For the case where the agglomeration is well-modelled by an isothermal 
sphere, the average density is three times the threshold density, so 
picking a link-length parameter $b=0.2$ identifies groups that are at 
average densities approximately 180 times the mean density \cite{CL96}.   
So, as a first step, we detect halos in every output timestep of the 
simulation with this algorithm and create a linked list of particles in 
each detected halo.  In practice, we allow $b$ to evolve with time such 
that the \textsc{fof} picks up structures at overdensity around 200 times 
the critical density.  It has been shown (KCDW) 
that the minimum size of groups that 
are actually dynamically stable in numerical simulations is 
around ten particles.  We employ, in our fiducial model, a more 
conservative minimum particle number of twenty.  This results in 
\textsc{fof} groups of minimum mass $1.65\E{11}\Msun$.  We find a 
total of $23\,000$ such groups in our final simulation output `snapshot', 
of which 3000 are not dynamically stable (see section~\ref{sec:falsies}).

\subsection{Properties}
\label{sec:halo_props}

\begin{figure*}
\centerline{\epsfxsize = 16.0 cm \epsfbox{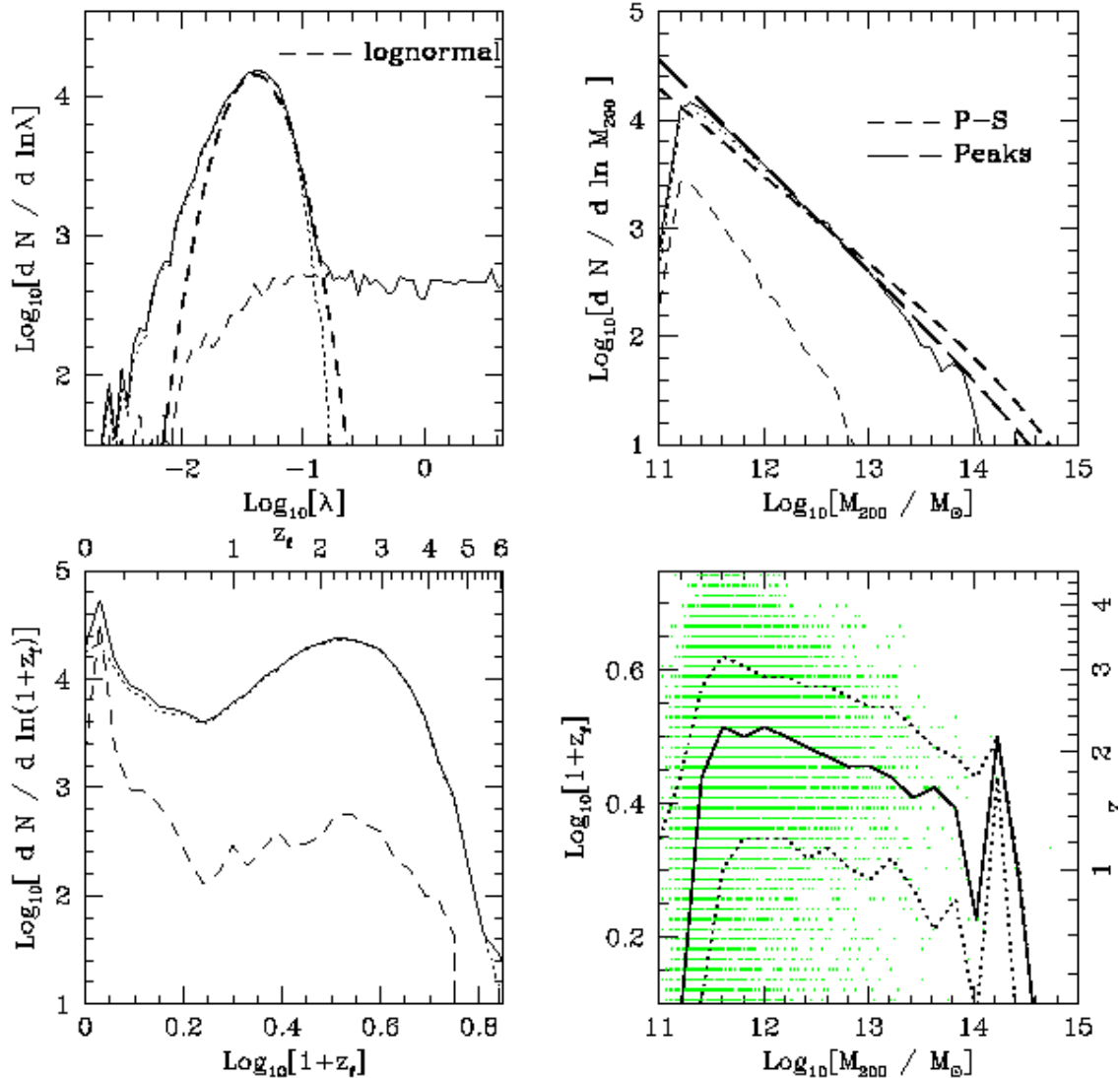}}
\caption{Distributions of several halo properties.  In each plot, the solid  
black line is for all groups identified by the friends-of-friends method, the 
dotted line is for halos with negative binding energy (\protect\ie bound objects), 
and the dashed line is for unbound systems.  
The top left panel shows the dimensionless spin parameter, compared to the 
empirical (log-normal) function of \protect\citeN{MMW98} (dashed line, 
with average $\bar{\lambda} = 0.04$, with arbitrary normalization).  
On the top right is the mass function (\ie distribution of halo virial masses), 
compared to Press--Schechter and Peaks model predictions (short- and long-dashed lines).  
Bottom left shows the formation time, defined in~\protect\ref{sec:tree}.  
Bottom right shows the dependence of formation time on the halo mass, showing the median 
and $\pm 1$-$\sigma$ spread of the distribution. }
\label{fig:halos_spin}
\end{figure*}

For each halo, we compute and record:  
\begin{itemize} 

\item the virial mass of the halo, $M_{200}$ which is assumed to be equal to the group mass, 
{\em i.e.} the number of particles multiplied by the particle mass.   

\item the position of the centre of mass of the halo, and its net 
velocity.

\item $a,b,c$, the principal axis lengths of the mass distribution.  

\item the thermal energy of the halo.  This is simply the sum of the squares of the 
particle velocities relative to the net velocity of the halo, multiplied by the halo mass.

\item the potential energy.  This is computationally expensive to measure, 
since it involves a sum over particle pairs.  In appendix~\ref{app:PE} 
we explain and justify our approximation for measuring this quantity. 

\item the virial radius $R_{200}$, which is deduced from the measured mass and 
potential energy, assuming a singular isothermal sphere density profile for the halo. 

\item the circular velocity of the halo, $V_{200}$.  
\[
V_{200}^2 = G M_{200} / R_{200}.
\]

\item the halo spin parameter, $\lambda$, defined by  
\[
\lambda = \frac{J \mod{E}^{1/2}}{G M^{5/2}_\mathrm{\textsc{fof}}}, 
\]
where $J,E$,and $M_\mathrm{\textsc{fof}}$ are the total angular momentum, energy and 
\textsc{fof} group mass respectively.  
\end{itemize}

\subsubsection{Halo spin distribution}

In the top left panel of Fig.~\ref{fig:halos_spin} we show the distribution 
of the spin parameter, for halos in our \LCDM, $N_\mathrm{part} = 256^3$ 
simulation.  We also plot the empirical lognormal distribution used by 
\citeN{MMW98}, which is found to be a good approximation to the distributions 
obtained from several other investigations (\eg \citeNP{BE87,CL96}).  Our 
distribution is slightly skewed relative to this model, with a tail down to 
low spin parameter, but it is peaked at $\lambda \approx 0.04$, close to the 
`fiducial' median value of $0.05$.  Our median value of $\lambda$ is $0.038$
for bound halos (see section~\ref{sec:falsies}), 
and the dispersion in $\ln \lambda$ is $0.52$.  This is similar to results in 
these other investigations.

\subsubsection{Halo mass function}

In the top right panel of Fig.~\ref{fig:halos_spin} we show the halo mass 
function, \ie the number of halos as a function of virial mass.  We compare 
this measurement with two analytic predictions.  The Press--Schechter 
approach \cite{PS74} uses a model for spherical collapse of initial 
perturbations to predict the final distribution of halo masses in a given 
cosmology.  This is plotted as a short-dashed line in Fig.~\ref{fig:halos_spin}.  

The peaks formalism of \citeN{BBKS} uses the general properties of Gaussian 
random fields, assuming initial peaks will be associated with future halos.  
Predictions for the number density of peaks, as a function of peak height
\cite{LS91,GHBM98}, 
thus translate to a number density of halos as a function of virial mass.  
The prediction from this model is shown by the long-dashed line in 
Fig.~\ref{fig:halos_spin}. 

It can be seen that neither these two models provides an 
accurate fit over the whole range of halo masses, with Press--Schechter overpredicting 
the number of massive halos, and the peaks formalism overpredicting the mass function 
at the light end, below $10^{12}$ solar masses.  

\subsection{Building the merger tree}

\label{sec:tree}

\begin{figure*}
\centerline{\epsfxsize = 15.0 cm \epsfbox{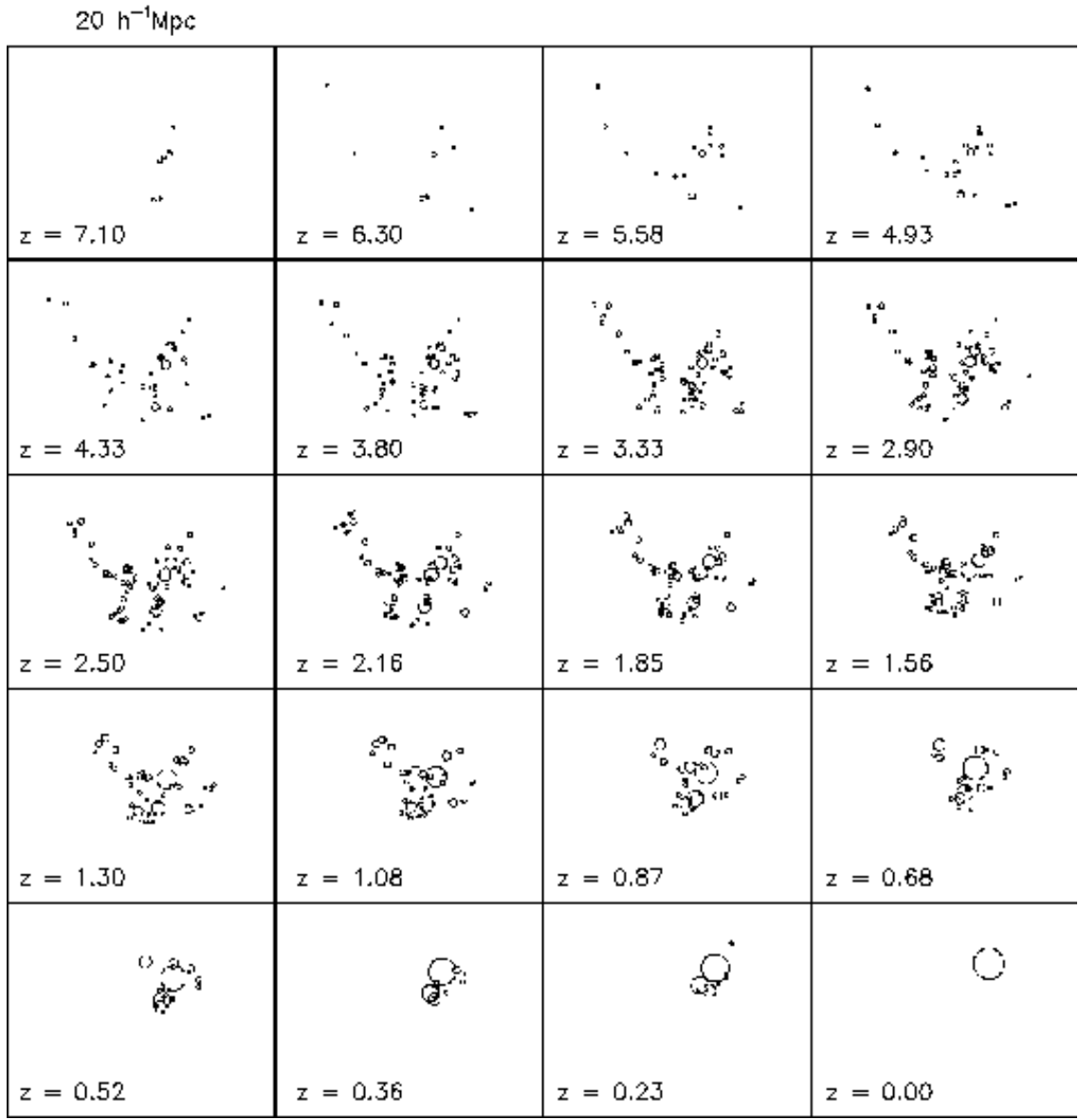}}
\caption{An example merging history for a dark matter halo with mass $1.9 \E{14}\Msun$ 
at redshift zero.  We show all the progenitors of the halo as circles with radii 
representing their virial radii on the same scale as the axes.}
\label{fig:halo_history}
\end{figure*}

In hierarchical models of structure formation, small objects form 
first, then merge together into more massive ones, or accrete matter 
from the background.  These processes can be represented with a tree,
the merging tree of the dark matter halos.
Our aim is to follow the evolution of halos detected with the group 
finder in the simulation outputs, and to save in a tree structure 
all the processes of accretion and merging between those objects, 
before subsequently modelling semi-analytically the evolution of 
baryons within these halos.  This hybrid approach has the advantage 
that it allow us to use all the spatial information from the simulations
as inputs to our models. 

Considering a halo $h_1$ at one timestep, we will identify a halo 
$h_2$ at the previous timestep as a progenitor if at least one 
particle is common to both halos.  

Having compiled a progenitor list for each halo, we compute the 
mass fraction of the halo coming from each progenitor.  This 
gives the basic structure of the tree.  This is a non-binary tree, 
in the sense that each halo can have several `sons'. 
In this general case, the sum of the masses of a halo's progenitors can be greater than 
the mass of the halo itself (fragmentation, evaporation, tidal stripping of a progenitor) 
and differs from the implementation of KCDW as each halo can have more than one descendent.  

It should be noted that an alternative method has been employed 
by other groups (SP99; CLBF).  
They have created a merger tree using random realizations 
of the Press--Schechter formalism.  
Their approach has theoretically infinite resolution, 
as the merging history can always be followed back to arbitrarily 
small progenitors.  
Our method, although resolution limited, has the advantage that 
we have full spatial information about the halos and their progenitors, 
and we can look at the `genuine' history of a halo rather than a 
statistical realization of that history.  

Having defined a list of progenitors for each halo, we can start to 
examine the history of how each halo acquired its present day mass.  
One particularly simple statistic that has been employed to parametrize 
this history is the formation redshift, $z_\mathrm{f}$ of the halo, 
defined as the time at which one half the halo's mass was present in a 
single object \cite{LC93}.  In practice this is calculated by recursively following 
the largest progenitor of the halo back through the tree, until it has 
one half the present-day mass.  If a halo has formed by fragmentation 
of another halo of mass greater than 
the current mass, we equate  the fragmentation time with the formation time.  
In the bottom left hand panel of  Fig.~\ref{fig:halos_spin} we plot 
the distribution of formation times, for halos in the final ($z=0$) 
output of our \LCDM\ simulation.

In the bottom right panel of this figure, we show the median and 
scatter ($\pm 1$-$\sigma$) of the formation time as a function of 
halo mass, for only those halos with negative binding energy.  
There is a clear trend that more massive halos tend to have formed 
more recently, as expected given the results of \citeN{LC93} and others.  
This relationship breaks down below 
$\approx 3 \E{11} \Msun$, showing that, although we resolve halos of this 
size, we cannot resolve their progenitors with half this mass, so the formation time 
becomes the time when the halo crossed the mass threshold to become detectable.  

In Fig.~\ref{fig:halo_history} we present the merging history of a typical 
massive halo in the final simulation output.  We show the locations and 
sizes of all the halo's progenitors in the preceding timesteps.  This 
clearly shows how virialized halos start to form in filaments, 
and then fall down these structures towards the centre, feeding the 
central massive halo.

\subsection{False halo identification}
\label{sec:falsies}
No group finder is totally accurate, as it is always possible 
to miss groups or to identify systems that are not in fact bound.  
Having determined the energy of our halos, we find that many, 
especially the lighter ones, have positive total energy, \ie they 
are not bound systems.  Leaving these halos in the sample can seriously 
skew the spin distribution, away from its usual log-normal 
behaviour.  This effect is illustrated in the top left panel 
of Fig.~\ref{fig:halos_spin}, where we plot the distribution for all the 
halos, and separately for those with positive and negative binding energies.  
It can be seen that excluding halos with positive energy immediately leads 
to an huge improvement in the symmetry of the distribution for high spins.

From the other panels of this figure, we can see that the mass function 
of these halos is steeper than that of `good' halos, and that the majority 
of them have extremely recent formation time.  
The recent formation time in particular suggests that the majority of halos 
with positive binding energies are transient collections of particles 
that have been misidentified as bound systems by the group finder.  
This could possibly be remedied by using a different group-finder (for 
instance \textsc{so}, the spherical overdensity algorithm), or by 
rejecting the `unbound' particles which contribute most to the halo 
thermal energy.   
 
However, it must be remembered that some of these halos could be 
made from pairs of similarly-sized objects that are currently going 
through a merger event and have not yet relaxed to virialization.  
In this case, we expect the kinetic energy to be rather higher than 
the virial theorem would predict, and the halo may have positive 
binding energy.  We would not want to reject these halos. 
Those objects with a longer formation time 
may be smaller halos in high-density regions that appear disrupted 
due to strong tidal forces, probably shortly before they merge 
with a more massive companion.  Clustering studies of these objects 
\cite{galics5} show that they are highly clustered, and concentrated 
in the densest regions of the simulation.

\subsection{Baryon tree}

Although we preserve all the information about particles shared, and hence can 
reconstruct the halo merging tree in full detail, for practical purposes we 
will be interested in following only the bulk of the dark matter.  
Two-body (resolution) effects on small scales imply that the exchange of one or 
two particles is not a guarantee that two halos are physically related -- in a 
higher resolution simulation, this exchange might not occur.  
For tracing the baryons and galaxies between timesteps, we will use a restricted 
version of the tree where we only consider merging between halos ({\em i.e.} objects 
with more than 20 particles). Ideally, one would like all the mergers to 
be binary (as in Monte-Carlo generated merging trees), because the order in which halos merge 
might slightly change the properties of the final halo. 
In practice, due to finite time resolution, it is not possible to avoid
multiple mergers between two outputs of the simulation, especially for the most massive halos at low redshifts.
However, we find that most of our halos do not undergo a merger 
between timesteps, and of the merger events that do take place, those where  
three or more progenitors can be identified in the preceding timestep are 
less numerous by a factor ten than the number of binary mergers \cite{Ninin99}.  
We will return to a more detailed study of time resolution effects in section~\ref{sec:tsres},
but in the mean time we assume, for simplicity, that halo mergers always take place 
exactly in the middle of two timesteps.

When considering a halo, we sort the list of all its progenitors in decreasing order
of contribution to its mass.  For each of these 
halos, we then see whether our halo is the main (\ie most massive) son.  If 
it is, it is assumed that the whole baryonic component (discussed in the next 
section) of the father is transferred to the son.
In this simplified 
description, then, each halo has zero (it just formed) or several progenitors, 
and zero or one descendent. 
We note that this simple tree is very similar to KCDW's as their halos are 
also allowed to have many progenitors but only one descendent.

\section{Baryon cooling in halos}
\label{sec:barycool}

Having compiled a list of halo properties based on their dark matter
content, 
and constructed a merger tree from the exchange of \nbody\ particles, 
we come to the point where we use semi-analytic recipes to add a 
baryon component to these dark matter halos.  The key assumption 
is that the overall behaviour of the matter distribution is 
determined  
by the dark matter density field, and that the baryon field 
thus represents only a small perturbation to the total density.  
Secondly, it is assumed that gravity is the only force operating 
between halos, so at a given timestep, each halo identified in the 
dark matter field can be considered completely independent from all 
other halos in the simulation.   
 
\subsection{Keeping track of baryons}
\label{sec:bary}
When a halo is first identified in the \nbody\ simulation, we assign 
it a mass of hot gas consistent with the primordial baryon 
fraction of the Universe, \ie 
\[
\mhot = M_{200} \frac{\Omegab}{\Omega_0}
\]
$\Omegab$ is a free parameter in our model, as the power spectrum 
used to construct the simulations has no dependence on the baryon fraction.  
For the results presented in this paper,  we adopt a value 
of $\Omegab = 0.02 h^{-2}$, suggested by observations of the deuterium 
abundance in QSO absorption lines \cite{TFB96}.  

As the halo subsequently accretes mass, we increase its stock 
of hot gas, assuming that the new mass accreted has the same 
primordial baryon fraction.  If the halo loses mass, for example 
through evaporation of dark matter due to two-body effects, or 
from fragmentation (when the halo's principal `son' is less 
massive than its progenitor), we discard some of its hot gas, based 
on the {\em current} baryon fraction of the hot gas in the halo.  

We will see in section~\ref{sec:feed} that galaxy formation processes 
can also lead to the ejection of gas and metals from the halo.  Whilst the 
physical processes are difficult to model, it seems reasonable to assume 
that a certain portion of this material will remain in the local environment 
of the halo, and may later be re-accreted.  
For this reason, we keep track of the mass of gas, metals and dark matter 
that have been lost from the halo through these processes, storing them 
in a halo `reservoir'.  It will be assumed that, when the halo subsequently 
accretes dark matter from the background, some of this 
dark matter will be `primordial' (\ie with baryon fraction $\Omegab$, and zero 
metallicity), and the rest will have the same baryon fraction and metal 
content as that of the halo reservoir, up to the point 
where the reservoir is fully used up.  We parametrize this effect 
with the halo recycling efficiency, $\zeta$, where $\zeta =0$ implies 
permanent loss of ejected material, and $\zeta=1$ is maximally efficient re-accretion, 
where all the reservoir is used up before the accreted dark matter is 
assumed to be primordial.  

There have been previous attempts to deal with this issue.  \citeN{Benson99} 
and KCDW both compare the standard model of CLBF in 
which gas is re-accreted after the parent halo has doubled in mass, with a 
model in which re-accretion is immediate (enriched gas can cool back onto the 
disk immediately after it has been ejected).  
Both of these models lack strong physical or observational support.  
SP99 adopt the extreme 
case where ejected material is forever lost from the halo.  This corresponds 
to our model with $\zeta =0$.  For the rest of this paper, we will adopt 
a standard model of $\zeta =0.3$.

\subsection{Cooling the hot gas}
\label{sec:cool}

At each time step we allow the hot intra-halo medium to cool 
onto a disc at the centre of each halo.  As described in 
section~\ref{sec:falsies}, there are some halos which have 
positive binding energy, either due to false identification by 
the group-finding algorithm, or extreme tidal disruption.  We 
do not allow cooling in these halos, or in halos with spin parameter 
$\lambda > 0.5$, the threshold for complete rotational support.  

The characteristic cooling time for the hot gas, as a function of radius, is given by
\[
t_\mathrm{c}(r) = \frac{3}{2} \frac{ \mu m_\mathrm{p} k T(r) }{ \rho_\mathrm{hot}(r) \Lambda(T)},  
\]
where $\mu m_\mathrm{p}$ is the molecular mass of the gas,
$\rho_\mathrm{hot}$ the hot gas mass density, and $\Lambda(T)$ is the 
cooling function, which depends on the metallicity and temperature 
of the gas.  We use the model of \citeN{SD93} to find $\Lambda$ for 
our halos.  
We will make the assumption that the gas has the same temperature 
throughout, which is known to be a good approximation for clusters, 
at least outside the central cooling flow region (\eg \citeNP{DJF96}).  
This isothermal approximation, applied to the equation 
of hydrostatic equilibrium, results in halo  temperature 
(in Kelvin); 
\[
T = \frac{\mu m_\mathrm{p}}{2 k} V_\mathrm{c}^2 = 35.9 \left (\frac{V_\mathrm{c}}{\kms}\right)^2.  
\]
This latter result comes from fixing $\mu$ by assuming 
the gas is totally ionized, consisting of 
one part helium to three parts hydrogen (by mass).  
We will also assume that the dark matter distribution follows an isothermal 
profile, in which case $V_\mathrm{c}$, the halo circular velocity, is identical 
to $V_{200}$, defined in section~\ref{sec:halo_props}.  

Given the cooling rate, we can calculate the mass of gas cooled 
during a time interval $\Delta t$:
\[
\label{eqn:mcool}
m_\mathrm{cool} = \int_{0}^{r_\mathrm{max}} \rho_\mathrm{hot}(r) d^3 r
\]
In equation \ref{eqn:mcool}, $r_\mathrm{max}$ is the minimum between the infall radius 
$r_\mathrm{inf}$, the cooling radius $r_\mathrm{cool}$, 
(defined respectively as the radii beyond which the gas has not had enough time to fall/cool onto 
the central galaxy in $\Delta t$), and the virial radius $R_{200}$.
We now need a model for the distribution of the hot gas in the halo. 
Observationally, measurements of profiles of X-ray gas are often 
translated into a $\beta$-model \cite{CFF76}, where the density follows 
\[
\rho_\mathrm{hot}(t) = \rho_0 (t) \left(1 + \frac{r^2}{r_c^2} \right)^{-3\beta/2}.  
\]
Several numerical studies \cite{ENF98,NFW_xray} suggest that 
clusters are well fit by a profile with $\beta = 2/3$, \ie a non-singular 
isothermal sphere with core radius $r_\mathrm{c}$. Here, we assume that the hot gas is in 
hydrostatic equilibrium within the virial radius of our isothermal dark matter haloes, which 
yields a hot gas distribution very similar to that of the $\beta$-model.
The central density $\rho_0$ of the gas is computed at each timestep by integrating
over the density profile given by the hydrostatic solution 
and equating the result to the current mass of hot gas in the halo (it is therefore time dependent). 
We can also define a 'time integrated cooling radius', 
$r_\mathrm{cool}^\mathrm{int}$, as 
the radius that would contain the current cold mass of gas, if all the gas (hot and cold) were 
distributed according to this profile. This radius is close to the one used in pure semi-analytic models
to compute the quantity of gas which has cooled since a halo has formed.

\subsection{Overcooling}
\label{sec:overcool}

A key problem with semi-analytic models is the 
overprediction of the bright end of the galaxy luminosity function.  
In the hierarchical framework, this can be interpreted as meaning that 
the central galaxies in massive dark matter halos are generally too
luminous, either because they have too much `fuel' for star formation, 
or because the merger rate in the halo is too high.  
A number of methods have been employed or suggested to remedy this situation:  
\begin{itemize}
\item The approach of KCDW has been to prevent gas from cooling 
altogether in halos with circular velocity greater than a certain limiting 
value. Unfortunately, this very simple assumption, introduced to account for the observational fact that
disc galaxies with rotational velocities in excess of $350 \mathrm{ km s^{-1}}$ are 
extremely rare, has no physical justification.  
\item CLBF point out that large halos are formed from small halos that have already undergone 
cooling, and thus they will lack the low entropy gas that cools most easily, 
so one should expect the hot gas profile to be rather broader than that 
of small halos, and cooling will be suppressed.  Although this model has 
some theoretical basis, there seems to be no observational evidence that 
large clusters have flatter gas density profiles.   
\item \citeN{NF95} point out that if the cooling time of some gas is 
longer than its freefall time to the centre of the halo, a cooling flow 
may occur, in which case the gas may remain in a compact cold cloud 
without forming stars, or it may form stars with a non-standard IMF, 
producing an excess of brown dwarfs that will not be visible.  
\item \citeN{WFN00} show that the energy in the hot X-ray gas is such that 
there needs to be significant non-gravitational (feedback, AGN) heating 
of this gas. In other words, if we do not include these heat sources 
we may be significantly underestimating the temperature from which our hot 
gas must cool.  However, it seems that the importance of this effect is likely to 
diminish with the mass of the halo, so this may not in itself solve the problem.  
\item Competitive cooling, whereby satellite galaxies can accrete cold gas as well as 
the central galaxy in a halo, is generally neglected in semi-analytic models, 
and would again help to prevent the assembly of over-luminous galaxies in the 
centres of halos. However, numerical simulations of satellites moving in 
a hot intra-cluster medium strongly suggest that RAM pressure stripping is
the dominant effect (\eg \citeN{QBB00}). 
\item SP99 point out that mergers of halos may shock heat the gas within, 
again resulting in the suppression of cooling onto the central galaxy.  

\end{itemize}

The failure to reach a consensus between these different approaches, and 
the absence of one well-motivated, observationally and theoretically 
justified model, must regrettably be regarded as one of the failings of 
the semi-analytic treatment, even though it probably finds its roots in a 
poor theoretical understanding of cooling flows. 
Therefore, in what follows, we simply decide to 
take advantage of the observed correlation between AGN and bulge mass
\cite{MTRB+al98}, and prevent gas from cooling in a halo when  
the bulge mass locked in its host galaxies becomes equal to 10$^{11}$ M$_\odot$.

We defer a more detailed modelling and investigation of the observational consequences
of some of the previously mentioned models 
to future work, as this is clearly beyond the scope of the present paper.

\section{Galaxies}
\label{sec:galaxies}
As hot gas cools and falls to the centre of its dark matter halo, 
it settles in a rotationally supported disc.  If the specific 
angular momentum of the accreted gas is conserved and starts off with 
the specific angular momentum of the dark matter halo \cite{MMW98}, 
we assume it forms an exponential disc with scale length $\rd$ given by:
\[
\label{eqn:disc_size}
\rd = \frac{\lambda}{\sqrt{2}} R_{200}.
\]
As gas builds up over a period of time to form a massive central disc
we recompute its scale length at each timestep, by
taking the mass-weighted average gas profile of the disc and that of
the new matter arriving, whose scale-length is also given by ~\ref{eqn:disc_size}. 
However, the half-mass radius of the disc (see definition below) is never allowed to exceed 
the virial radius of the halo, or to become smaller than the half-mass radius of the bulge, provided a bulge 
is present. We do not recompute the disc scale-length of galaxies which become satellites.
We also obtain the circular velocity of the disc, $V_c$ by enforcing rotational equilibrium  
at the disc half-mass radius, including all the mass enclosed within, {\em i.e.} baryons and dark matter
core. We assume that the DM core is depleted and/or flattened by a rapidly rotating disc as we would  
otherwise overestimate the amount of dark matter present within Milky-Way discs \cite{BE01}. More specifically
the fraction of dark matter within the half-mass radius of a pure disc $f_\mathrm{DM}$ is taken to be:
\[
\label{eqn:dmf_disc}
f_\mathrm{DM} = 1 - \left( \frac{V_c}{500 \; \mathrm{km/s} } \right)^2 \;,
\] 
so that a typical Milky-Way disc only has 80 \% of its DM core left.
This correction should be considered as a first attempt to model the complex issue 
of interaction between dark matter and baryons, which we plan to address in 
more detail in future work.  
Finally, we point out that this derivation of disc velocity differs from that of 
CLBF as these authors use a NFW profile for the density of the dark matter halo
and include fully self-consistent disc gravity.    

We will be concerned 
with a number of properties of galaxies, including their sizes, 
and we will generally use the radius 
containing one half the galaxy mass to define a size, since this 
is more physically relevant than the scale-length.  
For an exponential disc, the half-mass radius is given by 
\[
r_{1/2} = 1.68 \, \rd .  
\]

Galaxies remain pure discs if their disc is globally 
stable ({\em i.e.} $V_c < 0.7 \times V_{tot}$ where $V_{tot}$ is the 
circular velocity of the disk-bulge-halo system ; see e.g. \citeN{vdB98}), 
and they do not undergo a merger with another galaxy. 
In the case where the latter of these two events occurs, 
we employ a recipe which will 
be described in section~\ref{sec:merge} to distribute the stars 
and gas in the galaxy between three components in the resulting, 
post-merger galaxy, that is the disc, the bulge, and a starburst.  
In the case of a disc instability, we simply transfer the mass of gas and stars 
necessary to make the disc stable to the burst component, and compute the properties of 
the bulge/burst in a similar fashion as that described in section~\ref{sec:merge}
for galaxy mergers.
Bulges are assumed to have a density profile given by the \citeN{Hq90} 
model, 
\[
\rho(r) = \frac{M}{2\pi} \frac{\rb}{r(r+\rb)^3}.  
\] 
The half-mass radius is given by 
\[
r_{1/2} = (1+\sqrt{2}) \, \rb .  
\]
The bulges are assumed to be pressure supported with a characteristic 
velocity dispersion $\sigma$, computed at their half-mass radius.  
The material forming a starburst will be assumed to concentrate in a 
dense nucleus at the centre of the bulge.  We assume that its geometry 
is also modelled by Hernquist's equation, but the scale radius, $\rs$, 
is given by 
\[
\label{eqn:burstradfac}
\rs = \burstradfac \, \rb 
\]
where $\burstradfac$ is a free parameter to which 
we give the fiducial value $0.1$.  Note that gas does not cool onto bulges or 
starbursts: after a merger, a new disc is formed around the bulge 
by the cooling gas.  

Given a passively evolving galaxy, there are four processes that 
determine its make-up in terms of gas, stars, and metals, these being 
cooling, star formation, feedback and metallicity.  Although the 
metallicity of a galaxy varies more slowly, as it takes time for stars to 
process hydrogen into metals and release it back into the gas content, 
the other three can vary on quite short timescales, and are highly coupled.  
For this reason, we use an adaptive time resolution for calculating 
all quantities, breaking each simulation timestep (or each interval between 
mergers, whichever is shorter) into multiple substeps.  
Essentially, this fine resolution is a brute-force numerical way of solving the 
differential equations that relate cooling, star formation and feedback.

We now describe the interlinked processes responsible for making galaxies.

\subsection{Star formation}
\label{sec:SF}
At each timestep, a certain amount of the cold gas in a galaxy is allowed 
to form stars.  The rate of this star formation is given by 
\[
\Psi_\star = \frac{ \Mcold}{\beta \, \tdyn}.  
\label{eqn:SF}
\]
$\tdyn$ is the dynamical timescale of the galaxy, for which we use the time taken 
for material at the half-mass radius to reach either the opposite side of the galaxy or its 
center for a disc and a bulge respectively, and is given by:
\[
\tdyn = r_{1/2} \times \cases{ \pi \vc^{-1} & for discs \cr  
                               \sigma^{-1} & for bulges \cr}
\]
where $\vc$ is the circular velocity of the disc, and the material is assumed to 
have purely circular orbits, and $\sigma$ is the bulge velocity dispersion, where 
we assume the matter in the bulge has only radial orbits.  
$\beta^{-1}$ is the star formation efficiency.  Based on data for 
a sample of bright galaxies \cite{KTC94} it has been shown by \citeN{GHBM98} 
that a value $\beta \approx 50$ is capable of reproducing the mean Roberts 
times \cite{Roberts63} for discs.  We therefore use $\beta = 50$ as the fiducial value of this parameter.  
One can wonder why we define a star formation rate for the bulge as gas cannot cool on 
this component and stars form in the burst component during a merger. This is simply because 
stars eject gas and metals back in the ISM which 
can be trapped in the bulge triggering star formation there. 

As mentioned earlier, starbursts have the same geometry as bulges, so their 
star formation law is the same. The only difference is that the 
radius of the burst being smaller, the characteristic timescale for star formation  
is much shorter, giving birth to an almost instantaneous 'burst' of star formation.   
 
\subsection{Feedback}
\label{sec:feed}
For every mass of stars formed, a certain fraction 
will be contained in massive, fast-evolving stars that quickly become 
supernovae.  When a supernova occurs, some of the energy released 
will be imparted to the ISM, `blowing off' some of this gas.  If this 
ejected gas has high enough kinetic energy, it may even be blown out 
of the host halo altogether.  This process thus inhibits future star 
formation, and hence is termed feedback.   

The feedback is given by \cite{S03}: 
\[
\dot{m} = 2 \Psi_\star \frac{\epsilon_w \, \eta_\mathrm{SN} E_\mathrm{SN}}{\vesc^2}
\label{eqn:feed}
\]
where $\epsilon_w$ is the efficiency of the supernova-triggered wind which 
is proportional to $\vesc^2$ and depends both on the porosity of the ISM (see \citeN{S01} for details) 
and the mass--loading factor. This latter accounts for entrainment of interstellar gas 
by the wind and can be considered as a free parameter whose value is around 10 \cite{MKH02}.
This is modelled by our efficiency parameter $\epsilon$, and as a result, the outflow rate
given by equation~\ref{eqn:feed} for a starburst is of the order of the star formation rate.
Note that in the previous equation, $\eta_\mathrm{SN}$ is the number
of supernovae per unit star-forming mass, which is a prediction of the 
Initial Mass Function (IMF) chosen, and $E_\mathrm{SN}$ is the energy of a supernova, assumed 
to be $10^{51} \erg$.  
 
Equation~\ref{eqn:feed} is applied to find the fraction of 
gas in the ISM that is lost by the galaxy and ejected in the intra halo medium. 
We then equate the fraction of this gas that is completely ejected from the halo, 
to the galaxy/halo escape velocity ratio. The gas ejected from the halo is added 
to the halo reservoir where it may subsequently be accreted, as discussed in 
section~\ref{sec:bary}.   
 
The escape velocity is given by:   
\[
\vesc^2 = 2 v^2  \times \cases{ 1+\ln 2       &  for discs  \cr 
                                   2            &  for bulges \cr
                                 1+\ln(R_{200}/r) &  for halos  \cr}
\]
where $v$ is the characteristic velocity (\ie the circular velocity for 
discs and halos, the dispersion velocity for bulges), and, in the case of halo 
ejection, $r$ is either the galaxy's orbital radius 
in the halo, or, for the central galaxy in the halo, the half mass radius of the 
galaxy.  

\subsection{Metallicity}

The baryonic gas in halos and galaxies, initially composed solely 
of hydrogen and helium, acquires a metal content due to the processing 
of these light elements by the stellar population, and the subsequent 
release of the heavy elements synthesized to the inter-stellar medium at the end of 
a star's life.  
To track the metallicity, we need two components: a model for the 
rate at which metals are produced inside the stars, and a model for 
the amount of material released by a given stellar population.  

Using these models, we are able to use the tree to track the metals 
ejected by stars at each time step, over the whole merging history 
of the galaxy, and let new stars form out of the enriched 
gas, assuming instantaneous mixing in the ISM.  This is in contrast 
to the assumption of instantaneous recycling, the approach 
generally used in semi-analytic modelling (CLBF, KCDW, SP99).  
Correct modelling of the ejection is likely to be especially important for 
elements such as iron, nitrogen and possibly carbon, for which the production 
delay can be significant (\eg 1 Gyr for Fe produced by SN Ia).  

Given a mass $M$ of stars formed at some time $t_0$, we can calculate the current 
contribution to the stellar ejection during a timestep from $t_1$ to $t_2$ as 
\[
\Delta M_\star = - M \int_{m(t_1)}^{m(t_2)} [m - w(m)] \phi(m) dm
\] 
where $m(t)$ is the mass of a star having lifetime $t$, $w(m)$ is the mass 
of the remnant left after the star has died, and $\phi(m)$ is the IMF.   
Our fiducial model uses a Kennicutt IMF, and in Paper II we will explore the 
effects of using a Scalo or Salpeter IMF (see \citeNP{Kennicutt83,Salpeter55,Scalo86}).  
For all IMFs considered, we form no stars lighter than $0.1\Msun$ or heavier 
than $120 \Msun$.  

This formula can be extended to include the entire star formation history, 
producing a rate of increase in the gas mass due to ejection from the stellar 
population, 
\[
\label{eqn:yield}
{\mathcal E}(t) = \int_{m(t)}^\infinity \Psi_\star(t-t_m) [m - w(m)] \phi(m) dm
\]
where $t_m$ is the lifetime of a star of mass $m$ (\ie the inverse of $m(t)$).   

Equation~\ref{eqn:yield} can be adapted to predict the amount of metals produced, 
by making the replacement:
\[
[m - w(m)] \longrightarrow [m - w(m)] Z_\mathrm{cold}(t-t_m) +  m Y_Z(m)
\]
where the first term on the right hand side represents the re-introduction of 
the metals that were originally in the stars when they formed, and $Y_Z(m)$ 
is the fraction of the initial stellar mass transformed via stellar 
nucleosynthesis into metals, known as the stellar yield.  Our models for the 
remnant mass and the yield come from the \textsc{stardust} code of DGS, and 
are calculated self-consistently with the models for spectral evolution we 
will discuss in section~\ref{sec:spectra}.

Throughout this work, we assume chemical homogeneity (instantaneous mixing), 
such that outflows caused by feedback processes are assumed to have the same 
metallicity as the inter-stellar medium, though in reality the material in the outflow 
could be metal-enhanced \cite{Pagel98}.

\section{Merging}
\label{sec:merge}
In the hierarchical picture of structure formation, mergers 
are clearly crucial in understanding the properties of galaxies 
observed at the present day.  In our models, mergers are responsible 
for the formation of massive spheroids, and are assumed to 
trigger starburst activity.  The detailed modelling of the physics 
of individual mergers between galaxies is currently beyond the scope of what can 
be achieved in a cosmological simulation.  Only the highest resolution
\nbody\ simulations of \citeN{Springeletal01} are able to dynamically follow the merger 
of 0.2 $L_\star$ dark matter sub-halos within a single cluster.
Therefore, the best we can 
hope for is to approximately model the rate and effects of merging 
in a global sense.  To deal with the rate of merging, we model two 
effects, the gradual tendency of satellite galaxies to lose orbital 
energy and sink towards the centre of the cluster potential well, and 
the likelihood of occasional encounters between satellite galaxies in 
a cluster, based on probabilistic, cross-section arguments.  

\subsection{Halo mergers}

\label{sec:orb}

We identify mergers between dark matter halos using the halo tree, 
as described in section~\ref{sec:tree}.  When a merger occurs, the 
properties of the dark matter itself are obtained directly from the 
properties of the new halo in the \nbody\ simulation.  We apply 
recipes, however, to deal with the baryonic component.  Firstly, 
the ICM, \ie the hot cluster gas, of the progenitors are added 
together proportionally to the dark matter mass which ends up in the 
descendent halo (the constant of proportionality being the current 
baryon fraction of each progenitor), and given the virial 
temperature of the new halo.  Then, the properties of the two halo 
reservoirs are simply added.  Note that the new halo contains all the galaxies 
that were present in its progenitors (even if fragmentation occurs 
and the halo is less massive than all/some of its progenitors) and that the fraction 
of the progenitors' ICM which does not end up in the new halo is put in its reservoir, 
ensuring conservation of metals.  

When two halos merge (sometime in between two output times), there is a 
discontinuity between the centre of the new halo and the centres of mass 
of the progenitors (linearly extrapolated from their positions and 
velocities in the previous timestep).  We measure this `jumping' distance, 
$R_\mathrm{j}$, for each of the progenitors directly from the 
\nbody\ simulations and use it to assign each galaxy an orbital radius 
in the new halo, using the cosine law: 
\[
r_\mathrm{new} = \sqrt{r_\mathrm{old}^2 + R_\mathrm{j}^2 - 2 r_\mathrm{old} R_\mathrm{j} \cos \theta},  
\]
where $r_\mathrm{old}$ is the orbital radius of the galaxy in the progenitor 
halo and $r_\mathrm{new}$ is its orbital radius in the new halo.  
This assumes that the galaxy was initially at an angle $\theta$ to the 
vector joining the centres of the two halos.  We select $\cos\theta$ randomly 
from a flat probability distribution between $-1$ and $1$, taking account 
of the spherical symmetry.  

To illustrate how this works in practice, we consider two cases.  The first 
is a merger between a small halo and a much larger halo.  It is clear that 
the centre of mass of the large halo is perturbed only slightly by the encounter.  
$R_\mathrm{j}$ is thus small, and the galaxies that were previously in the large 
halo have $r_\mathrm{new} \approx r_\mathrm{old}$.  For the small halo, $R_\mathrm{j}$ 
is close to the virial radius of the large halo.  Thus $r_\mathrm{new} \approx R_\mathrm{j}$ 
for the galaxies that were in this halo, in other words they are placed close to the virial 
radius of the new halo.  
For a collision between equal mass halos, $R_\mathrm{j}$ is approximately the 
progenitor virial radius, so the central galaxies are placed at this distance, 
while non-central ones are placed randomly throughout the new halo.  

Any galaxy whose orbital radius after the merger is less than its own 
half-mass radius becomes the new central galaxy of the halo.  If there is more 
than one of these objects, they are merged with each other in order of 
descending mass.  It may well happen that there is no central galaxy 
after a merger, in which case we cool gas onto the galaxy closest to the 
halo centre.  
Although this scheme is really only a na\"{\i}ve geometrical model, we 
use it to approximately replicate the sort of scatter in galaxy positions 
we expect to see when halos collide.  It has the advantage that it 
reproduces the desired behaviour for the extreme cases of equal or 
very unequal mass mergers in a natural, continuous way without free parameters.   

\subsection{Dynamical friction}
\label{sec:dynfric}
Dynamical friction causes satellite galaxies to sink gradually to the 
centre of their host halo, resulting in a merger if there is already a 
galaxy at the centre.  After a merger between halos, the galaxies are 
given orbits as prescribed in section~\ref{sec:orb}.  In subsequent 
timesteps, the radii of the orbits are decreased by an amount obtained 
from the differential equation \cite{BT}:
\[
r \frac{dr}{dt} = -0.428 \frac{G m_\mathrm{sat}}{\vc} \ln \Lambda
\label{eqn:dynfric}
\]
where $r$ is the orbital radius, $m_\mathrm{sat}$ the mass of the satellite
(including its tidally stripped dark matter core, see next section), 
$\vc$ the circular velocity of the halo and $\ln \Lambda$ the 
Coulomb logarithm, approximated by 
\[
\Lambda = 1 + \left( \frac{\mhalo}{m_\mathrm{sat}} \right ) 
\]
(see \eg \citeNP{Mamon95}).  
Calculating the amount of orbital decay at each timestep seems far 
more natural than the approach favoured by other authors (CLBF, KCDW)
who assign each galaxy a dynamical friction timescale, and assume the merger 
occurs after that fixed time, unless a major merger occurs, at which point this 
timescale is recomputed.  By modelling the true radial coordinate, albeit in a 
na\"{\i}ve way, we automatically take into account possible evolution of the halo 
properties, and avoid having to differentiate between `major' and `minor' 
mergers.  
 
As soon as the orbital distance of a galaxy becomes lower than the sum of 
its half-mass radius and the half-mass radius of the central galaxy in the 
halo, it is assumed to merge with this central galaxy.  

Other workers (CLBF, KCDW, SP99) have taken into account the fact that 
galaxy orbits are not in general circular.  This fact alters the extent of 
dynamical friction, and \citeN{NFW_hierarch} have shown that applying the formula 
for circular orbits results an average overprediction by a factor two of the time 
taken for an infalling galaxy to reach the centre.  A proper consideration of this 
effect requires a modified form of the dynamical friction as a function of orbit 
eccentricity \cite{LC93}, and an assumption for the distribution of ellipticities.  
However, careful examination of \citeNP{NFW_hierarch} (especially their figure~8), 
shows that one can obtain the true dynamical friction time more accurately by 
simply halving the time for circular orbits, than by using the actual orbital 
eccentricity and applying the modified form of \citeNP{LC93}.   We thus 
take into account the effect of non-circular orbits by multiplying the 
right hand side of equation \ref{eqn:dynfric} by a factor two.

\subsection{Tidal stripping}

When a merger between halos occurs, the new satellite galaxies will, 
to some extent, retain the dark-matter cores of their previous host 
halos, and this dark matter will continue to dominate their dynamics.  
However, as the galaxy slowly descends to the centre of the new host 
this sub-halo will be stripped by tidal interactions.  Eventually, as the 
mass of the remnant becomes comparable to the baryonic mass of 
the galaxy itself, it will no-longer dominate over the effects of the 
self-gravity of the galaxy, and the dynamics may undergo a serious change.  
Following SP99, we assume that the sub-halo is stripped 
down to a radius $r_t$ such that $\rho_c(r_t) = \rho_h(r_0)$, where 
$\rho_c$ is the core density and $\rho_h(r_0)$ is the halo density
at the orbital radius of the remnant (for isothermal spheres this density 
is simply proportional to the mean density inside the same radius).  
Note that we do not attempt to model the stripping of 
baryons by tidal effects, only that of the dark matter.  

\subsection{Satellite-satellite mergers} 
\label{sec:ss_merge}
Satellite mergers are assumed to occur through direct collisions between 
galaxies in clusters.  In the absence of detailed numerical simulation 
of the precise positions and velocities of the galaxies within a halo, 
it is appropriate that some sort of cross-section argument be used to 
determine the merging rate.  There is no detailed study of how such 
cross-sections behave in a realistic halo environment; a first analytical estimate 
by \citeN{Mamon92} has been followed numerically by \citeN{MakinoHut97}, 
with \citeN{Mamon00} showing that these results are in good agreement.  
We assume the probability of a galaxy having a merger in time $\Delta t$ 
is given by:
\[
P = \frac{\Delta t}{\tau},
\]
where $\tau$ is the merger timescale, whose dependence on galaxy and halo 
characteristics is parametrized by \citeANP{MakinoHut97} as:
\[
\label{eqn:ss}
\tau^{-1} = \psi (N_\mathrm{g} -1) \left( \frac{r_{1/2}}{R_{200}} \right)^3 
\left( \frac{\vg}{\vc} \right)^3 \left( \frac{\vg}{r_{1/2}} \right).  
\]
In this formula, $N_\mathrm{g}$ is the number of galaxies in the halo, 
and $\vg$ is taken to be a mass-weighted average of the disc rotation speed and 
the bulge dispersion velocity. 
A direct transformation of the results of \citeN{MakinoHut97} leads to 
a value of $\psi =  0.017\, \mathrm{s} \gyr^{-1} \mathrm{km}^{-1} \mpc $, which is the value we shall take for our 
fiducial model.  However, their work assumes all galaxies 
are spheroids of the same size and mass, which is not the case in our models, 
and in Paper II we will investigate the effect of altering $\psi$ 
from this standard value. We are aware that our extrapolation of these authors' results 
is quite crude and should be, in the best of cases taken as a rough estimate
of the importance of satellite-satellite mergers. Finally we note that the results obtained 
by \citeN{Springeletal01} for sub-halo merging in dark matter clusters support the view that
this kind of mergers is negligible in such environments.

\subsection{Post-merger morphology}
\begin{figure}
\centerline{\epsfxsize = 8.0 cm \epsfbox{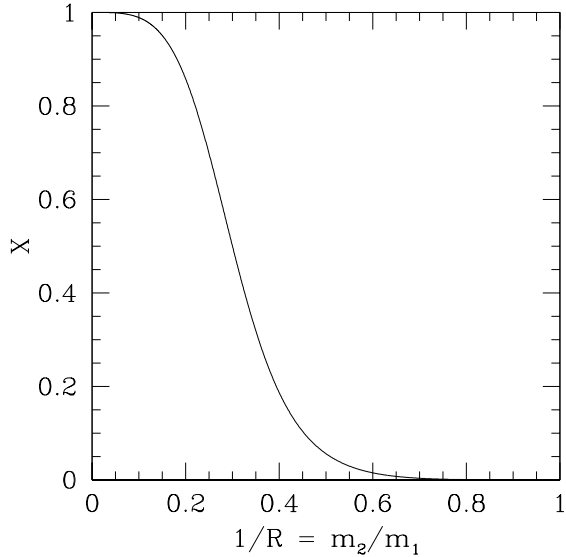}}
\caption{Our algorithm for calculating the fraction of disc 
material remaining in the disc after a merger, as a function of the mass ratio 
of the two progenitors.}
\label{fig:merge}
\end{figure}

In our model, mergers and disc instabilities are both responsible 
for creating galaxy bulges, and as such are drivers of morphological evolution.  
In the case of a disc instability, the mass transfer between galaxy
components has already been discussed in section~\ref{sec:galaxies}, and 
therefore we will only be concerned by mergers in this section.
The standard method for modelling a merger-driven morphology 
(SP99, CLBF, KCDW) 
has been to take the ratio of the masses of the two objects merging, 
and add the stars of the lighter galaxy to the disc of the heavier one 
if the mass ratio is less than $\fbulge \approx 0.3$, or destroy the 
disc and form a bulge if the ratio is higher.  

It has been shown \cite{WMH96} that a galactic disc can be 
completely disrupted by an encounter even when the interloper 
is rather less massive than the disc itself, and this simple 
method reproduces this behaviour.  However, it does not allow for 
any intermediate behaviour -- there is a sharp cut-off at 
$\fbulge$ between totally disrupting, or not disrupting, 
the morphology of the galaxy, which seems quite false.  

Our model is constructed in terms of the function $X$.  In general, 
this can be any function of the ratio of progenitor masses that maps 
the range one to infinity onto the space 0 to 1.  The step function 
used be other workers is one example of such a function, but we will 
consider a smoother function, drawing inspiration from Fermi-Dirac 
statistics, and defining 
\[
\label{eqn:fermidirac}
X(R) = \left[1 +  \left(\frac{\chi -1}{R-1}\right)^{\chi} \right] ^{-1}
\]

This function is shown in Fig.~\ref{fig:merge}, for the value 
$\chi = 3.333$.  In equation~\ref{eqn:fermidirac}, $R$ represents the 
mass ratio of the heavier to the lighter progenitor, and $\chi$ is 
the critical value of that ratio, \ie the value that gives $X = 0.5$.  
Since step functions with $\fbulge \approx 0.3$ have been found to give 
good results in the past, we set $\chi = 1 / 0.3$ as our fiducial value.  
In Paper II we will explore the effects of varying $\chi$.  

We then define two matrices, 
\[
\mathbf{\mathsf{A}}_\mathrm{star} = {\left( \begin{array}{ccc} 
 	 X   & 0 & 0 \\
 	 0   & 1 & 0 \\
 	 1-X & 0 & 1 	
                 \end{array}
 \right)} 
\]
\[
\mathbf{\mathsf{A}}_\mathrm{gas} = {\left( \begin{array}{ccc} 
 	 X   & 0   & 0 \\
 	 0   & 0   & 0 \\
 	 1-X & 1   & 1 	
                 \end{array}
 \right)} 
\]
Considering the vector $\mathbf{V} = (D,B,S)$ where $D,B,S$ refer to the mass 
in the disc, bulge and starburst respectively, we have: 
\[
\mathbf{V}^\mathrm{new}_\mathrm{gas} = \mathbf{\mathsf{A}}_\mathrm{gas} \mathbf{V}^1_\mathrm{gas} + \mathbf{\mathsf{A}}_\mathrm{gas} \mathbf{V}^2_\mathrm{gas}
\]
and similarly for the stars.  

In other words, during a merger a fraction $X$ of gas and stars originally sitting in the 
disc remain in the disc.  The rest of the gas from the disc goes into the starburst, as well 
as its stars.  Any stars that were already in the bulge stay in the bulge, 
but all its gas is put in the starburst.  
All the material (gas and stars) that was originally in a starburst remains in that starburst.  
Note that gas is never added to the bulge in this process, but that a small amount of gas 
is generally present in the bulges, coming from the evolution of its stellar content.  

\subsection{Properties of merger remnant}
Having ascertained how much mass goes where, we must decide 
on the properties of the resulting galaxy after the collision, 
\ie its rotation/dispersion speed and size.  For this, we adopt a similar 
model to that of CLBF.  

Under the virial theorem, the total internal energy is given by 
$E = - T$ where $T$ is the internal kinetic energy.  
Applying conservation of energy, 
\[
T_\mathrm{new} = T_1 + T_2 - E_\mathrm{int}, 
\]
where $E_\mathrm{int}$ is the interaction energy of the collision.  
For a dynamical friction encounter as well for satellite-satellite mergers, 
we use the total orbital energy, 
\[
E_\mathrm{int} = - \frac{G m_1 m_2 }{r_1 +r_2}
\] 
($r_1$ and $r_2$ are the half-mass radii of the two progenitors). Note that  
$m_1$ and $m_2$ stand for all the mass (dark matter core and baryons) contained
in each progenitor respectively. 

These energy considerations, coupled to mass conservation enable us to obtain $r_\mathrm{new}$, the 
half mass radius of the merger product:  
\[
r_\mathrm{new} = \frac{(m_1+m_2)^2}{ - \frac{E_\mathrm{int}}{0.4 G} + \frac{m_1^2}{r_1} + \frac{m_2^2}{r_2} },
\]
where the factor $0.4$ depends (not sensitively) on the exact density profile (Hernquist in our case).
As for the pure disc case, we do not allow $r_\mathrm{new}$ to become bigger than $R_{200}$.
We then assume that $r_\mathrm{new}$ is the half mass radius of the bulge and 
hypothesize that the largest disc is the most likely to survive the merger. Following  
the method presented in section~\ref{sec:galaxies}, we can calculate the characteristic 
velocities of our galaxy components. In particular, we use equation~\ref{eqn:burstradfac} 
to calculate the radius of the starburst after merging and assume the starburst 
has the same velocity dispersion as the bulge.  

In the case of a satellite-satellite merger, the orbital radius of the merger remnant is 
obtained through a simple mass-weighting of the orbital radii of the two progenitors.

\section{Galaxy luminosities}
\label{sec:spectra}

Having decided where the baryons are, and what their physical properties 
are, we apply a set of models to calculate the amount of light they 
produce.  Luminosities in visual and near-infrared (rest-frame) wavelengths 
are calculated from stellar synthesis models, which take into account the 
metallicity and age of the stellar population.  We then add geometry- and 
metallicity-dependent models for the absorption and re-emission of this 
light by the dust and gas in the inter-stellar medium.  

\subsection{Stellar Spectra}

For each time output of the simulation, we compute the stellar contribution 
$F^\star_\lambda (t)$  to the galactic flux at time $t$, which can be written:
\[
F^\star_\lambda (t)= \int^t_0 \int_{m=0}^{\infty} \Psi_\star(t-\tau)
\phi(m) f_\lambda(m,\tau,Z_\star) \mathrm{d}m \, \mathrm{d}\tau  \, \, ,
\]
where $f_\lambda(m,\tau,Z_\star)$ is the flux at wavelength 
$\lambda$ of a star with initial mass $m$, initial metallicity 
$Z_\star$, and age $\tau$ (\ie $\tau = 0$ corresponds to the zero age 
main sequence, and $f_\lambda(m,\tau,Z_\star) = 0$ if $\tau > t(m)$).  
$\phi(m)$ is once more the IMF.  We neglect the nebular component.  
For $f_\lambda(m,\tau,Z_\star)$ we use the \textsc{stardust} model of 
DGS, and full details can be found in that paper.

We use the full merging history of the galaxy to compute the stellar 
spectrum, summing up the contribution to the present day spectrum 
from all the stars formed in the previous timesteps in all the progenitors 
of the galaxy.

\subsection{Dust Absorption}

To estimate the stellar flux absorbed by the interstellar medium in a 
galaxy, one first needs to compute its optical depth.  
As in \citeNP{GRV87}, we assume that the mean perpendicular optical depth of 
a galaxy at wavelength $\lambda$ is:
\[
\tau_\lambda^{\mathrm{z}}  = \left( \frac{A_\lambda}{A_{\mathrm{V}}} \right)_{\zsun}
\left( \frac{Z_\mathrm{g}}{\zsun} \right)^s  \left( \frac{\avg{N_\mathrm{H}}}{2.1\E{21} \mathrm{atoms \, cm^{-2}}} 
 \right) \, ,
\label{eqtau}
\]
where the mean H column density (accounting for the presence of
helium) is given by:
\[
\avg{N_{\mathrm{H}}} = \frac{\Mcold} {1.4 m_p \pi (a r_{1/2})^2} \, \, {\mathrm{atoms \, cm^{-2}}} , 
\]
and $a$ is calculated such that the column density represents the average 
(mass-weighted) column density of the component, and is $1.68$ for discs,  
$1.02$ for bulges and starbursts.  The extinction curve depends on the gas 
metallicity $Z_\mathrm{g}$ according to power-law interpolations based on 
the Solar neighbourhood and the Large and Small Magellanic Clouds,
with $s = 1.35$ for $\lambda <2000 \AA$ and $s = 1.6$ for $\lambda >2000  \AA$ 
(see \citeNP{GRV87} for details).  The 
extinction curve for solar metallicity $({A_\lambda}/{A_{\mathrm{V}}})_{\zsun}$ 
is taken from \citeNP{Mathis83}.

For the spherical components we use the generalization given by \citeN{LDGB89} 
for the analytic formula giving obscuration as a function of optical depth
$\tau_\lambda^{\mathrm{sph}}$ \cite{O89} to the case where scattering is
taken into account via the dust albedo, $\omega_\lambda$ (we use the model 
of \citeNP{DL84} for the albedo): 
\[
A_\lambda (\tau) = -2.5 \log_{10} \left[ \frac{\al}{1 - \ol + \ol \al}\right],
\]
where
\[
\al(\tau) = \frac{3}{4\tl} \left [   1 - \frac{1}{2\tl^2} + \left (\frac{1}{\tl} 
+ \frac{1}{2\tl^2}\right ) \exp(-2\tl)  \right ].  
\]
We assume dust and stars are distributed together uniformely and thus 
there is no need to average over inclination angle, since the bulge is 
spherically symmetric.

For discs, the situation is more involved, due to inclination effects.  
We use a uniform slab model \cite{GRV87} for the extinction as a function of 
the inclination angle, $\theta$:
\[
A_\lambda(\tau,\theta) = -2.5\log_{10}\left[ \frac{1 - \exp(-\al \sec\theta)}
{\al \sec\theta}\right], 
\label{eqn:dust}
\]
where $\al = \sqrt{1 - \ol} \tl^z$.  
To calculate this quantity, we pick a random inclination of the 
galactic disc to the observer's line of sight.  
For comparisons of the same galaxy at different 
epochs, and for examining inclination-corrected statistics (\eg  
the Tully--Fisher relation), we also compute and store the face-on 
magnitudes of the discs ($\sec\theta = 1$).   
 
Our final result is the extinguished stellar spectrum:
\[
F_\lambda =F^\star_\lambda \times \dex [-0.4 A_\lambda].  
\]

\subsection{Dust Emission}

Dust absorption in bulges is spherically symmetric, so the emission is trivially 
calculated, but for discs it is anisotropic, so to calculate the total energy budget 
available for infrared emission by dust one must include contributions from all directions.  
We use the angle-averaged version of equation~\ref{eqn:dust}, 
\[
\bar{A}_\lambda(\tau) = -2.5\log_{10}\left[ \int_0^1 \frac{1 - \exp(-\al/x)}{\al/x} dx
 \right].  
\]
The value of the integral on the right hand side of this equation is: 
\[
\frac{1}{2\al}\left [ 1 + (\al -1) \exp(-\al)  - \al^2 \ei(\al) \right ], 
\]
where $\ei$ is the first-order exponential integral.

Given an unextinguished stellar spectrum $F^\star_\lambda$, the total bolometric 
infrared luminosity of the galaxy is:  
\[
L_\mathrm{IR} = \int F^\star_\lambda \times (1 - \dex [-0.4 \bar{A}_\lambda] ) \, d\lambda.  
\]

We assume the emission is isotropic, since dust itself is generally optically 
thin to infrared light (except in heavily obscured starbursts).  
To compute the IR emission spectrum, one needs to model both the 
size distribution and the chemical composition of dust grains in the ISM. 
The model we use is based upon the MW model of \citeN{DBP90}, which 
includes contributions from polycyclic aromatic hydrocarbons, very small 
grains and big grains.  Our chief refinement to this model is to allow a
second population of big grains, closer to the star forming region, to be
in thermal equilibrium at a higher temperature for galaxies undergoing
massive starbursts.  Our emission model makes use of the colour/luminosity  
correlations observed by {\em IRAS}, and is detailed in DGS.  
We stress that a key weakness of this model is that it is based on a local 
sample and therefore assumes that dust properties do not evolve with time.

\section{Free parameters}
\label{sec:freepara}
The semi-analytic recipes described here contain a number of 
free parameters that will affect our results.  We detail them 
in this section.   

\begin{itemize}
\item $\Omegab$ is the baryon fraction of the Universe.  Whilst a high 
baryon fraction can have a serious effect on both the shape of the initial power 
spectrum of density perturbations (\eg \citeNP{Sugiyama95}) and the internal dynamics 
of halos, providing it remains small it can be treated as a 
free parameter of the galaxy formation recipes.  Increasing $\Omegab$ 
results in more fuel for the star formation process, and should thus 
produce brighter galaxies.  

\item $\beta^{-1}$ is the star formation efficiency.  This is the parameter 
that appears in equation~\ref{eqn:SF}.  Increasing $\beta$ will clearly 
result in less cold gas being turned into stars, moving the peak in the 
star formation history closer to the present day.  

\item $\epsilon^{-1}$ represents the efficiency of mass--loading during 
the trigerring of a galactic wind by SN explosions (section~\ref{sec:feed}).  
Decreasing $\epsilon$ produces more feedback, heating more cold 
gas, ejecting more hot gas from halos, and thus reducing the amount of 
gas that can potentially form stars.  

\item $\psi$ is the normalization for satellite--satellite merging law, 
equation~\ref{eqn:ss}.  From \cite{MakinoHut97} we expect its value to 
be around $0.017$, but as stated in section~\ref{sec:ss_merge}, 
we will treat it as a free parameter.  
Satellite--satellite merging competes with dynamical friction merging (especially 
for massive galaxies, see \citeNP{Mamon00}), 
and energetics of these collisions are different, so increasing $\psi$ produces more 
diffuse galaxies, slowing the global star formation rate. 

\item $\chi$ parametrizes our merging law in equation~\ref{eqn:fermidirac}, 
it is the critical mass ratio, \ie the one at which half the disc is disrupted.  
Our fiducial value is $1/0.3$, raising $\chi$ means that a larger disc 
will be disrupted by a smaller interloper. It therefore affects the morphological mix
of our galaxy population. 

\item $\zeta$ is the efficiency with which gas stored in the halo reservoir 
is admitted back to the halo during accretion, as explained in section~\ref{sec:bary}.  
Turning $\zeta$ right down will mean that metals and gas ejected from a halo 
are only returned very slowly, restricting the supply of metals to the 
cooling gas.  

\item $\burstradfac$ is used to determine the starburst radius from the bulge radius, 
as shown in equation~\ref{eqn:burstradfac}.  Smaller 
$\burstradfac$ produces smaller starburst regions, where star formation is therefore 
faster, brighter, and more heavily extinguished.  

\end{itemize}

\begin{table}
\centering
\centerline{
\vbox {\halign {$\hfil#\hfil$&&\quad$\hfil#\hfil$\cr
\noalign{\hrule\vskip 0.03in}\cr
\noalign{\hrule\vskip 0.08in}
\hbox{name} & \hbox{meaning}  & \hbox{fiducial value} & \hbox{range}   \cr
\noalign{\vskip 0.03in}
\noalign{\hrule\vskip 0.08in}
\Omegab  & \hbox{Baryon fraction}                     & 0.02 \,h^{-2} & 0.01\hbox{--}0.03 \, h^{-2} \cr
\beta    & \hbox{Inverse star formation efficiency}   &  50     & 10\hbox{--}100    \cr
\epsilon & \hbox{Inverse of mass--loading for feedback}         &  0.1    & 0.01\hbox{--}1.0  \cr
\psi     & \hbox{S-S merging normalization} &  0.017  & 0.01\hbox{--}0.03  \cr
\chi     & \hbox{galaxy merger power-law}   &  3.333  & 2.0 \hbox{--}5.0   \cr
\zeta    & \hbox{recycling efficiency}      &  0.3    & 0.0\hbox{--}1.0   \cr
\burstradfac & \hbox{ratio of burst to bulge radius}& 0.1    & 0.05\hbox{--}0.2   \cr
\noalign{\vskip 0.08in}
\noalign{\hrule}
}}}
\caption{Summary of free parameters in the models}
\label{tab:free_params}
\end{table}

These parameters are summarised in table~\ref{tab:free_params}.  
In addition, there are various other inputs to the models, including 
cosmology, initial stellar mass function (our fiducial model uses the 
\citeN{Kennicutt83} IMF), and dust recipe, that will 
have an impact.  We will consider the impact of changing some  
of these inputs, and varying the above free parameters within the range
quoted in table~\ref{tab:free_params}, in paper II.

\section{Results}
\label{sec:results}

\begin{figure*}
\centerline{\epsfxsize = 16.0 cm \epsfbox{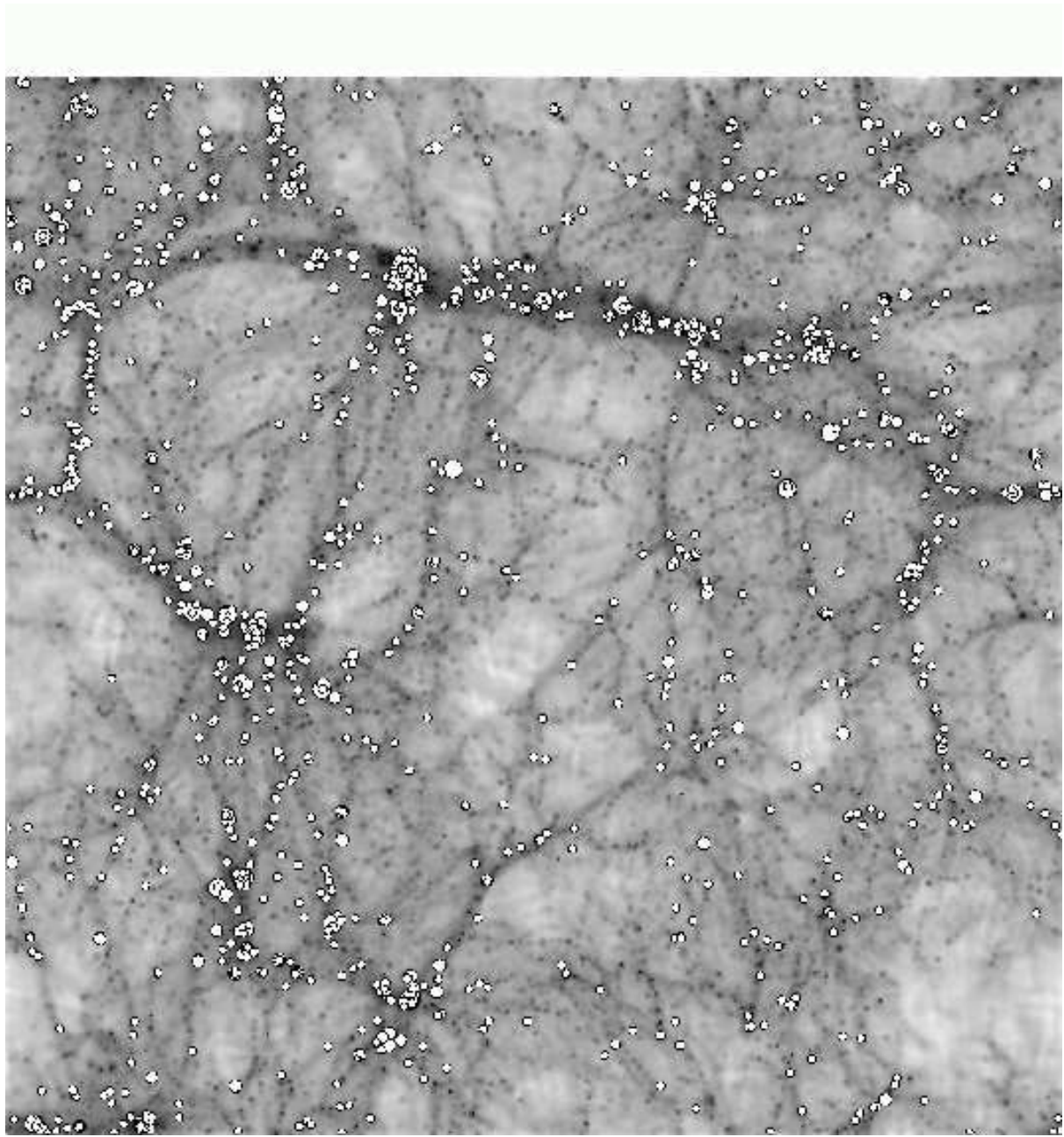}}
\caption{A slice through our simulation at redshift 0.  Dark matter 
density is represented by a grey-scale, whilst white circles mark galaxies, 
with their sizes depending on the \protect\bband\ luminosity.  Only bright 
galaxies ($M_B < -19.5$) are shown.  The slice 
is $100\hmpc$ on a side (the entire simulation cube) and $10\hmpc$ in thickness.  }
\label{fig:prettypic}
\end{figure*}

In order to test our models against observations, we will use a number 
of constraints in the literature.  These fall into two main categories; 
intrinsic properties, \ie relationships that apply for individual galaxies 
such as the Tully--Fisher relation or the dependence of metallicity on 
galaxy mass; and global properties, such as luminosity functions and 
clustering statistics.  In this paper we will just present work concerning 
the local (\ie low-redshift) Universe, and we will not use any spatial 
information.  Paper II will include results from higher redshift, and the 
effect on the results of changes in the parameters described in 
section~\ref{sec:freepara}.  Papers III  and IV will be concerned with the 
evolution of many of the statistics presented here, including the 
luminosity function and predictions of faint galaxy counts.  We will also 
employ the spatial information from the dark matter simulations 
to investigate the clustering properties of our galaxies in paper V.   

In Fig.~\ref{fig:prettypic} we present a slice through the simulation at its 
final output time, with 
the dark matter density shown as a grey-scale.  The slice is of side length 
$100\hmpc$, the whole simulation box, but is $10\hmpc$ in thickness.  
We overplot the locations of bright ($M_B < -19.5$) galaxies, with 
the sizes of the points representing the galaxy \bband\ magnitudes.   
One can clearly see the way that the galaxies trace the high-density 
regions of the simulation, with most of the light concentrated in filaments 
and clusters.  
The power of the hybrid approach is immediately apparent when one considers that 
state-of-the-art attempts to model the sub-grid physics with smoothed particle 
hydrodynamics produce around 2000 galaxies in a cube with one-third the volume 
of ours \cite{Pearce_etal00}, 
when we have $30\,000$ galaxies in total in our final timestep, thus giving us 
access to a much broader range of galaxy mass and merging history.  

\subsection{Effect of resolution}
\label{sec:res}

\begin{figure}
\centerline{\epsfxsize = 8.0 cm \epsfbox{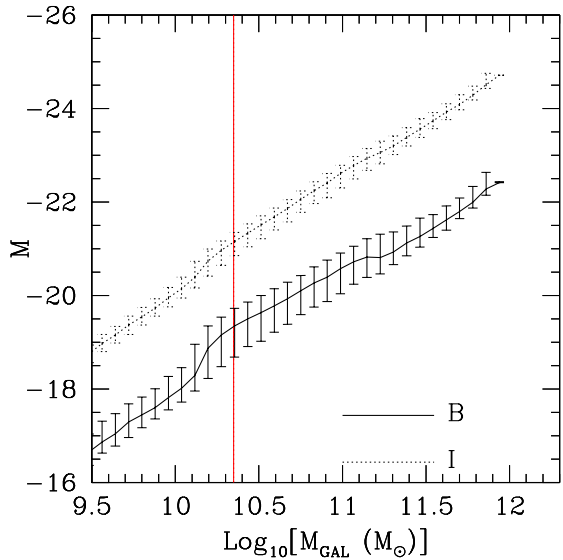}}
\caption{Galaxy absolute magnitudes as a function of total baryonic 
mass in the \iband\ and \bband.  The solid vertical line shows the 
effective completeness limit as defined in section~\protect\ref{sec:res}.  
Results for the completeness limit in a number of wavebands are presented 
in table~\protect\ref{tab:maglims}.}
\label{fig:resol}
\end{figure}

The identification of halos with a given cut-off mass produces an effective 
resolution limit on our results.  We define  the galaxy resolution limit 
as the smallest halo mass times the baryon fraction of the Universe.  
Galaxies of baryonic mass less than this exist in our models, but no galaxy 
of mass greater than this could exist in unresolved halos.  This is thus a 
`safe' limit for completeness 
in our sample, and produces a formal completeness limit 
of galaxy baryonic mass $\mgal = 2.2 \E{10}\Msun$ for the \LCDM\ simulation.  

\begin{table}
\centering
\centerline{
\vbox {\halign {$\hfil#\hfil$&&\quad$\hfil#\hfil$\cr
\noalign{\hrule\vskip 0.03in}\cr
\noalign{\hrule\vskip 0.08in}
\hbox{Waveband} & \hbox{Completeness limit}  \cr
\noalign{\vskip 0.08in\hrule\vskip 0.08in}
\mathrm{U} & -18.8 \mag\cr
\mathrm{B} & -18.9 \mag\cr
\mathrm{V} & -19.7 \mag\cr
\mathrm{R} & -20.4 \mag\cr
\mathrm{I} & -20.9 \mag\cr
\mathrm{J} & -21.6 \mag\cr
\mathrm{H} & -22.2 \mag\cr
\mathrm{K} & -22.7 \mag\cr
12\mic     & 0.41 \E{9} L_\odot\cr
25\mic     & 0.25 \E{9} L_\odot\cr
60\mic     & 0.79 \E{9} L_\odot\cr
100\mic    & 1.34 \E{9} L_\odot\cr
\noalign{\vskip 0.08in}
\noalign{\hrule}
}}}
\caption{Resolution limits in a variety of optical and infrared 
bands (magnitudes for Johnson--Mould filters, $\lambda L_\lambda$ in solar
luminosities for the \IRAS\ bands).   
Magnitudes are absolute, in the Vega system, and calculated 
using the process described in section~\protect\ref{sec:res}.}
\label{tab:maglims}
\end{table}

In Fig.~\ref{fig:resol} we show the dependence of galaxy luminosity 
on  total baryonic mass for galaxies in our fiducial model, defined 
in section~\ref{sec:freepara}.  The 
tight relationship between these two quantities allows us to 
estimate the magnitude completeness limit corresponding to the mass 
completeness limit we just defined.  For each waveband, we perform a  
linear fit of magnitude on luminosity, and take the intercept of this 
fit with the line  $\mgal = 2.2 \E{10}\Msun$.  The results of this treatment 
for a number of optical and near-infrared filters are presented in 
table~\ref{tab:maglims}.  It will be noted that, in the \bband\, 
the resolution limit is about 1.5 magnitude fainter than the 
value of $M_\star = -20.5$ (assuming our value of $H_0$) from the 
Stromlo--APM survey \cite{Loved92}.  We are therefore complete to 
galaxies around one-fourth as luminous as an $M_\star$ galaxy.  

Although we are effectively complete down to these magnitude limits, 
we do not fully resolve the merging history of the faintest objects.  
This is an advantage of the method employed by \citeN{Benson99}, 
where halos identified in a dark matter simulation have their 
merging histories computed with the Press--Schechter formalism with 
theoretically infinite resolution, so even the smallest objects have 
a fully resolved history.  The smallest galaxies in our models are 
in halos that have only recently been identified in the simulations, 
and hence have a high gas content and star formation rate, but low 
metal and stellar content, and an inevitable spiral morphology.  
We expect the effects of this lack of information to propagate 
over the formal mass completeness limit.  
We consider resolution effects in rather more detail in 
section~\ref{sec:resolu}.

\subsection{The halo ICM}

\begin{figure*}
\centerline{\epsfxsize = 16.0 cm \epsfbox{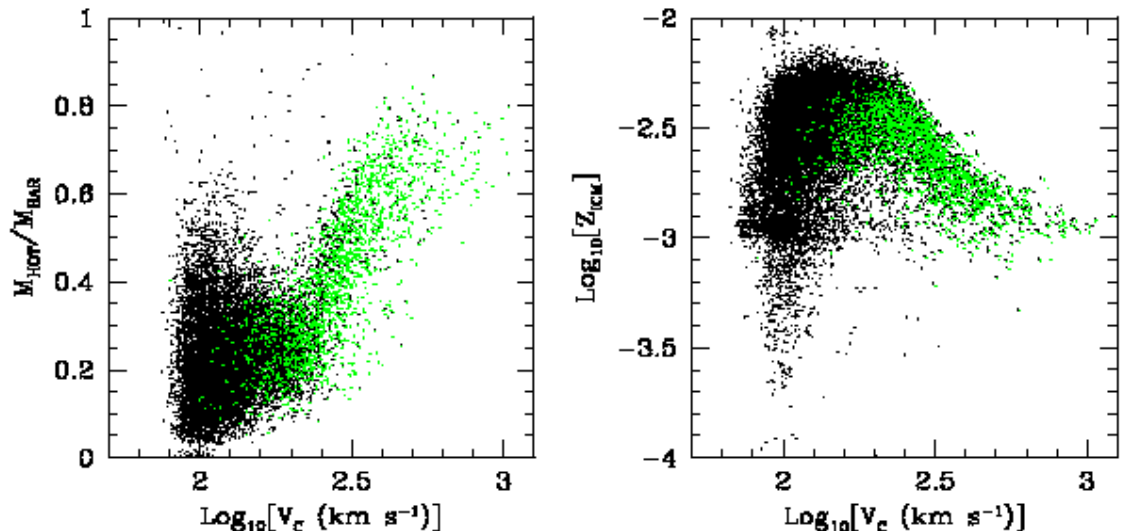} \vspace{-8.5cm}}
\caption{Baryonic properties of halos, as a function of halo circular velocity.  
On the left we show the ratio of hot gas in the halo to the total mass of baryons.  
On the right, the metallicity of this hot phase.}
\label{fig:vc_halos}
\end{figure*}

Before examining the galaxies themselves, we will look at the overall behaviour 
of the baryons in the dark matter halos.  In the left panel of Fig.~\ref{fig:vc_halos}, 
we show the ratio of hot gas to total baryonic mass for dark matter halos as a 
function of their circular velocity.  Note that for isothermal spheres there 
is a one-to-one mapping from circular velocity to halo dark-matter mass, $\vc \propto M^{1/3}$.  

We notice that: 
\begin{itemize}
\item the most massive objects have around seventy per cent of their baryonic mass in the form 
of hot intra-cluster medium (ICM), with a weak dependence on circular velocity.  
\item for objects with lower circular velocity, $\vc < 300 \; \kms$, the gas fraction 
starts to fall off rapidly as the halo mass decreases, this scale defining the 
difference between a galaxy and a cluster.  
\item finally, around $100 \; \kms$, the scatter becomes very large.  
This is due to the finite resolution of our approach, in that the 
smallest objects do not have well-resolved merging histories, 
so cooling does not start until the timestep they are discovered, resulting in an 
overestimate of the hot gas fraction.  The cooling in these 
objects is rapid enough, however, that by the time the mass has increased slightly, 
they have joined the main locus on the plot.
\end{itemize}

In the right hand panel of Fig.~\ref{fig:vc_halos} we show the behaviour 
of the metallicity (defined as the mass in metals relative to the  
mass in gas) of the hot ICM.  The resolution effects are clearly obvious again at 
low halo mass, with many objects of extremely low metallicity, since their stars are 
very young and have not had time to contribute via nucleosynthesis 
and ejection.  Above $100 \; \kms$, the metallicity is a decreasing function of halo 
mass, suggesting that larger halos have a higher proportion of primordial gas, 
and galactic winds are less important in clusters than in galaxies.  
This result is in contrast to the observational results of \citeN{Renzini99}, who finds 
for the largest clusters ($\vc > 1000 \; \kms$) that the metallicity, as measured 
from the iron abundance, has a constant value of 
around one third solar.  This is between five and ten times more metals than 
the amount we find in our halos.  

We will show in section~\ref{sec:physical} that our galaxy metallicities 
appear quite accurate.  Could the missing metals be hiding in the galaxies, or 
possibly in the halo reservoir?  We have done the same analysis as in Fig.~\ref{fig:vc_halos},
but added to the ICM all the metals contained in the halo reservoir, and all the 
metals locked up in the cold gas inside galaxies.  This thus represents the maximum 
metallicity we can attain without somehow lowering the mass of non-metals in the 
ICM.  Although this has a large effect for small halos, there is little change in 
the metallicity of the largest clusters.  We conclude 
that we are not simply `misplacing' the heavy elements.  

There are several potential explanations for this discrepancy: 
\begin{itemize}
\item observations measure only the abundance at the centre of the 
cluster, assuming the hot gas is chemically homogeneous.  In fact, 
metallicity gradients have been observed in many clusters (\eg 
\citeNP{KTT91} for the Virgo cluster, \citeNP{WDHH94} for several 
cooling flow clusters), and the observations thus yield an upper limit on the total 
metallicity. 
\item the metallicity is measured using the iron abundance, whereas the 
metallicity we quote is the total mass of metals in the cluster.  In fact, 
our models pay no attention to the energy injected to the ISM from type Ia 
supernovae (though these supernovae do contribute to the metallicity).  
The overall contribution of these objects is likely to be small, 
but, since these stars contain an iron core, the relative abundance 
of iron can be strongly enhanced, especially from evolved, gas-poor, early-type galaxies, 
such as are found in clusters.  Thus the simple assumption used to convert Fe abundance 
to total metallicity, $Z/Z_\odot = Z^\mathrm{Fe}/Z^\mathrm{Fe}_\odot$, may be invalid 
in clusters, and will produce an upper limit on the total metallicity.  
\item one should also consider the existence of population III stars in the early Universe 
which may have reionized the Universe homogeneously and likewise introduced metals 
into the primordial gas.  This could be simulated by allowing a non-zero primordial 
metal abundance in our models. 
\item the yields we use could be too low especially if the IMF is biased towards
massive stars in the early stages of cluster formation.
\item Although these effects could all play a role in the metallicity underestimate, 
it seems more likely that the major contribution we are unable to model comes 
from objects below our resolution limit. Indeed, it is sensible to argue that the majority of 
metals seen in the ICM today came from old dwarf galaxies \cite{Renzini99,Renzini2000}, 
in shallow potential wells, and we are a long way from achieving the resolution 
needed to see these galaxies.  
\end{itemize}

\subsection{Mass to light ratios}

\begin{figure}
\centerline{\epsfxsize = 8.0 cm \epsfbox{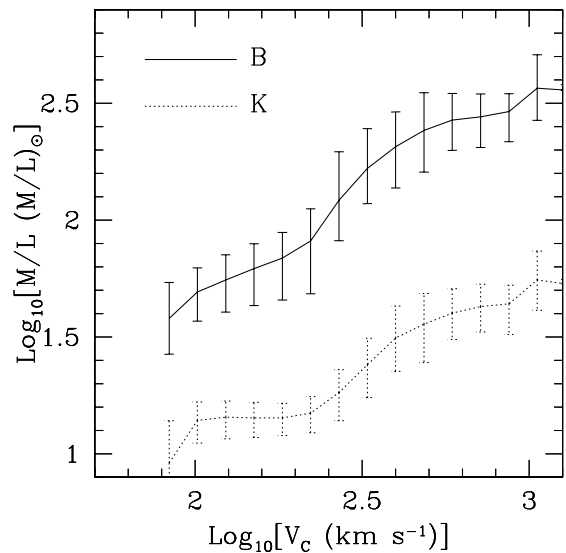}}
\caption{Mass to light ratios of our halos, in units of the solar 
mass to light ratio.  The upper line is for the \protect\bband, 
the lower for the \protect\kband.  The lines are the medians, with \errorbar s 
showing the $\pm 1$-$\sigma$ scatter.  We only include contributions from 
galaxies with baryonic mass greater than the formal completeness limit 
(section~\ref{sec:res}). }
\label{fig:m_over_l}
\end{figure}

The mass to light ratio of dark matter halos tells us important information about measuring the 
dark mass from observations, and about which environments are most efficient at
producing energy.  In Fig.~\ref{fig:m_over_l} we show mass to light ratios, in 
solar units, for our halos.  We only include contributions 
to the luminosity from galaxies above our resolution limit.  
We compare results in two wavebands, the \bband, 
where the energy is dominated by young stars, and the \kband, which has been 
shown to reliably estimate the total stellar content \cite{KC98}.  
One can see that the distributions are qualitatively similar in shape and that  
low mass, galactic halos are brighter per unit mass than the larger, 
cluster-sized objects.  In the \bband, MW type halos have a mass-to-light 
ratio of around 60, rising to 300 for the biggest clusters in the sample
whereas in the \kband the mass-to-light ratio is rather constant around 30 for galaxy size
halos ($V_c < 300 \; \kms$) and rises to 60 for the biggest objects.

Our failure to fully resolve the merging history of halos with circular 
velocity less than $100 \; \kms$ makes it difficult to draw any firm conclusions about the shape 
of the mass-to-light ratio at the low-mass end.  

\subsection{Assorted galaxy properties} 
\label{sec:physical}

\begin{figure*}
\centerline{\epsfxsize = 16.0 cm \epsfbox{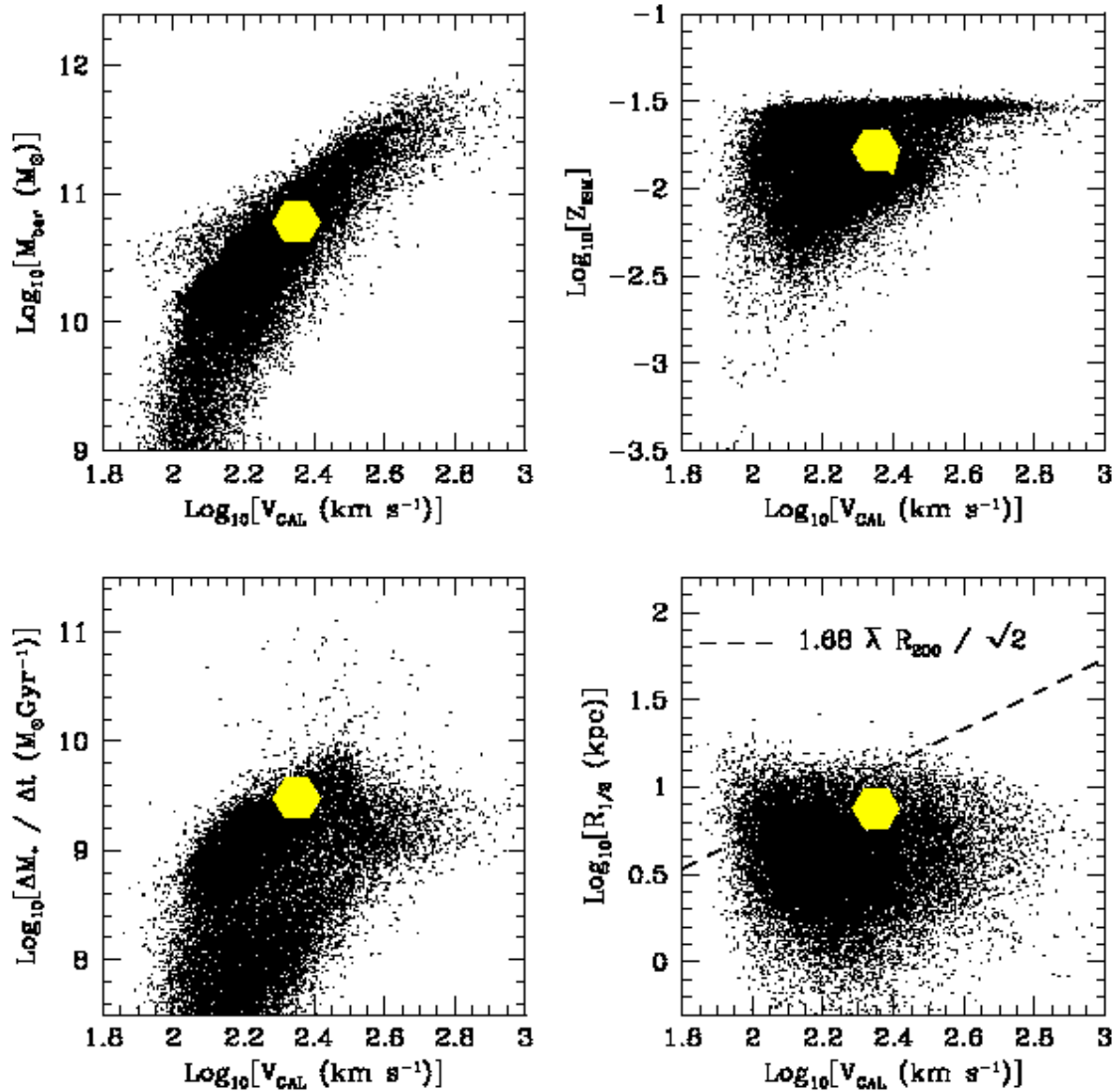}}
\caption{Key galaxy properties, as a function of characteristic velocity, which 
is a mass-weighted average of the disc circular velocity and the bulge dispersion 
velocity.  The top-left panel shows the total baryonic mass of the galaxy, \ie stars 
plus cold gas.  Top right is the metallicity of the interstellar medium, the ratio 
of the mass in heavy elements to the total gas mass.  In the lower-left panel we 
show the star formation rate, and in the lower-right the galaxy radius, defined as 
the mass-weighted average of the disc, bulge and burst radii.  For clarity, on this 
last figure we also plot the half mass radius of a disc forming in a halo with 
velocity $V_\mathrm{gal}$ if its spin parameter were equal to $0.04$ (equation~\protect\ref{eqn:disc_size}).  
In each panel, the grey hexagon shows 
the approximate location of the Milky Way, from data given in tables~\protect\ref{tab:mw1} 
and~\protect\ref{tab:mw2}.  }
\label{fig:vc_gals}
\end{figure*}

In Fig.~\ref{fig:vc_gals} we show scatter plots for some of the intrinsic properties 
of our galaxies as a function of galaxy characteristic velocity.  
This is the circular velocity for discs, three-dimensional dispersion velocity for 
bulges, and a mass-weighted average of the two for objects of mixed 
morphology.  

Firstly, we show the total mass of baryons, \ie cold gas plus stars, 
for our galaxies.  The main locus of these points is seen to approximately follow the relationship 
$M_\mathrm{bar} \propto V_\mathrm{GAL}^{5/2}$, although there is a substantial 
scatter to low baryon mass for galaxies with low velocity.  This is 
another example of resolution effects entering our results, in that 
these are galaxies in small halos that have not yet had time to cool.  
Compared to halos, whose virial masses go as the cube of the circular velocity, 
we already see a difference for galaxies: the baryonic mass depends  
more weakly on the velocity.  
We can already see the effects of the cooling and merging, in that the galaxy 
mass and velocity are obviously not scaling linearly with those of its parent halo.

Secondly, we present the metallicity of the cold gas in the galaxy.  This is
a slowly varying function of circular velocity.  One can contrast this 
with the results for the halo metallicity presented in Fig.\ref{fig:vc_halos}: 
although the  metallicity of the cluster ICM decreases with increasing 
cluster mass, the metallicity of the central object in the cluster increases.  
There is also a clear upper cut-off in metallicity around twice the solar value, which corresponds 
to galaxies having consumed almost all their gas and therefore having a very low, if any star formation rate at all, 
and the very few metal rich young stars produced fail to enrich the ISM much more.
It is worth noting that the main locus of our points is around a flat line with $Z \approx Z_\odot$,
indicating that for our mass resolution the metallicity of the Milky--way is quite typical.  

In the bottom left panel, we show the star formation rate as a function of 
characteristic velocity.  We note that the majority of galaxies lie close to 
the line defined by the star formation rate being proportional to the velocity 
squared, but that there is a small population with very high star formation rates 
of up to 100 solar masses per year, presumably representing galaxies which 
are undergoing merger-driven starbursts.  There is also considerable scatter at the low and high 
circular velocity end, due to galaxies that have exhausted their gas supply.  

We leave a detailed consideration of galaxy sizes 
(bottom right panel) to the next section.  

\subsubsection{Predictions for the Milky Way}
\begin{table}
\centering
\centerline{
\vbox {\halign {$\hfil#\hfil$&&\quad$\hfil#\hfil$\cr
\noalign{\hrule\vskip 0.03in}\cr
\noalign{\hrule\vskip 0.08in}
\hbox{property} & \hbox{actual MW}  & \hbox{reference} & \hbox{range adopted}    \cr
\noalign{\vskip 0.08in\hrule\vskip 0.08in}
m_\mathrm{gas}/m_\mathrm{bar} & 0.10     &  1 & \pm 0.05     \cr
M_K                           & -23.7\,\mathrm{mag}    &  2 & \pm 0.3 \,\mathrm{mag} \cr
V_\mathrm{c}                  & 220 \,\kms &  3 & \pm 20\, \kms \cr
\noalign{\vskip 0.08in}
\noalign{\hrule}
}}}
\caption{Observed parameters of the Milky Way, used to define a MW type galaxy in the simulation.  
The references are: 1.~\protect\citeNP{PA95}; 2.~\protect\citeNP{KDF91} ; 3.~\protect\citeNP{BT}.}
\label{tab:mw1}
\end{table}

\begin{table}
\centering
\centerline{
\vbox {\halign {$\hfil#\hfil$&&\quad$\hfil#\hfil$\cr
\noalign{\hrule\vskip 0.03in}\cr
\noalign{\hrule\vskip 0.08in}
\hbox{property} & \hbox{actual MW}  & \hbox{reference} & \hbox{Simulation}    \cr
\noalign{\vskip 0.08in\hrule\vskip 0.08in}
m_\mathrm{bar} & 0.6  \E{11}\Msun                     & 4 &  (0.5   \pm 0.1 ) \E{11} \Msun     \cr
\Psi_{\star}   & 2\hbox{--}4 \, \Msun \mathrm{yr}^{-1} & 1 &  (1.5   \pm 0.4  ) \, \Msun \mathrm{yr}^{-1} \cr
r_\mathrm{D}   & 4\hbox{--}5 \, \mathrm{kpc}           & 2 &  (2.8   \pm 0.6  ) \, \mathrm{kpc}  \cr
Z_\mathrm{ISM} & 0.016                                & 5 &  (0.012 \pm 0.002) \cr
\noalign{\vskip 0.08in}
\noalign{\hrule}
}}}
\caption{Observed properties of the Milky Way compared with our predictions 
from a set of galaxies derived by applying the conditions presented 
in table~\protect\ref{tab:mw1}.  The references are as that table except:  
4.~\protect\citeNP{BE01} and 5.~\protect\citeNP{Pagel98}.  The uncertainties in the fourth column are the 
$\pm 1$-$\sigma$ limits of the distributions.}
\label{tab:mw2}
\end{table}

To provide a consistency check for our models, we will compare 
MW-like galaxies in our simulations with the known properties
of our galaxy.  On each panel of Fig.~\ref{fig:vc_gals} we show a grey 
hexagon for the approximate location of the Milky Way itself.  The 
values adopted for the Milky Way are presented in tables~\ref{tab:mw1} 
and~\ref{tab:mw2}.  We also select MW candidates from our galaxies by 
applying the selection criteria shown in table~\ref{tab:mw1}, and 
requiring that the galaxy have spiral (or Sb) morphology (defined in 
section~\ref{sec:morph}). 
Selection of this kind produces around six hundred MW galaxies in the simulation.   
The properties of these galaxies are shown in table~\ref{tab:mw2}, where we 
quote the median and one standard deviation of the distributions, and compare 
to data for the Milky Way itself.  We find a generally good agreement, in that 
applying the selection of table~\ref{tab:mw1} produce galaxies that do look a 
lot like the Milky Way.  For most of the properties considered, the Milky Way 
would fall within $1$- or $2$-$\sigma$ from the median.  

\subsection{Galaxy sizes}

In the final panel of Fig.~\ref{fig:vc_gals}, we show the galaxy half-mass 
radius.  This is obtained from a mass-weighted average of the disc and bulge 
half mass radii.  The straight line is the radius of a galactic disc with the 
same velocity as its parent halo, if the radius is given by $1.68 \; \bar{\lambda} 
R_{200} / \sqrt{2}$, \ie equation~\ref{eqn:disc_size} with $\bar{\lambda} = 0.04$.

\begin{figure}
\centerline{\epsfxsize = 8.0 cm \epsfbox{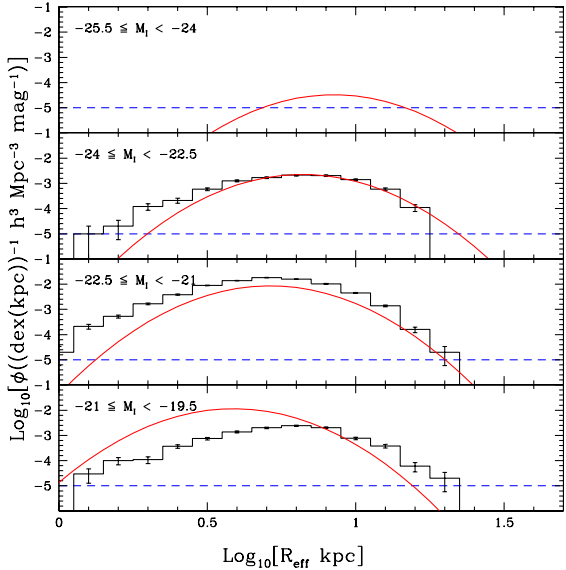} }
\caption{Bivariate density distribution of field spiral galaxies as a function of effective disc radii 
in different I-band absolute magnitude bins. Magnitude cuts are indicated in the top corner
of each panel. The solid histogram represents results from the hybrid model, and 
the solid line the best fit to the data of de Jong and Lacey~\protect\citeyear{DJL00} (their
eq. 7). The horizontal dashed lines in each panel represent our volume resolution.}
\label{fig:bivar}
\end{figure}

In Fig.~\ref{fig:bivar} we examine the sizes of field disc galaxies (\ie galaxies which are alone in their 
halos) emerging from our model.  We compare our predictions with the observed bivariate space density 
distribution of de Jong and Lacey~\protect\citeyear{DJL00} derived from a sample containing more 
than a thousand field spiral galaxies. As stressed by these authors, 
the main advantage of using this data is that it should be quite insensitive to surface brightness incompleteness effects.
Note that in the model we included all the spirals with bulge-to-total-light-ratio
less than 0.33, {\em i.e.} morphological types later than Sab, whereas the data does not 
contain types later than Sd. Indeed these very late type galaxies cannot be excluded from the model sample as 
the only morphological information we are able to compute is a bulge to disc ratio.
We would therefore expect an overprediction of the total number of galaxies, especially in the faint
magnitude bins and for small disc sizes. However, we do not see this excess of small faint galaxies but 
rather an excess of bright small galaxies with respect to the data.  
Moreover, as pointed out by \citeN{DJL00} the model distribution is (only slightly in our case) broader 
than the observed one.
As we assume specific angular momentum conservation during disc formation, the obvious culprit
is the too broad initial spin distribution of the dark matter halos.
However, we would like to draw the attention of the reader to the fact that unlike 
the CLBF model which overpredicts quite a lot the density of small discs, we seem to predict about 
the right number of small-sized bright galaxies. Although our mass resolution biases us against 
small-sized galaxies, we estimate (see table \ref{tab:maglims}) that our sample is almost complete for galaxies 
with absolute magnitudes brighter than -20.9 in the I band. Therefore the results displayed in the
first three panels of figure~\ref{fig:bivar} should not be too sensitive to an increase in resolution.  

Once again we emphasize that our recipe for disc sizes depends crucially on the assumption of 
specific angular momentum conservation, and it has been shown (eg. \citeNP{NFW_hierarch}) that
disspative effects can play a part in destroying some of the angular momentum of the gas, producing
smaller discs. \citeN{DTS98} show that this loss is less important for gas cooling onto 'evolved'
galaxies with stellar bulges, so these effects are likely to preferentially reduce the size of the smaller,
less resolved galaxies in our model.

\subsection{Morphologies}
\label{sec:morph}
We predict galaxy morphologies based on the ratio of \bband\ luminosities 
of the disc and the bulge components.  It has been found \cite{SdV86} that 
this ratio correlates well with Hubble type.  
We define a morphology index, $I = \exp(-L_\mathrm{B}/L_\mathrm{D})$, such 
that a pure bulge has value $0$, a pure disc $1$.  Following 
\citeN{Baugh_hub96}, we translate this index into a morphology 
by assuming ellipticals have $I < 0.219$, S0s have $0.219 < I < 0.507$, 
and spirals have $I > 0.507$.  

We first consider the morphological mix of galaxies in our simulation.  
Using the APM Bright Galaxy Catalogue (BGC, \citeNP{BGC96}), \citeN{Baugh_hub96} point out that 
in the local Universe, the ratio $\mathrm{E:S0:SP+Irr} = 13:20:67$.  
In fact, one cannot directly compare this to the ratio arising in our 
simulations, since the catalogue is a flux-limited sample of the Universe, 
whereas our samples are effectively volume limited, containing an entire 
simulation cube at $z=0$.  If there is any 
correlation between \bband\ luminosity and morphological type, 
some types may be seen over a larger volume in the BGC than others.  

For a better comparison, we take the Stromlo--APM 
redshift survey \cite{Loved96b} and construct from it a volume limited sample 
of galaxies containing all those with well-measured morphology, down to the 
absolute magnitude limit of our models, $M_\mathrm{B} < -18.8$.  We obtain 
$\mathrm{E:S0:SP+Irr} = 13:11:76$ (the sample contains just 120 galaxies, 
so the Poisson errors are $\pm 3:3:8$).  
Thus it appears that, since S0's are 
intrinsically brighter objects than spirals, the ratio from the BGC 
overestimates their abundance when this selection is made.
We compare to the ratio found for our models with a similar magnitude limit.  
We find $17:16:67$ for a sample of twenty thousand galaxies, with a total number density 
around twenty per cent higher than that measured in the volume limited Stromlo--APM sample.  
Thus it seems that for these galaxies we slightly over-represent Es and S0s and under-represent 
spirals, although it must be remembered that the statistics from the real data are poor.  

If we apply a more 
conservative limit, going $0.75$ magnitudes brighter (\ie a factor two in luminosity), 
we find the Stromlo--APM data give $14:13:73$ (250 galaxies in the sample, since the volume probed 
has increased, which implies Poisson errors $2:2:5$), and $21:17:61$ from our models.  
Again our number density is around twenty per cent higher than in the real data.
In other words, as we increase the absolute magnitude of the selected objects,  
the quantitative comparison with the data remains quite stable.    		
This effect is unsurprising, since in section~\ref{sec:res} we 
stated that our magnitude for completion in the B-band is around -18.9.

It is worth noting that, with more complete surveys such as 2dF and SDSS, 
volume limited samples of the local Universe with much better statistics 
will be available, and these numbers can be checked more accurately.  
As it is, we are thus broadly confident that our methods produce approximately the correct morphological 
mix at low-redshift.  
However, it must be appreciated that we are comparing a simple recipe for 
determining morphologies with what is observationally a sophisticated and 
subjective process (\ie assuming a one-to-one mapping between bulge-to-disc 
ratios and Hubble type), and one should not place too much emphasis on 
these results.

\subsection{Colours}
\label{sec:shapes}
\begin{figure}
\centerline{\epsfxsize = 8.0 cm \epsfbox{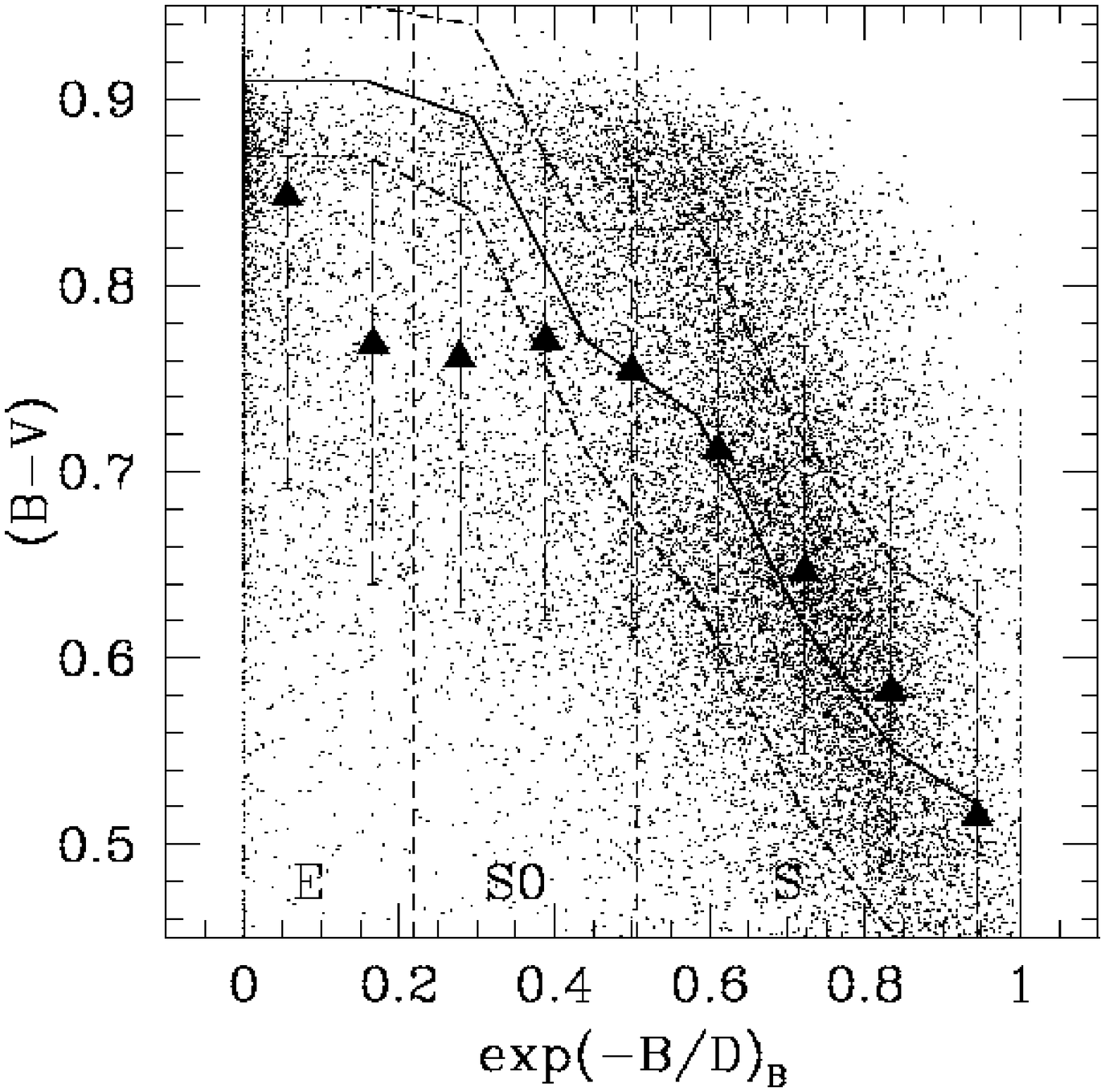} }
\caption{$B-V$ colours as a function of morphology for our galaxies brighter (than -18.8 
in the B-band) seen face-on.  The solid 
triangles show the median of the distribution, with the dashed \errorbar s giving the 
$1$-$\sigma$ range.  The solid line shows the measurements of \protect\citeNP{Buta_etal94}
(converted to a bulge-to-disc ratio index according to \protect\citeNP{SdV86}) with 
the observed $1$-$\sigma$ dispersion (dot-dashed lines).  
The short-dash vertical lines show where the divisions are drawn between elliptical, lenticular 
and spiral morphologies.}
\label{fig:shapes}
\end{figure}
We next consider the dependence of colour on galaxy morphology.  Fig.~\ref{fig:shapes} 
presents $B-V$ colours as a function of morphology for our galaxies.  The solid line represents 
data taken from \citeNP{Buta_etal94}.  
Our spiral galaxies closely follow the data both in average value and scatter, but more important 
is the dependence of colour on morphological type for early type galaxies which is not as strong as 
observed, leading to elliptical/SO galaxies which are too blue with too strongly scattered colors, 
independently of the exact criterion used to define these morphological types.
We will come back to this important issue in paper II, but we mention that this is due 
to recent overcooling leading to late star formation in early type galaxies. Indeed, if 
we remove ellipticals and S0s with more than 1 \% of gas in mass still present at $z = 0$ 
our colors (and dispersions) are in good agreement with the observations, although more 
than half of our early-type galaxies drop out of the sample. Finally, we note that even though 
this discrepancy is problematic, our median colors are still within 1-2 $\sigma$ of the observations, 
whatever the morphological type considered. Furthermore, RAM pressure stripping could help 
redden the colours of S0s and cluster ellipticals.

\subsection{Luminosity functions}

\begin{figure*}
\centerline{\epsfxsize = 16.0 cm \epsfbox{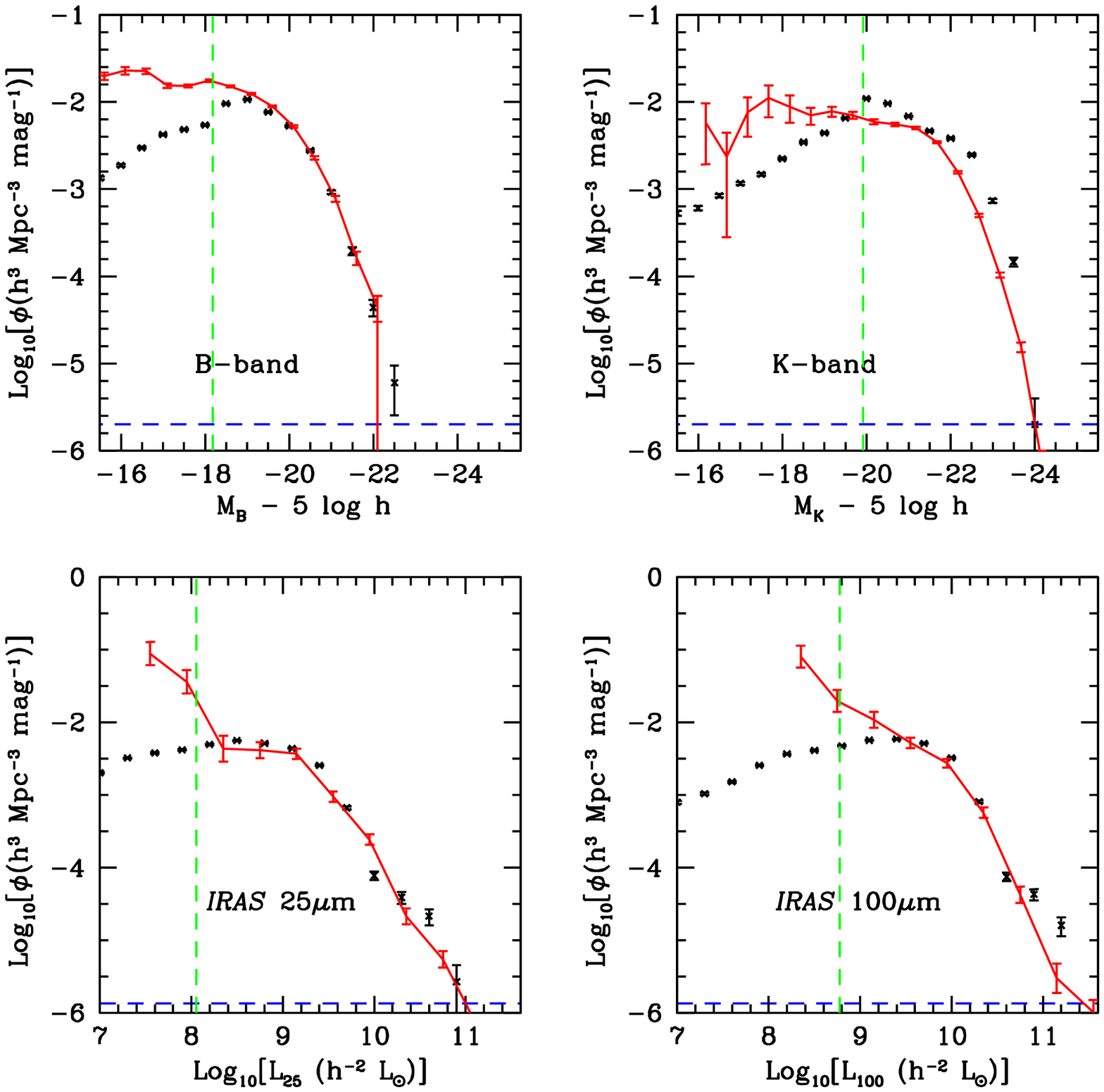}}
\caption{Galaxy luminosity functions.  Top left: the \protect\bband\ luminosity 
function (points with \errorbar s), with data from the 2dF galaxy redshift survey 
\protect\cite{CDC01} (connected, lighter curve with \errorbar s).  Top right: the 
\protect\kband\ luminosity function, with data from 2dF \protect\cite{Cole01}.
The \errorbar s are Poisson errors, and the magnitudes are given in the AB system.  
Bottom left: {\em IRAS} $25\mic$ luminosity function.  Bottom right: {\em IRAS} 
$100\mic$ luminosity function.  {\em IRAS} data comes from \protect\citeNP{SN91}.  
In each panel, the dashed vertical line is the formal resolution limit defined 
in section~\protect\ref{sec:res}, and the dashed horizontal line is the volume limit.  
(\protect\ie one galaxy per magnitude bin).}
\label{fig:lumfun}
\end{figure*}

The luminosity function, \ie the number density of objects as 
a function of magnitude, is a commonly measured observational 
statistic.  In Fig.~\ref{fig:lumfun} we show the results for 
the luminosity functions arising from our models in different 
wavebands.  We compare these with data coming from several different 
sources.  For the optical and near infrared (\bband\ and \kband) we 
use recent data from the 2dF galaxy redshift survey (\citeNP{CDC01} 
for the \bband, \citeNP{Cole01} for the \kband).  
For the {\em IRAS} wavebands ($25$ and $100\mic$), we compare with 
results from \citeN{SN91}.  

As noted in section~\ref{sec:res}, the finite mass resolution of 
our simulations produces a lower bound to the galaxy luminosity to 
which we are complete in any given waveband.  Although we have 
galaxies fainter than this limit, there will be galaxies of similar 
luminosity in halos that we do not resolve, and hence we expect to 
see some depletion of the counts lower than this limit.  We plot 
the magnitude limit corresponding to the mass limit with a vertical 
dashed line in each of the four plots in Fig.~\ref{fig:lumfun}.  
The resolution effect, \ie the depletion of counts faintwards of 
this line, is apparent in all four wavebands.  For the optical and 
near infrared, this depletion seems to occur quite close to the 
resolution limit, starting $0.5$ -- $1.0 \mag$ brighter.  This 
suggests that this is an accurate estimate of the resolution limit 
in these bands.  In the \IRAS\ bands, the turnover in the counts 
occurs at a luminosity of around $0.5 \dex$ (roughly $1\mag$) 
brigther than the formal limit.   

On the whole, the plots show quite good agreement with observed galaxy number 
densities over two orders-of-magnitude in wavelength.  In the \bband, the counts 
agree reasonably well with the shape and normalization of the 2dF data.  
This is a significant improvement on the hybrid model results 
of KCDW who found a fairly flat, power-law slope for the \bband\ 
luminosity function in a \LCDM\ Universe.  Our results are similar to those of 
CLBF, although they are able to get a slightly better amplitude of the luminosity 
function because they have a free parameter to control the fraction of 
stars forming brown dwarfs (which of course do not contribute to the luminosity 
function), that they fix using the \bband\ number density at 
$M_\mathrm{b_J} - \flh = -19.8$.  

In the \kband, we get a good (although on the high side) overall number density, and 
fit the data quite well at the `knee' of the luminosity function, 
representing most of the old stellar mass in the Universe.
Moreover, we note that our overprediction of intermediate luminosity galaxies with respect 
to the 2dF data might not be so much of a problem since a more recent survey with similar
number statistics \cite{HGCT03} seems to also find a higher number density of these galaxies.
Our luminosity function has an overall appearance which is similar to the 
Schechter function that well describes the data \ie we    
reproduce the exponential cut-off at the bright end that is present in the data.
This behaviour is due both to the prescription we employ to solve the overcooling 
problem (see section~\ref{sec:overcool}) and the more efficient feedback we implemented 
(see section~\ref{sec:feed}).
KCDW, with their different prescriptions for both these effects, obtain a 
power-law shaped \kband\ luminosity function which overestimates the number of bright objects.  
CLBF obtain a better fit to the data, which is insensitive to dust emission and 
absorption but once again relies on the free parameter to control the number of 
brown dwarfs formed. The conclusion is that overpredicting the bright end of the \kband\ luminosity 
function seems a rather natural feature of semi-analytic models and one has to invoke
at least a rather strong feedback (from supernovae and/or AGNs) to get rid of stars in the 
the brightest galaxies. 

For the {\em IRAS} number densities, we again find fairly good agreement 
with available data except at the faint end where resolution effects 
dominate.  In both wavebands the agreement is of excellent quality over the whole 
range of luminosities above the formal resolution limit.   

On the whole, we find the broad agreement with observations 
over several orders-of-magnitude in galaxy luminosity and two 
orders-of-magnitude in wavelength, a good confirmation of our models both 
for the semi-analytic recipes of the physics of galaxy formation, the 
determination of spectra, and the modelling of extinction effects.  

\subsection{Optical--IR colour magnitude relation}
\label{sec:iras_colours}
\begin{figure*}
\centerline{\epsfxsize = 16.0 cm \epsfbox{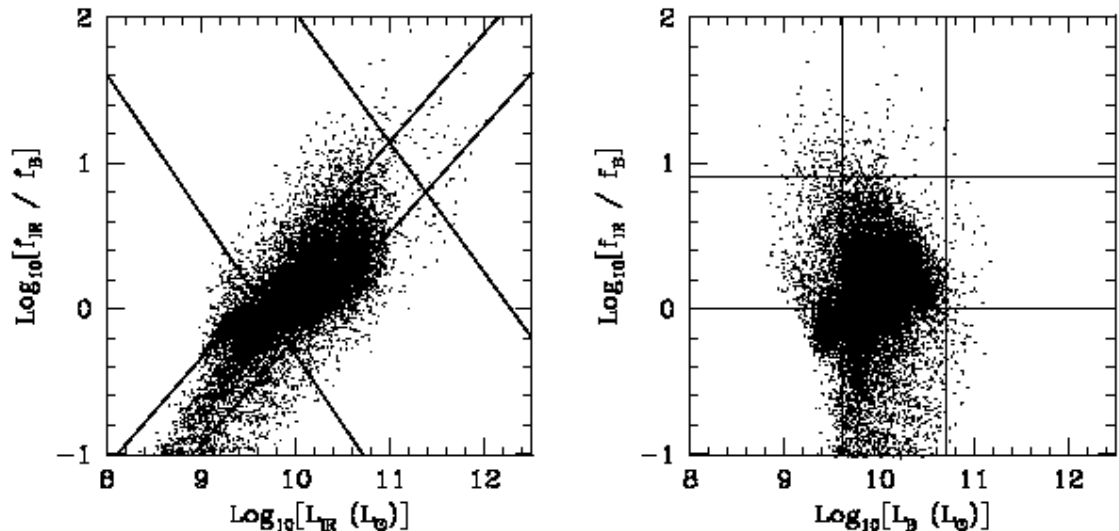} \vspace{-8.5cm}}
\caption{The relationship between the optical--IR colour, defined as the ratio 
of infrared to blue luminosity, and the absolute bolometric luminosity in the 
infrared (left) and \bband\ (right).  All galaxies with baryonic masses 
greater than the formal mass limit are plotted (section~\ref{sec:res}).  
Solid lines represent lines excluding about 1 $\sigma$ of the data plotted in figs~5a and 5b 
of \protect\citeNP{Soifer87}. } 
\label{fig:iras_colours}
\end{figure*}

In Fig.~\ref{fig:iras_colours} we examine the behaviour of the optical--IR 
`colour', \ie the difference in luminosity between these wavelength ranges, 
as a function of galaxy luminosity.  
We include all galaxies with baryonic masses greater than the formal mass limit, 
as defined in section~\ref{sec:res}.  \citeN{Soifer87} find that the colours of 
bright galaxies in the \IRAS\ sample are closely related to their total IR 
luminosities, but have no correlation with the optical luminosity, 
implying that galaxies have similar optical properties no matter what their IR 
emission is.  It can be seen from Fig.~\ref{fig:iras_colours} that our galaxies 
follow a similar trend.  On the left hand side we show the dependence of the 
ratio of bolometric IR luminosity to \bband\ luminosity against the IR emission.  
For low to moderate luminosities, this relationship 
is close to linear.  A linear regression reveals that the slope of this 
correlation is $0.70$.  Fitting a linear relationship ``by eye'' to the data of Soifer \etal (their fig.~5a) 
yields a slope of 0.75, so the results are quite comparable.  Our distribution 
is somewhat skewed, with a population of galaxies of quite high 
IR flux densities and extremely red colours, departing from the slope that fits 
the rest of the galaxies quite well.  Since galaxies in Soifer's plot need both 
IR and \bband\  luminosities to be measurable, galaxies that are intrinsically 
faint in the \bband\ are discriminated against in this sample.  It is thus likely 
that many of the galaxies in the upper part of our distribution would not 
be included when this sort of selection is applied.  This is especially clear
in the right hand panel of Fig.~\ref{fig:iras_colours} where we plot the 
colour against the \bband\ luminosity.  No correlation is visible, except 
for the handful of extremely red galaxies, which all have \bband\ luminosities less than 
$3 \times 10^9 \Lsun$.  Note that in \citeNP{Soifer87} (their fig.~5b) there are very few 
galaxies indeed with \bband\ luminosities this low. Furthermore, the \IRAS\ sample is flux limited 
whereas ours is volume limited, and therefore the observations should be biased towards 
IR bright and distant objects, 
which partly explains why our population of galaxies seem to populate the low-IR part of both 
diagrams. The other, and more fundamental reason is that although we only plot galaxies above the formal mass 
resolution limit, we do not satisfactorily resolve the star formation history of galaxies just above this 
threshold, which obviously leads to bluer colors and lower IR emission because according to the simulation, 
most of them have formed very recently.   

Bearing these limitations in mind, we conclude that our model is fairly successful at reproducing 
the observed correlations between luminosity and colour in the infrared and blue wavebands.

\subsection{Tully--Fisher relation}
\label{sec:tf}
\begin{figure*}
\centerline{\epsfxsize = 16.0 cm \epsfbox{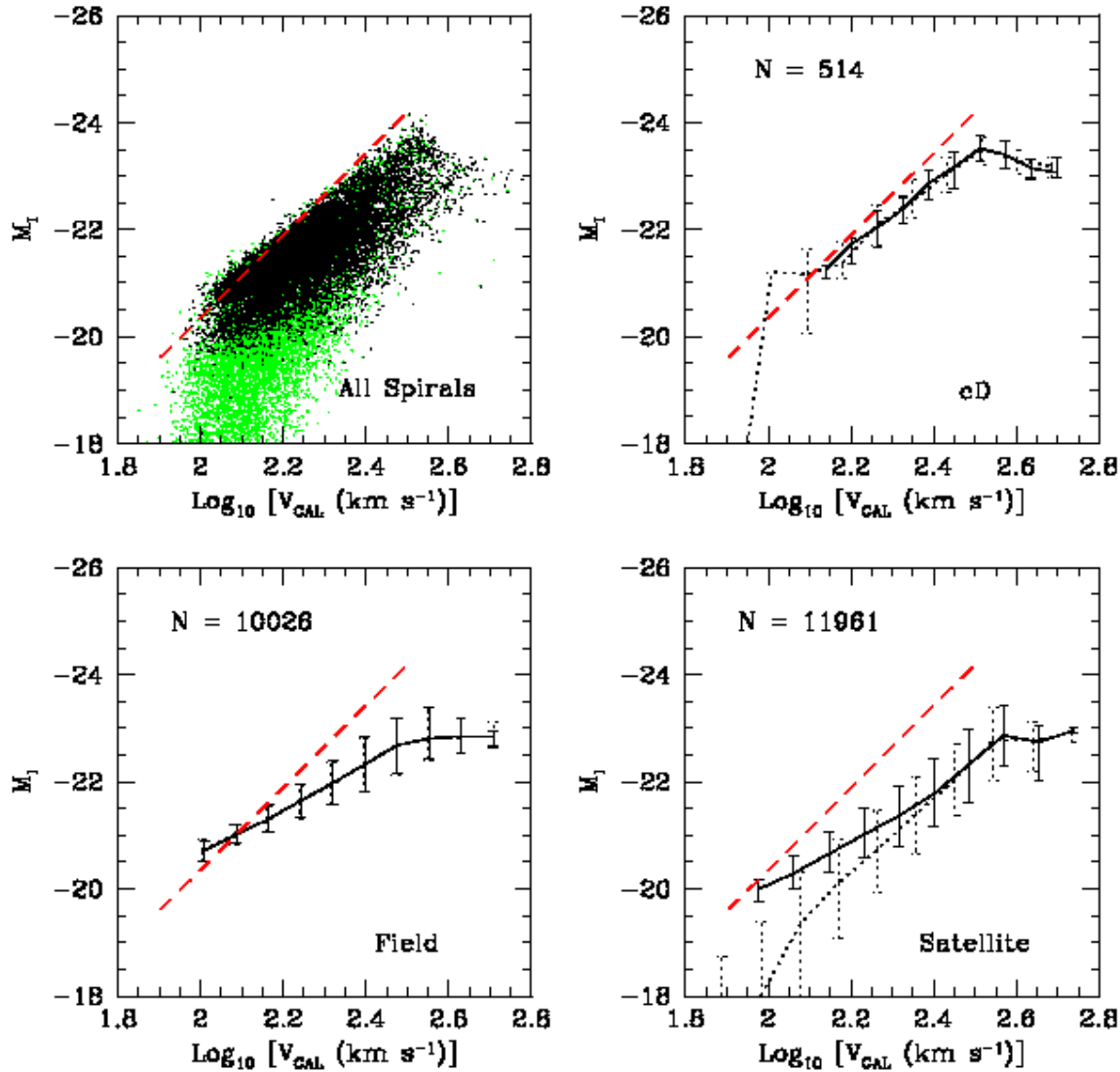}}
\caption{The Tully--Fisher relation in the \protect\iband\ for all our 
spiral galaxies above the resolution limit (top left).  The other three 
panels show the same data split into subsamples by galaxy environment, 
with the median and $\pm 1$-$\sigma$ scatter.  In each panel, the dashed  
diagonal line is the observational Tully--Fisher relation derived by 
\protect\citeN{Gio97}.  We also show the distribution 
both before (dashed line) and after (grey solid line) making the gas 
column density selection of equation~\protect\ref{eqn:gascolden}. }
\label{fig:TF2}
\end{figure*}

There is a well-established empirical relationship 
between the luminosity and rotation velocity of galactic 
discs.  This effect is known as the Tully--Fisher 
relation \cite{TF77}.  In the top left hand corner of Fig.~\ref{fig:TF2} we present a scatter 
plot for the \iband\ magnitude against circular velocity for all spiral 
galaxies in our simulation.  The dashed line on the plot 
represents the best fit to the \iband\ Tully--Fisher relation found by 
\citeN{Gio97}, where we have assumed that the 21cm line width is simply 
twice the galaxy circular velocity.  We plot the relation only for the 
range of velocities over which Giovanelli \etal have data.  
We note that the value of the Hubble constant derived from this dataset 
is $h = 0.69 \pm 0.05$, which is consistent with our value of $2/3$.  
It is clear from this plot that the majority of galaxies 
in our sample do indeed fall in a single locus on these axes, although 
there is significant scatter in the relation and a population of galaxies 
with rather lower luminosities for a given circular velocity.  The main 
locus itself is close to the relation of \citeN{Gio97} although our galaxies appear
to be somewhat fainter, and the slope of the correlation is shallower.

We proceed to perform a more detailed analysis by splitting our galaxy sample into three 
subsamples.  These are:
\begin{itemize}
\item `field' galaxies, that are the sole occupants of their dark-matter 
halos, 
\item `\cD' galaxies,  which are defined as central galaxies in halos with 
more than one occupant, 
\item `satellites', all non-central inhabitants of multi-member halos.  
\end{itemize} 
In the other three panels of Fig.~\ref{fig:TF2} we show the Tully--Fisher 
relation for each of these subsamples.  In each case, we plot the line 
connecting the median of the binned distribution, and add \errorbar s 
to show the $\pm 1$-$\sigma$ range.  We again compare to the 
line from \citeN{Gio97}.  It can be seen from the plots that our 
\cD\ galaxies follow the observed Tully--Fisher relation quite well, 
although these galaxies are slightly too faint (or rotate 
too fast).   

It is clear that the observational Tully--Fisher relation is not 
followed nearly so well by our field and satellite galaxies (the solid lines 
in the panels of Fig.~\ref{fig:TF2}).  The median 
magnitude as a function of circular velocity is well below the 
observed trend, and the $1$-$\sigma$ scatter is very large compared 
to that for the \cD\ galaxies. Also the satellite galaxies are 
largely responsible for the large scatter in the plot in the top left.  
This is especially interesting 
because the sample used by \citeNP{Gio97} to measure this relation 
is actually a sample of spiral galaxies in 24 galaxy clusters, \ie corresponding 
to our satellites.  There are two key reasons for this discrepancy: first is that 
satellites or galaxies with massive bulges are no longer being `fed' with hot gas cooling onto the 
discs, with the result that they use up their store of cold gas, and 
consequently have low star formation rates and luminosities. 
In other words, galaxies that have been satellites for different lengths of time can therefore 
have quite different gas fractions, and this is responsible for the large scatter in the 
relation. Second, and more fundamental 
is that the circular velocity of a galaxy results from the complex interaction between baryons and 
dark matter, and as a result our simple treatment of the disc dynamics is probably too crude, especially
for the fast rotating (massive) systems which are also likely to have a complex merging history.   

As pointed out by CLBF, the data in fact are subject to a strong selection effect in 
that galaxy rotation velocities are measured using HI or optical emission 
lines, and a significant column density of interstellar gas must be 
present in order to make those measurements.  In fact, many of our satellite 
spirals are gas-poor, and so would not be included in the \citeANP{Gio97} 
dataset.  To briefly demonstrate this selection effect, we applying a cut-off 
mimicking that which must exist in the data, 
\[
\label{eqn:gascolden}
\Mcold / \pi r_{1/2}^2 > 10^{7} \Msun \kpc^{-2},  
\]
which corresponds to a column density of approximately 
$N_\mathrm{H} = 1.25\E{21} \mathrm{atoms \, cm^{-2}}$.    
This selects only the satellites that still have a significant 
amount of gas, and the effect is clearly visible on the 
Tully--Fisher relation (solid line in Fig.~\ref{fig:TF2}).  The 
median now has similar behaviour to that of the field galaxies, 
although the satellites are still between 0.25 and 0.5 magnitude fainter 
than field galaxies depending on circular velocity. 
The scatter has been vastly diminished, 
although it is still larger than that of the field galaxies.  
This also reduces the number of satellite spirals by about a factor of $2.5$, in keeping 
with a number of observational results (\eg \citeNP{Solanes01}) 
demonstrating that spirals in cluster environments are often 
gas-depleted.  

We can also compare the scatter in the Tully--Fisher relation with the 
intrinsic scatter estimated from the observations of \citeANP{Gio97}.  
We compare the scatter for a MW-type galaxy with circular velocity 
$220 \kms$, using their equation~11.  Their prediction is a scatter of 
$\epsilon_\mathrm{int} = 0.22$ magnitudes.  We find a slightly higher scatter 
for our field galaxies, by interpolating a spline fit to the error bars 
shown in Fig.~\ref{fig:TF2}, $\epsilon_\mathrm{int} = 0.33$.  The `cD' 
galaxies have a similar scatter, $\epsilon_\mathrm{int} = 0.24$, 
and the cluster spirals have much larger scatter than the observations 
predict, $\epsilon_\mathrm{int} = 0.97$ for the whole sample and 
$\epsilon_\mathrm{int} = 0.49$ after applying the column density selection 
of equation~\ref{eqn:gascolden}.  
Whilst it is encouraging that our cD galaxies have similar scatter 
than the data, it seems that even with selection, our model predicts a 
wider variety of spirals in cluster environments than are observed.  
These results imply that, in reality, either gas cooling plays the dominant 
role in determining the properties of spirals in both cluster and field 
environments, and our suppression of cooling onto satellite and massive field 
galaxies is responsible for their variety, or our estimates of galaxy 
velocities are systematically offset towards higher values and more dispersed
than in reality.

\subsection{Faber--Jackson relation}
\begin{figure}
\centerline{\epsfxsize = 8.0 cm \epsfbox{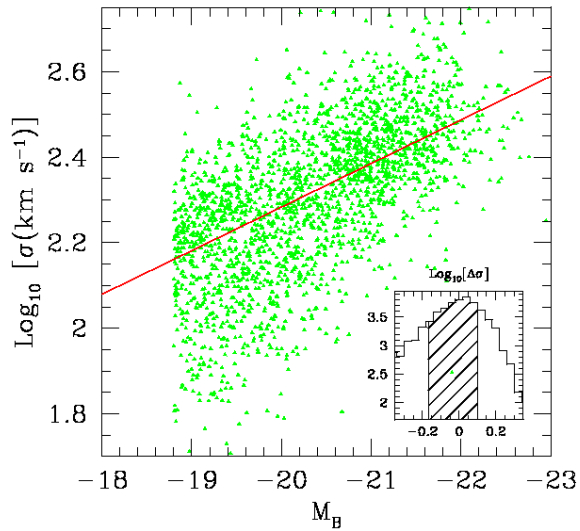} }
\caption{The Faber--Jackson relation for our elliptical and S0 cluster galaxies.
We only plot galaxies above our resolution limit M$_B$ = -18.8.  
The solid  line shows the fit found by \protect\citeN{FP99} for local ellipticals.  
The inset shows the scatter of our galaxies from the observed relation.  }
\label{fig:FJ}
\end{figure}

Analogous to the Tully--Fisher relation for spiral galaxies, observations 
show a correlation between the luminosities and dispersion velocities for 
elliptical and S0 galaxies, known as the Faber--Jackson relation \cite{FJ76}.  
In Fig.~\ref{fig:FJ} we show the Faber--Jackson relation for our galaxies, 
and compare with that found by \citeN{FP99} for cluster ellipticals (shown 
by the solid line).  We mimic the observational selection by plotting only 
elliptical and S0 cluster ({\em i.e.} halos containing more than 10 galaxies) galaxies
but note that the match is qualitatively similar if we use all the galaxies we
classify as elliptical or S0s.  

It will be seen that the galaxies sit around a line which is slightly fainter 
(or at a slightly higher velocity dispersion) than that of Forbes \& Ponman,  
once again consistent with too high velocity dispersion estimates.
However we consider this a fairly good imitation of the observed Faber--Jackson 
relation for our galaxies because of the scatter in the relation, which is shown as an inset on 
this figure.  This has been obtained from the differences between the straight 
line and the individual velocities as a function of \bband\ magnitude.  We have shaded the 
central area containing 68.3 per cent of the data, and we can thus read off the 
1-$\sigma$ limit as $+ 0.12$ and $- 0.15$ dex on each side of the observed relation, corresponding 
to approximately $\pm 0.13$ dex
of scatter if we were to define a Faber-Jackson relation directly from our model galaxies.  
Estimates of the scatter by Forbes \& Ponman yields a 1-$\sigma$ limit of $\pm 0.10$ dex, 
so we are confident that even though our relation looks slightly offset from the data
our models still are in fairly good agreement with observations.   

\subsection{Fundamental plane}  

\begin{figure*}
\centerline{\epsfxsize = 16.0 cm \epsfbox{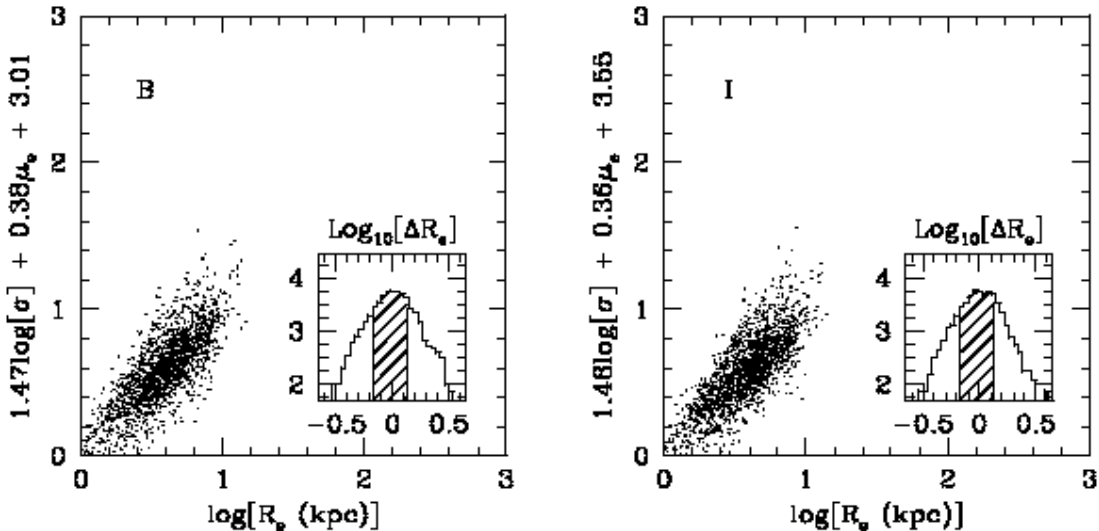} \vspace{-8.5cm}}
\caption{The fundamental plane for our elliptical and lenticular galaxies, in the \protect\bband\ 
(left) and \protect\iband\ (right), above our resolution limit.  The coefficients 
on the $y$-axis are found by minimizing the absolute deviation from the line 
$y = x$, as explained in the text.  The insets show the distribution of 
the residuals about the plane.  }
\label{fig:FP}
\end{figure*}

It has been found observationally that the scatter in the Faber--Jackson 
relation is rather larger than what would be expected simply from measurement errors, 
leading to the idea \cite{DJ87,Dressler_etal87} that there is a further 
variable involved.  The so-called Fundamental Plane relates the galaxy 
luminosity, velocity dispersion, and radius, and is found to apply to 
elliptical galaxies, and, to some extent, the bulges of spirals as well.

Following a similar procedure to that employed by observers, we assume that 
the plane is of the form:
\[
\label{eqn:fp}
\log R_\mathrm{e} = a \log\sigma + b \mu_\mathrm{e} + c
\]
In this equation, $R_\mathrm{e}$ is the effective radius of the spheroid, 
defined as the radius that contains one-half the galaxies light when viewed 
in projection (for our bulges this is $R_\mathrm{e} = 1.815 \; r_\mathrm{B}$, see \citeNP{Hq90}).  
$\sigma$ is the one-dimensional velocity dispersion, and $\mu_\mathrm{e}$ is the average 
surface brightness within $R_\mathrm{e}$.  

We determine the coefficients $(a,b,c)$ in equation~\ref{eqn:fp} 
by minimizing the total absolute deviation of all our elliptical and 
S0 galaxies above the formal resolution limit from this plane.  
In order to do this, the variables 
are first transformed into a `normalized' co-ordinate system by 
subtracting off their mean value and dividing by the standard deviation, \ie
\[
\sigma_1 = (\sigma - \bar\sigma) / \sqrt{\avg{\sigma^2}}
\]

We show our Fundamental Plane relation in Fig.~\ref{fig:FP}, in both 
the \bband\ and the \iband.  There is clear evidence for a fairly tight 
correlation of these three bulge properties, in both wave-bands, with the 
relation given by:
\[
(a, b) = \cases{(1.47, 0.38) & \bband \cr
                (1.48, 0.36) & \iband }
\]

For a sample of 74 early-type galaxies towards the Coma cluster, \citeN{Scodeggio_etal98} 
find that the fundamental plane that minimizes the absolute deviation is given by (their table 6): 
\[
(a, b) = \cases{(1.40, 0.35) & \bband \cr
                (1.70, 0.33) & \iband }
\]
These results for the coefficients $(a,b)$ are in good agreement with those we find 
(and reasonable agreement with other observations, \eg \citeNP{PD97}).  
It will be noted that we also succeed in replicating 
a trend that exist in the data as well, namely that the 
parameter $b$ is quite independent of the waveband. The steepening of the slope $a$ of the dependence 
on the velocity dispersion as a function of increasing wavelength,
is not so marked in our simulated galaxies.  

In the insets to the two panels in Fig.~\ref{fig:FP} we show the distributions 
of the residuals about the plane, \ie the differences between the actual 
$\log R_\mathrm{e}$ values and the prediction from the derived FP relation.  
The width of these distributions are approximately $\pm 0.15$ and $\pm 0.12$ 
for the \bband\ and \iband\ respectively. An estimate from Scodeggio \etal fig.~7 shows the 
1-$\sigma$ scatter in the data to be approximately $\pm 0.10$ dex in both the \bband\ and the \iband.
Thus we not only produce fundamental planes with a similar tilt to that observed, but also obtain 
 a good match for the residuals.  
This also suggests that much of the scatter in the observed FP relation is intrinsic, 
due to the different merging histories of the objects examined, rather than being 
caused by measurement errors.   

This is the first time that semi-analytic models have been pushed as far as 
looking at the fundamental plane, and we find the results an encouraging 
vindication of our methodology.

\section{Resolution tests}
\label{sec:resolu}

The issue of resolution enters our models in several ways, and 
it is important to check that we understand how resolution 
affects the results presented here.  In this section, we 
will endeavour to show that our results are quite stable to 
an increase in resolution.  This will generally be done by 
comparing with a simulation of slightly {\em worse} resolution, 
and showing that the results have already converged.  

In the following sections, we consider the effects of finite 
force, time, and mass resolution.  We will use the abbreviations 
HTR and LTR to stand for high and low time resolution, and HMR and LMR 
for high and low mass resolution.   Thus the HTR (or HMR) is generally 
the original, high resolution simulation.

\subsection{Spatial resolution}
Initially, in the \nbody\ simulations, there is a spatial resolution 
set by the softening length used in computing the smoothed gravitational 
potential.  In our code, this softening length is set at one twentieth 
the mean interparticle separation.  For our \LCDM\ simulation, at redshift 
zero, this is around $30\, \mathrm{kpc}$.  We identify halos with a density 
contrast of $200$ relative of the critical density, which thus have an 
average particle-particle separation 
length of the mean separation divided by $(200/\Omega_\mathrm{t})^{1/3}$, 
where $\Omega_\mathrm{t}$ is the ratio of matter density to critical density 
as a function of time.  The mean distance 
between particles inside halos at $z=0$ is thus twice the softening length.  
Improved resolution may then have an effect on the central, dense regions of 
halos, where the separation length is smaller than the softening length, but 
is unlikely to influence the virial mass itself.  Thus, attempts to fit 
density profiles to these inner regions would be affected by the softening, 
but we do not do this, measuring only virial masses and energies, and assuming an isothermal 
profile for all our halos.  The spatial resolution limit set by gravitational 
softening is thus not a relevant concern for this work.

\subsection{Time resolution}
\label{sec:tsres}

To calculate the particle trajectories in the \nbody\ simulations, we use 
around twenty thousand timesteps (table~1), equally spaced in expansion 
factor from $z \approx 35$ to the present day.  To construct the halo 
merging tree, a subsample of these steps is used, spaced logarithmically 
in expansion factor.  This produces seventy timesteps between the present 
day and the first identification of a virialized object with twenty 
particles, at redshift $z \approx 10$.  The spacing between the penultimate and 
the final timesteps corresponds to sixty of the simulation timesteps.  
It is thus the choice of output times for constructing the merging tree 
that really dictates time resolution.  This especially 
has an effect on the baryon cooling tree described in section~\ref{sec:tree}, 
since we will only be interested in binary mergers in this section.  If a massive halo has accreted 
two smaller halos in the course of one timestep, we only follow 
the galaxies in the larger of the two halos.  The occurrence of more-than-binary 
mergers obviously increases with the length of the timestep, so more galaxies 
are lost with worse resolution.  

To check our sensitivity to this timestep resolution, we compare results 
with those from a tree made using only one half the number of simulation 
outputs (\ie with a factor two worse time resolution).  There 
is a one-to-one correspondence between halos (but not galaxies) in the final timestep 
of the two trees, so we can look to see how resolution has had an effect 
on the properties of individual halos.  

\begin{figure}
\centerline{\epsfxsize = 8.0 cm \epsfbox{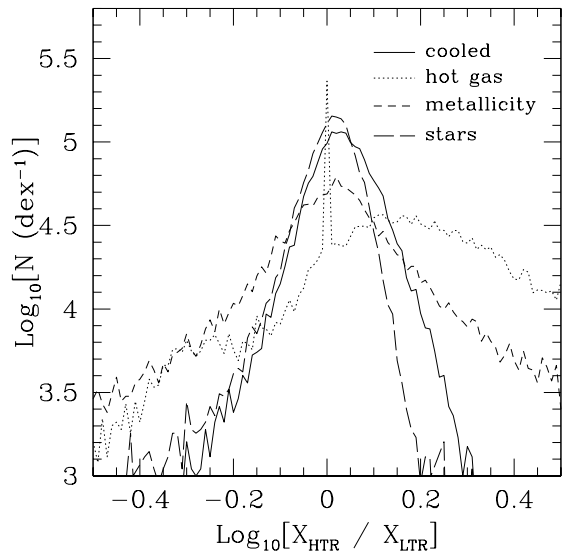}}
\caption{The difference between several halo properties (computed on an individual 
basis) between halos in the full resolution (HTR) run and halos in the run only using 
half the number of timesteps (LTR), as explained in section~\protect\ref{sec:tsres}. }
\label{fig:tres_halo}
\end{figure}

\begin{figure}
\centerline{\epsfxsize = 8.0 cm \epsfbox{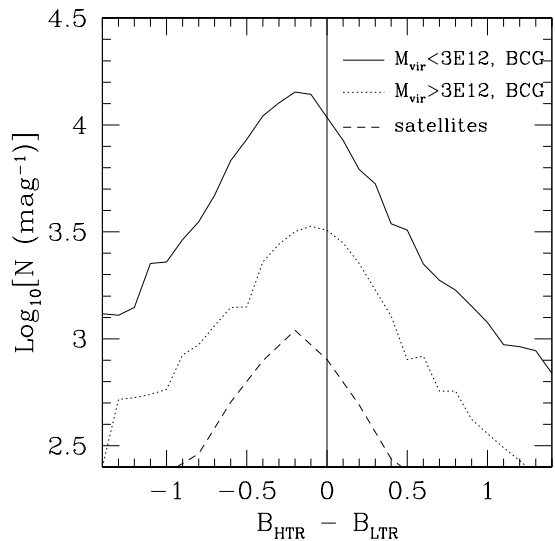}}
\caption{Differences in \bband\ magnitudes for individual halos in the 
full and half time resolution runs.  The upper (solid) line is the 
magnitude difference of the brightest halo member (BCG) for halos with mass less 
than $3\E{12} \Msun$.  The middle (dotted) line is the same 
statistic for the other, more massive halos.  
The lower (dashed) line is the difference in satellite magnitude 
(\ie summed over all galaxies in the halo excluding the brightest one) 
for each halo having more than one member galaxy.  A vertical line has been 
plotted at zero difference to guide the eye.  }
\label{fig:tres_lums}
\end{figure}

Fig.~\ref{fig:tres_halo} shows a halo-by-halo comparison for several key halo 
properties.  The solid line shows the number of objects with a given fractional difference 
between their cold gas masses in the full simulation and in the low time 
resolution version.  It can be seen that there is a very small systematic 
offset between halos in the two simulations, with halos in the HTR having slightly 
more cold gas, and there is some scatter (the curve has a FWHM of the order $\pm 20$ per cent).  

For hot gas, the picture is rather different.  Although there is a sharp 
peak at zero difference, this is due to halos that have just been discovered 
in the final timestep, and so have not cooled at all in either version.  For 
the rest of the halos, the peak in the distribution occurs when halos in the 
HTR run have around forty per cent more hot gas than those in the LTR.  
However, the total amounts of hot gas in the two simulations are similar, 
with the HTR run having around eight per cent more gas than the LTR.  
This implies that the halos containing large amounts of the hot gas are 
the same in the two simulations, but that a large number of halos containing 
small amounts of gas are discrepant, and their precise hot gas mass is 
sensitive to the time resolution chosen. 
It is interesting to note that this effect does not have a noticeable impact 
on the metallicity of the hot gas, shown by the short-dashed line in 
Fig.~\ref{fig:tres_halo}:  the metallicity difference is peaked 
close to zero, showing that whatever mechanism is responsible for losing 
gas in the LTR run also loses a similar fraction of metals.  With the long-dashed 
line, we show the difference in total stellar mass for each halo.  Happily, this is peaked 
close to zero, so it seems that time resolution has little effect on the total 
stellar masses.  

In Fig.~\ref{fig:tres_lums} we present three curves comparing luminosities 
between the two runs.  The upper curve is the \bband\ magnitude difference 
between the brightest galaxy in the halo (BCG) in the two different runs, for 
halos of virial mass less than $3 \E{12}\Msun$, although the majority 
of halos in this mass range in fact only contain a single galaxy.  It can be seen that 
this is peaked to the left of zero, implying brighter BCG's (by around $0.2$ 
magnitudes) in the HTR run.  This is despite the fact that we know from above 
there is no offset in the total mass of stars between the two runs.  Since 
it is the \bband\ magnitudes we are comparing, which are dominated by contributions 
from the youngest stars, this implies that the long timestep in the LTR 
smoothes out the star formation over too long a period.  
The second (dotted) curve compares the BCG's in the other, more massive halos.  
The offset is reduced ($\approx 0.1$ magnitudes) so this effect is less important 
for massive halos.  
In the lower curve, we show the difference in total halo magnitude when 
summed over all galaxies in the halo except the brightest one.   
Again, there is more luminosity in the HTR run.  In fact, the final 
timestep in the HTR run contains a total of 6355 satellite galaxies, 
as opposed to 5096 in the LTR.  This is because galaxies are lost 
due to more-than-binary merging in the LTR run, as explained in at the 
beginning of this section.  This twenty per cent difference in galaxy numbers  
is largely responsible for the ten per cent difference in satellite luminosity 
in this plot.  

In summary, there would have 
been little difference had we run our simulation with worse time resolution, 
except for the depletion of low-mass satellites and slightly fainter 
field galaxies.  Since these effects are all of order $0.1$--$0.2$ magnitudes, 
which is quite small compared to the systematic errors due to uncertainties in 
the modelling techniques, we conclude that our time resolution is close to convergence.  
Full convergence could be achieved with better time resolution, though we have 
chosen in the previous parts of this paper to gain improvement by  
allowing the code to cope with multiple merging of halos.

\subsection{Mass resolution}
\label{sec:spatres}

There are two ways in which mass resolution enters our results.  
Initially, in the \nbody\ simulations, there is a resolution limit 
set by the finite mass of each particle.  Subsequently, when 
identifying dark matter halos, there is a minimum number of particles
needed.  We will consider the impact of both these effects.

\subsubsection{\nbody\ resolution}

\begin{figure*}
\centerline{\epsfxsize = 16.0 cm \epsfbox{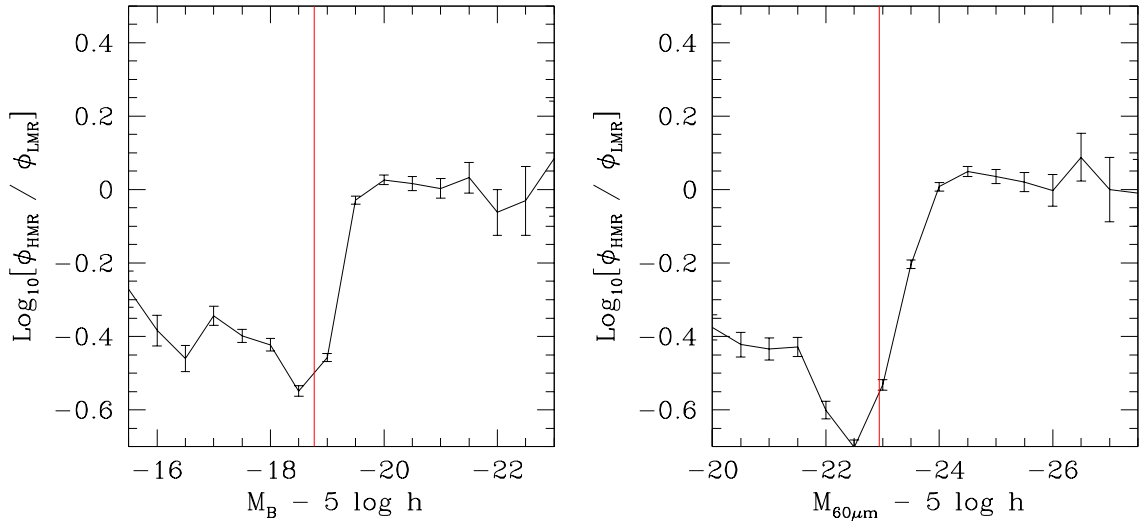}\vspace{-8.5cm}}
\caption{Differences in the \bband\ (left panel) and \IRAS\ $60\mic$ (right panel) 
luminosity functions for the full resolution and half resolution runs.  The vertical 
line is plotted at the estimated resolution limit of the LMR simulation 
(approximately $0.75$ magnitudes brighter than values for the HMR simulation 
quoted in table~\protect\ref{tab:maglims}).  }
\label{fig:lcomp}
\end{figure*}
\label{sec:degrade}

To investigate the effect of \nbody\ resolution, one would ideally run two 
simulations of different resolution and compare the results.  Although we 
have runs of $64^3$ and $128^3$ particles for the same box size and initial conditions 
as the standard ($256^3$) simulation, there is very little overlap in the 
halo mass function between them, making a direct comparison rather 
unproductive.  Instead of comparing the simulations, we degrade the 
resolution of the HMR run by selecting, at random, one half the particles, 
assigning them double their original mass.  We then run the halo-finding, 
tree-building and galaxy formation codes on this LMR (degraded) simulation.  
In Fig.~\ref{fig:lcomp} we show how the luminosity functions (in the \bband\  
and \IRAS\ $60\mic$ filter) 
of these simulations differ.  For faint galaxies, the LMR run underestimates 
the number density by a factor $\approx 2.5$.  For bright galaxies, the 
estimates converge.  It is the behaviour between these two regimes that is 
especially interesting.  The vertical line in the plot is the formal resolution 
limit ($M_{B} - \flh < -18.8$ or $M_{60\mic} - \flh < 22.9$, see section~\ref{sec:res}) of the LMR run, 
and it will be seen that it is right in the middle of the transition between 
these two regimes, and that the LMR simulation does not fully converge with 
the HMR until approximately three-quarters of a  magnitude brighter than 
this limit.  As stated in section~\ref{sec:res}, we expect to have problems 
for slightly brighter galaxies than this limit, since we cannot accurately follow 
the merging histories of the lightest objects.
  
Assuming that this behaviour is similar whatever the resolution limit, we 
have therefore shown that our galaxy `selection' is complete for 
galaxies around three-quarters of a magnitude brighter than the formal 
resolution limit.  Hence, our results are reliable for galaxies down 
to one magnitude fainter than $L_\star$.

This is not a direct test of the effects of \nbody\ resolution, since we 
have compared with a degraded version of the same simulation rather than 
a simulation actually run with fewer particles.  In order to check the 
validity of this degrading approach, we sample the $256^3$ simulation at 
a much lower rate, selecting one particle in $4^3$, thus giving it the 
same resolution as the $64^3$ simulation.  Making a similar comparison,  
we find no significant differences in the luminosity functions.  
Having verified this approach with a factor of $64$, it is reasonable to 
expect that sampling with a factor two will produce a good imitation 
of a simulation run with half the mass resolution.

\subsubsection{Minimum particle number}

\begin{figure}
\centerline{\epsfxsize = 8.0 cm \epsfbox{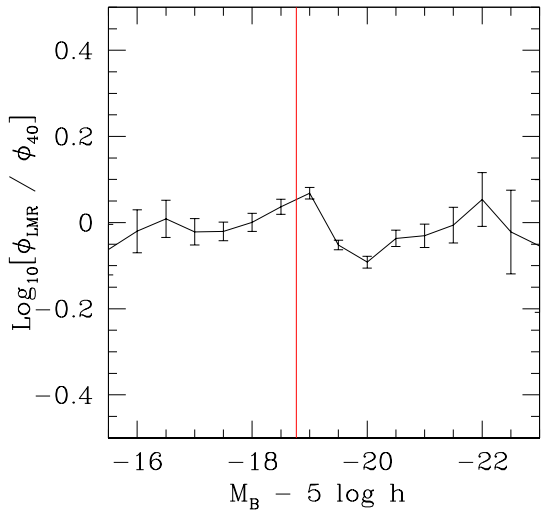}}
\caption{As Fig.~\protect\ref{fig:lcomp}, but this time we compare the 
luminosity function from a run with a minimum particle number per halo 
of forty, with that of the half-resolution sampled run.}
\label{fig:lcomp2}
\end{figure}

Increasing the minimum particle number for halos will mean that 
the building blocks of our galaxy formation process are more massive, 
so the effect should be similar to that of degrading the mass resolution 
of the simulation.  However, the halos will be better resolved in this 
new case, so there may be some differences.  To investigate this, we 
compare the LMR (degraded) simulation of section~\ref{sec:degrade} 
with the HMR simulation when the minimum particle number is increased 
from twenty to forty.  In Fig.~\ref{fig:lcomp2} we compare the \bband\ luminosity 
functions of galaxies in these two runs.  As can be seen, there is no 
significant difference between the two.  

The overall conclusion is that the limit of our resolution, in terms 
of noticeable effects in the local Universe, depends only on the mass 
of the smallest halos identified, and is not 
individually sensitive to the actual 
numerical resolution of the simulation or the number of particles per halo.  

This also suggests that it would be valid to improve the resolution of these 
runs by extending the treatment to halos with less than twenty particles in them,   
for example going down to as few as seven, the limit quoted by KCDW for 
dynamical stability.   This alone should result in roughly a one magnitude 
improvement in resolution.

\section{Discussion}
\label{sec:discuss}

\subsection{Summary}

In this paper, we have presented our \nbody/semi-analytic hybrid model for 
studying galaxy formation, and demonstrated its application to a reference 
model within a \LCDM\ cosmology.  Here, we summarize the assumptions and weaknesses of the model,
emphasize some of its results, and comment on several aspects of the 
programme. 

\subsubsection{Implementing new recipes}

At root, our approach is entirely independent of pre-existing work 
as we use different \nbody\ codes and different methods 
for tracing the history of dark matter halos.  
However, most of the recipes used to model physical processes (\eg gas 
cooling, star formation, feedback, dynamical friction) are 
similar to those that have appeared elsewhere.  
We include a number of innovations 
that have not previously been used in hybrid models:
 
\begin{itemize}

\item We have measured the halo properties directly from 
groups identified in the \nbody\ simulation, 
rather than assuming distributions \eg for the spin parameter.

\item We have noticed that the friend-of-friend group finder identifies 
a number of (mainly low-mass) halos that are not bound systems.  We 
have dealt with these objects by suppressing cooling in them.  If they are 
simply random collections of particles, misidentified, they will thus not 
affect our results.  

\item In general, we have endeavoured to use `natural', continuous models 
as opposed to the all-or-nothing methods applied by other workers.  Examples  
include the model for halo reservoirs, where metals and hot gas ejected from 
the halo are retained in the local environment and gradually re-accreted, 
and the recipe for merging, where the effects are a smooth function of 
the mass ratio of the progenitors, rather than the discontinuous 
step function.  

\item We have modelled disc instability as another possible mechanism 
for the formation of bulges.

\item We allow for mergers between satellite galaxies in clusters, and gradual 
tidal stripping of dark matter cores of sub-halos.  These follow schemes that 
have been applied to pure semi-analytic models by SP99.  

\item We use the detailed chemical and stellar evolution models described in 
DGS, giving us self-consistent, metallicity-dependent spectra and cooling.  
We do not make the common assumption of instantaneous recycling of metals. 

\end{itemize}

A few key weaknesses of our recipes present themselves:

\begin{itemize}

\item The process used to subdue cooling in massive halos, and in 
small halos after recent mergers, is critical to avoid overcooling 
on to the central object, and hence overly bright central galaxies. 
Our approach has been to treat this problem phenomenologically but to point out 
in details where it affects our final results. The main reason behind this choice
being that whilst an approach like that 
of CLBF can be qualitatively justified, and is perhaps more reasonable than the simple 
approximation of forgetting about cooling in halos above a certain circular 
velocity (KCDW), there is no quantitative or observational 
justification for it. 

\item The treatment of the interactions between dark matter and baryons is 
simplistic. Clearly this problem needs to be tackled by high resolution hydrodynamical 
plus N-body simulations. We plan to undertake a detailed study of the impact 
of various astrophysics processes on this interaction in the future. In particular, this
will affect our predictions for correlations involving the dynamics of galaxies (Tully--Fisher,
Faber--Jackson and fundamental plane).

\item The modelling of dynamical friction relies on Chandrasekhar's formula, 
which is correct on average but does not reflect the diversity of cases that appear in 
numerical simulations \cite{Springeletal01}. 

\item Whilst we estimate the radial positions of galaxies inside halos, the 
angular coordinates are not determined in this model.  

\end{itemize} 
In particular, these points call attention to the need for more 
detailed/accurate modelling of what actually occurs to both gas and 
galaxies within cluster environments.  

\subsubsection{Addressing the optical/IR luminosity budget of galaxies}

We have introduced an efficient technique for modelling the distribution, composition 
and effects of dust in order to provide a full, panchromatic view of galaxy 
formation and evolution processes.  For this purpose, we use the ingredients of the 
\textsc{stardust} model of chemical and spectrophotometric evolution of 
stellar populations (DGS).  This model was specifically designed to be 
implemented within a semi-analytic model of galaxy formation and evolution
\cite{DG00}.   The star formation rate and cold gas
metallicity histories are followed along the merging history trees of galaxies,
and, once an IMF is chosen, the model gives theoretical spectra of the stellar 
populations.  Given the galaxy radius and column density, the extinction and the 
IR/submm spectral energy distribution are computed without additional free parameters.
Computing the IR properties of galaxies is particularly important in the picture of 
hierarchical galaxy formation since observations show that mergers radiate mostly in the IR.  

\subsubsection{Studying resolution effects}

We provide some simple tests of the sensitivity of our results to time 
and mass resolution.  This enables us to put confident bounds on the 
regime where the results should be trusted.  This is a very important step 
in the development of hybrid modelling, and resolution effects should be 
studied in more detail.  
The actual limit due to the baryonic mass of the lightest halos 
makes us incomplete fainter than $M_B  = -18.8$ at $z=0$.  This limitation 
is well understood and we simply have to appreciate that our predictive power 
is restricted to volume limited samples above this magnitude.  However, the 
inability to resolve merging histories of small galaxies within this limit 
is a subtler problem, and requires more work to deal with.  For instance, 
we attribute the underestimate of the metallicity of the intra cluster medium 
to the absence of dwarf galaxies. Morover, the results of 
section~\ref{sec:spatres} suggest that we are only completely accurate 
down to galaxies around three-quarters of a magnitude brighter than this $M_B  =-18.8$ limit. 

\subsubsection{Mimicking selection effects}

Our models produce a wide diversity of galaxies, but observational data 
only probe the emerging part of this iceberg.   
We have stressed several times in this paper the need to understand 
observational selection effects when comparing models to data. 
Examples include the Malmquist bias in using the APM catalogue to 
measure the relative abundance of different galaxy types, 
and the surface brightness selection that is generally present in all 
astrophysical datasets. We have roughly tested the influence of 
these selection effects on the comparison of the theoretical and 
observational samples in these two cases. As a result, we manage to 
imitate the datasets in a plausible way.  This deserves more work, and will be 
addressed in forthcoming papers.  

\subsection{Main results} 

While we have the flexibility to alter the free parameters governing the 
star formation rate, feedback efficiency, etc, we find that our 
observationally 
motivated initial choice of reference parameters is well able to match a number 
of observational constraints:

\begin{itemize}

\item We find a wide variety of galaxy properties, among which the Milky Way 
appears to have `normal' properties for its circular velocity, 
in terms of baryonic mass, ISM metallicity, star formation rate and size.

\item The luminosity functions in the optical and IR are well reproduced
by the model, above the resolution limit. 

\item The optical/IR luminosity budget in $z=0$ galaxies is well reproduced 
by the model, as it appears from the fit of the luminosity functions around 
$L_\star$ at various optical/IR wavelengths, and from the existence of an
increasing IR/optical luminosity ratio with IR luminosity as observed in the \IRAS\ 
data.  This is particularly interesting as the optical/IR properties of galaxies 
are closely linked to the process of hierarchical galaxy formation, since mergers 
radiate most of their luminosity in the IR.  These first results are thus very 
encouraging.  

\item Because the definition of morphological types in usual catalogues of 
bright galaxies lies on a mix of `objective' and somewhat `subjective'
classification criteria, it is not easy to compare 
the predictions of this kind of model with data. With our simple choice of 
the bulge--to--disc luminosity ratio as a tracer of the type, we fairly
reproduce the type fractions in a volume-limited sample. 
We also reproduce the overall behaviour of the colour-type correlation, 
and find that the $B-V$ colour monotonically reddens from the late to the 
early types. However we predict too slow an evolution when going from a morphological type to another,
and therefore $B-V$ colours which are too blue for the S0s/ellipticals.
We are also unable to model the transition of the low-luminosity spirals 
to the `dwarfs', that is expected to occur around $M_B  = -17$, which is well 
below our resolution limit. 

\item The sizes of high surface-brightness discs are in the same range 
as the observational data, but we predict slightly too many large bright discs 
than are observed.  We recall that the sizes 
are computed under the assumption of specific angular momentum conservation 
during the dissipative collapse of baryons, and in the absence of dynamical 
effects of the feedback mechanism on the structure of the galaxies.
Given the crudeness of the recipes which are used, 
we consider that this fit is still quite satisfying. 

\item We obtain an arguably decent match of fundamental structural properties: 
the Tully--Fisher relation for the spiral galaxies, and the Faber--Jackson 
relation for the ellipticals. We compute the 
dispersion in these relations, and find good agreement for the TF of cD 
galaxies, but the scatter is too high and the slope of the relation itself 
too shallow for cluster satellites and field galaxies.  
The FJ scatter and slope for cluster early-type galaxies are in better agreement 
with those obtained from the data.  

\item Finally, we obtain a fair fit to the fundamental plane in two 
wavebands, reproducing the observed parameters and the scatter.  
This is the first time semi-analytic 
models have been applied to look at this important observational test.   

\end{itemize}

Thus the impression that emerges from this first set of results confirms
the overall robustness of the main statistical features predicted for local 
galaxies, with respect to previous works involving semi-analytic 
models or the hybrid approach, although the details of the current 
implementation in \galics\ are fully new and original. Hierarchical
galaxy formation does provide us with a natural explanation of many patterns
of our local universe, at least qualitatively : the rough percentage of
each morphological type along the Hubble sequence, the 
fact that early-type galaxies are redder on an average, the correlation 
between luminosity and a characteristic internal velocity, the 
existence of an upper limit for the luminosities of galaxies, and the
overall luminosity budget between the optical and the IR. 
These patterns appear very naturally from the interplay of a limited 
number of physical processes. 

The quantitative fit to local galaxies is generally satisfactory, given the crudeness of the 
assumptions, but some failures remain. More interesting perhaps is the 
inner consistency of these successes and failures. The massive galaxies
seem to rotate too fast and be too blue, with a lot of ellipticals 
still having discs. This can be attributed to an improper modelling of cooling and/or to possible
changes in the IMF, but it can be anticipated that a modification of the recipes
will have simultaneous consequences on these three features.
More work is needed from our reference model to clarify this point.

The well-known influence of selection effects on the observed properties 
of galaxies receives a new emphasis from hierarchical galaxy formation. 
This theory generically predicts that galaxies scatter within a wide
range of properties, and that observations always unveil only the tip of the
iceberg. Such a pattern considerably complicates the comparison of predictions
with published data for which the exact observational selection criteria are
not always fully documented. Large galaxy samples produced by extensive surveys,
with a careful control of selection effects,
should provide us with a better basis for the comparison with the predictions 
of the models.

Other papers in the series will continue to detail the successes and
failures of the current model at higher redshifts, so that a consistent view 
on the evolution of most of the patterns can be obtained.

\section{Conclusions}

The semi-analytic approach has often been criticized for the number of 
free parameters it implies, and the differences between the various models. 
The impression after this fully original work here, when it is compared \eg
to KCDW, is the overall robustness of many results, despite  
differences in the details of the recipes that are used to model the 
astrophysical processes, and to handle halos and galaxies.

After a brief description of the \nbody\ simulation, and the way we build halo 
merging history trees, we have detailed the semi-analytic recipes that are 
used to model the baryons.  We have then shown a series of predictions for 
galaxy properties in the local Universe 
coming from a reference model.  The model produces a wide range of properties of the
local galaxy population, and gives reasonable fits to the optical/IR luminosity 
functions, the correlation of the IR/optical luminosity ratio with IR 
luminosity, the morphological fractions, and the colour-morphology relation.  
Satisfactory fits of the disc sizes, the Tully--Fisher and Faber--Jackson 
relations, and the fundamental plane, are also obtained.  
We have tried to test the influence of various selection effects on the comparison of 
predictions to data.  We have also shown that the influence of 
time and mass resolution on the results can be estimated, and should not 
affect our conclusions.

One of the ambitions of the \textsc{galics} research programme is to make 
predictions for a full set of observational data, at various redshifts. 
Clearly such a programme will produce a host of results that will be covered 
by several papers in this series.  Consequently, it is too early to form a 
full judgement of the ability of the model to capture the main 
statistical properties of galaxies, and their time evolution.  
In this first paper, we have presented our \textsc{galics} hybrid
model for the study of galaxy formation. 
Forthcoming papers will complement the picture given here.  
In paper II, we will study the influence of the cosmological and 
astrophysical parameters on our results, and we will predict the redshift evolution of a set
of statistical galaxy properties. Paper III will be devoted
to Lyman-break galaxies.
Paper IV will present multiwavelength faint counts and angular correlation functions, as well as 
the construction of mock images that are the most fundamental basis for the 
study of selection effects.  In paper V, we will study the 3D correlation 
function, and examine the behaviour of the bias parameter for a variety of 
galaxy samples as a function of scale and time.  
After this global overview, subsequent papers will address more specific issues.

\section*{Acknowledgements}
The authors gratefully acknowledge the support extended by the
IFCPAR under contract 1710-1. They thank J. Blaizot, B. Lanzoni, 
G. Mamon and F. Stoehr for help and criticism all along this project as  
well as the anonymous referee and Simon White
for their very careful reading of this manuscript and their
constructive remarks.
The \nbody\ simulation used in this work 
was run on the Cray T3E at the IDRIS supercomputing facility.  
SJH acknowledges support of the EU 
TMR network `Formation and Evolution of Galaxies'.  
The research of JEGD at Oxford  has been supported by a major 
grant from the Leverhulme Trust.

\bibliographystyle{mnras}

\bibliography{all_refs}

\appendix

\section{The Treecode}
\label{sec:appA}
\subsection{The hierarchical approach}

When computing 
the gravitational potential acting on a particle, it is possible 
to ignore the details of the internal structure of a group of 
distant particles.  Replacing several particle-particle 
interactions by one particle-distant group interaction 
is much more efficient than a particle-particle code,
and one can still control the accuracy of the computations.  
This is the key simplification of the hierarchical approach.  

The first codes of this kind were written by 
\citeANP{Appel_81} \citeyear{Appel_81,Appel_85}, \citeN{Jernigan_85}, 
and \citeN{Porter_85}, who were trying to follow 
the spatial distribution of mass as closely as possible.   
\citeN{Barnes_Hut_1} used a simpler 
algorithm, employing an hierarchical decomposition of space, 
allowing them to measure the accuracy of 
the force estimation.  This code was vectorized by 
\citeN{Hernquist_87}, \citeN{Makino_1}, and \citeN{Barnes_90}
and checked on vectorial architectures like Cray X-MP.  
The adaptation for cosmological cases, especially the adding of
periodic boundary conditions (using the Ewald summation method)  
was performed by \citeN{Hernquist_Bouchet_88},
\citeN{Hernquist_et_al_1}, and \citeN{Hernquist_et_al_2}.  

The tree-code uses a multi-polar expansion to compute the 
interaction of a particle and a group of particles.  
We can do this if the size of the group is small enough compared to
the distance of the particle to this group: a variety 
of different criteria are in practice used for that condition.  

This concept is implemented by a hierarchical division of space,
usually called a tree, which, in our case is a binary tree.  
The tree is constructed by starting from the entire box and dividing 
the space along one direction (say $x$ first) in two equal cells.  
Each of those cells is divided along another direction ($y$), 
and this process (division along $x$, $y$, $z$, alternately) 
continues until each cell has one or zero particles inside.  

The computation of the force for one particle starts from the `root' 
(that is the entire simulation box) of the tree.  The selection 
criterion depending on the size of the cell as compared to the 
particle-cell distance is applied, to determine whether we can 
replace the cell by its centre of mass.  If so, we add to the 
`interaction list' the term corresponding to this interaction.  
If not, the same check is done, recursively, on sub-cells, 
until each term of the force is computed.  

This is the basic idea for the `sequential' version of the code.
We now describe the implementation of our parallel approach.  

\subsection{Parallelization}

Our code is written specifically for a T3D or T3E Cray architecture, 
which has a distributed memory type.  Fast Cray libraries like 
\textsc{ShMem} allow us to use shared memory routines,
like remote reads or remote writes, which are rather faster than the 
usual message passing routines of \textsc{PVM} or even \textsc{MPI}.  
The code could be easily translated into \textsc{MPI-2} routines.

At each timestep, a global sort of particles is performed on all
processors.  This global sort is based on four keys, which are 
defined by interleaving the binary digits of the $x$-, $y$-, and 
$z$-positions of particles, in such a way that two physically close 
particles are, most probably, on the same 
processor, except for particles at the borders.  This sorting 
operation thus amounts to a `pre-building' of the tree, and also 
helps to decrease communications between processors during the force 
computation.  The tree is then built, simultaneously on all 
processors.  The binary tree is a shared one, with root on processor 
$0$.  To decrease communications, cells that contain particles on the
same processor will also be on this processor, most of the time;  
that is to say, going down along branches of the tree, 
we find that cells tend to be on the same processors 
as the particles they `control'.

The code also takes into account periodic boundary conditions
by using the Ewald summation method, and copying the grid of 
periodic replicas onto each processor.

The force computation is finally done in a parallel way using the 
shared memory library of the Cray T3E.  Once forces are computed, 
we can derive accelerations, and then update particle
velocities and positions, using a standard integration scheme.

We describe a few finer details of the code below.

\begin{itemize}

\item {\bf Sharing of particles on nodes.  } 
First of all, to balance load, each processor is assigned approximately 
the same number of particles.  More precisely, if $N_{\mathrm{part}}$ 
is the total number of particles, and $N_{\mathrm{procs}}$ the number 
of processors, we perform the Euclidean division:
\[
N_\mathrm{part} = q N_\mathrm{procs} + r
\]
Each processor has at least $q$ particles, and the first $r$ 
processors have an additional one.

\item {\bf Sorting particles.  }
Each particle is then assigned a key, composed of four integer numbers.
Those keys are based on the idea of the bijection between $R^3$ and $R$.
For computers, this is actually a bijection between $N^3$ and $N$,
because of the limited precision of real numbers.  
The building and sorting of particles is essentially based 
on Cantor's construction of this bijection, which we briefly 
review here.  We obtain the first three integers by interleaving the 
binary digits of the three spatial coordinates of the particles, where 
all the coordinates are normalized between 0 and 1.
Let those coordinates be written as, for example:
$x = 0.x_1 x_2 x_3 x_4 x_5 x_6 x_7 x_8 x_9$, 
$y = 0.y_1 y_2 y_3 y_4 y_5 y_6 y_7 y_8 y_9$,  
$z = 0.z_1 z_2 z_3 z_4 z_5 z_6 z_7 z_8 z_9$.  
The first three integer keys will be: 
$k_1 = x_1 y_1 z_1 x_2 y_2 z_2 x_3 y_3 z_3$,
$k_2 = x_4 y_4 z_4 x_5 y_5 z_5 x_6 y_6 z_6$,
$k_3 = x_7 y_7 z_7 x_8 y_8 z_8 x_9 y_9 z_9$.  
The fourth number used is just the original particle number, such that 
in the rare event that two particles have the same position,
they will not have the same keys.

The keys are used to sort particles on the processors in increasing 
lexicographic order.  This is a global sort: after sorting, particles 
on processor number $n+1$ will all have keys greater than those of the 
particles on processor $n$.

\item {\bf Building the tree.  } 
A global binary tree shared on all processors is then built.
The entire box is the root of the tree.  We then divide recursively 
all cells of the tree alternately along the ($x$, $y$, $z$) directions,
until each cell is either empty or contains only one 
particle.  As the tree is shared on all nodes, the top of the tree is 
divided on all processors.  So, the beginning of the tree-building 
process requires a certain amount of communication between them.  
But, as the tree building proceeds, most of the sub-cells of a cell 
will be on the same processor (as the one controlling the cell), 
and this will finally decrease considerably inter-processor 
communication.

\item {\bf Tree properties.  }
We compute the mass and centre of mass positions of each cell.
This is done by starting from the lowest cells in the tree,
just above the particle level.  A synchronisation of the computation 
is made between each level of the tree, because the results of the 
previous level are required to compute the mass
and centre of mass position of this level.

\item {\bf Force computation.  } 
The force computation is done by starting from the top of the tree,
going down recursively until each term of the force has been computed:
each cell is either `opened' or accepted in the interaction list,
if the opening criterion is satisfied.

\item {\bf Particles update.  }
A conventional leapfrog algorithm, taking into account the expansion of the 
Universe, is used to update positions and velocities of particles.

\item {\bf Optimisation.  } 
Several tricks have been used to optimize the code.  Firstly we 
`derecursified' most of the routines of the code.  A copy of the 
top of the tree (the first $N \approx 12$ levels), is also saved 
on each node, thus decreasing communications in the tree descent. 
To decrease cache misses, we also modify the tree, so that the right 
pointer previously used for the `right child' is instead used,
in the force computation, as a `sibling pointer': this modification 
goes with the derecursification of the tree descent.  

\end{itemize}

\section{Halo potential energies}
\label{app:PE}

\begin{figure}
\centerline{\epsfxsize = 8.0 cm \epsfbox{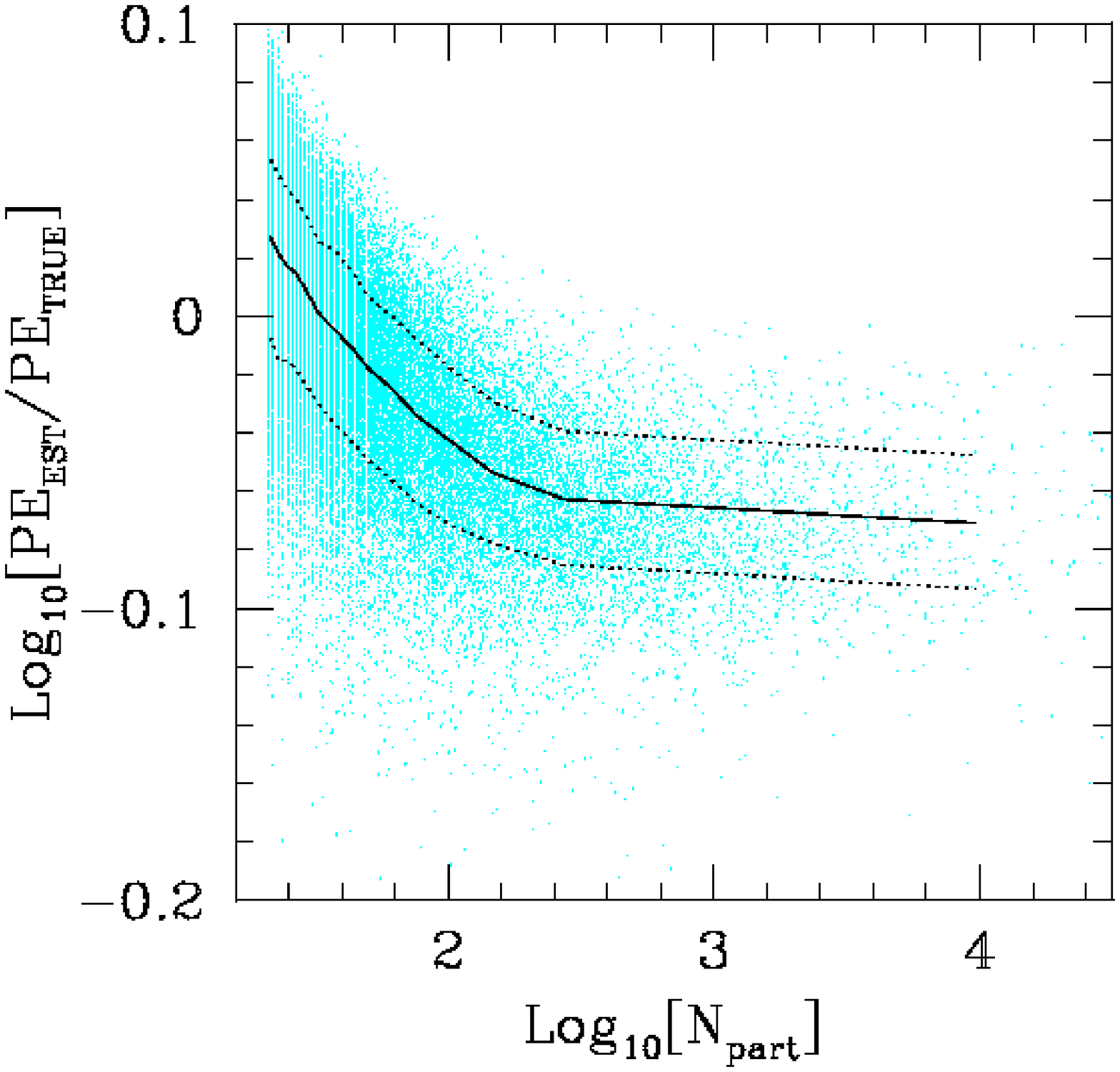}}
\caption{Relationship between the true, measured potential energy from pair counting 
and that found by applying equation~\protect\ref{eqn:PE_ellipoid} to the mass distribution.  
The faint dots show the relation for each halo, while the solid line is the median and the 
dotted lines the scatter at plus and minus one standard deviation. }
\label{fig:PE_compare}
\end{figure}

As mentioned in section~\ref{sec:halo_props}, the potential energy of 
large dark matter halos is computationally expensive using 
a simple sum over pairs.  We compare the measured potential with that 
from applying the formula used by \citeN{SB95} 
for the potential energy of an ellipsoid,  
\[
\label{eqn:PE_ellipoid}
W = - \frac{3}{5} G M^2 R_\mathrm{F} (a^2,b^2,c^2), 
\]
where $R_\mathrm{F}$ is Carlson's form for the elliptic integral of the 
first kind (see \eg \citeNP{NUMREC}), 
\[
R_\mathrm{F} (x,y,z) = \frac{1}{2} \int_0^\infty \frac{dt}{\sqrt{(t+x)(t+y)(t+z)}}.
\] 
This procedure does not correct for the non-inverse square nature of 
the softened gravitational potential in the simulations, but, as shown 
in figure~\ref{fig:PE_compare},  there is a good correlation between 
the true potential energy and the energy estimated in this way.  
We note from this figure that the scatter in the relation is 
fairly small, of order five per cent.  It will also be seen that 
the median depends on the number of particles in the halo.  It 
ranges from being a slight overestimate of the true potential 
energy at low halo mass, to a direct correspondence at around 25 particles, 
before flattening off towards large particle number with an underestimate 
of the magnitude of the potential energy of around fifteen per cent.  

In order to use an unbiased recipe, whilst still speeding up the computation, 
we apply this formula,  
to halos of more than one thousand particles, using the full double-sum method 
for smaller halos.  The potential energies of these halos are then accurate only 
to five per cent.  Since the spin parameter depends on the square root of the 
total energy, this additional scatter will have little effect on the distribution 
of spin parameters.  

It should be noted that this cut-off means that the energies are computed 
properly (\ie without this approximation) for the majority of halos (given the 
mass distribution of Fig.~\ref{fig:halos_spin}), and for almost all the halos that were found 
to have positive total energy (since their mass function has a steeper slope).  

\end{document}